\documentclass[fleqn,usenatbib]{mnras}
   \usepackage[british]{babel}             
   \usepackage{newtxtext}                  
   \usepackage[slantedGreek]{newtxmath}    
   \let\la=\lesssim 
   \usepackage[T1]{fontenc}                

\usepackage{graphics,graphicx,rotating}
\usepackage{natbib}
 \def\emerlin{$e$-MERLIN}
\usepackage{xspace}
\usepackage{amssymb}
\usepackage{amsmath}
\usepackage{epstopdf}
\usepackage{subfigure}
\usepackage{makecell} 
\usepackage{pdflscape}
\usepackage{rotating}
\usepackage{xcolor}
\usepackage{xspace}
\usepackage{amssymb}
\usepackage{amsmath}
\usepackage{epstopdf}
\usepackage{subfigure}
\usepackage{xcolor}
\usepackage{hyperref}
\usepackage[percent]{overpic}
\usepackage{longtable,lscape}
\usepackage{xcolor}
\usepackage{pict2e}
\usepackage{float}
\let\la=\lesssim %
\let\ga=\gtrsim

\begin{document}

\title[LeMMINGs. VI. Nuclear activity and bulge properties]{LeMMINGs. VI. Connecting nuclear activity  to
 bulge properties of  active and inactive  galaxies: radio scaling  relations and galaxy environment}  
  \author[Dullo et al.]{B.\ T.\
 Dullo$,^{1}$\thanks{E-mail: bdullo@ucm.es} J.\ H.\ Knapen$,^{2,3}$ R.\ J.\
 Beswick$,^{4}$\
 R.\ D.\ Baldi$,^{5,6}$\
D.\ R.\  A.\ Williams$,^{4}$\
I.\ M.\ McHardy$^{6}$\
            \newauthor
D.\ A.\ Green$,^{7}$\
A.\ Gil de Paz$,^{1}$\
 S.\ Aalto$,^{8}$
A.\ Alberdi$,^{9}$\
M.\ K.\ Argo$,^{10}$\
 H.-R.\   Kl\"ockner$,^{11}$
I.\ M.\ Mutie$,^{12,4}$\
      \newauthor
     D.\ J.\ Saikia$,^{13}$\
    P.\ Saikia$,^{14}$\
            and
I.\ R.\ Stevens$^{15}$\\
$^{1}$Departamento de F\'isica de la Tierra y Astrof\'isica, IPARCOS, Universidad Complutense de Madrid, E-28040, Madrid, Spain\\
$^{2}$Instituto de Astrof\'isica de Canarias, V\'ia
  L\'actea S/N, E-38205, La Laguna, Tenerife, Spain\\
$^{3}$Departamento de Astrof\'isica, Universidad de
  La Laguna, E-38206, La Laguna, Tenerife, Spain\\
$^{4}$Jodrell Bank Centre for Astrophysics, School of
  Physics and Astronomy, The University of Manchester, Alan Turing
  Building,\\ Oxford Road, Manchester, M13 9PL, UK\\
$^{5}$Istituto di Radioastronomia - INAF, Via P. Gobetti 101, I-40129 Bologna, Italy\\
$^{6}$School of Physics and Astronomy, University of Southampton, Southampton, SO17 1BJ, UK\\
$^{7}$Astrophysics Group, Cavendish Laboratory, 19 J. J. Thomson Avenue, Cambridge CB3 0HE, UK\\
 $^{8}$Department of Space, Earth and Environment, Chalmers University of Technology, 412 96 G\"oteborg,
Sweden\\
$^{9}$Instituto de Astrofísica de Andalucía, IAA-CSIC, Glorieta de la Astronomía s/n, E-18008 Granada, Spain\\
$^{10}$Jeremiah Horrocks Institute, University of Central Lancashire, Preston, Lancashire, PR1 2HE, UK\\
 $^{11}$Max-Planck-Institut für Radioastronomie, Auf dem Hügel 69, 53121 Bonn, Germany\\
$^{12}$Technical University of Kenya,  P.O. box 52428-00200, Nairobi, Kenya\\
$^{13}$Inter-University Centre for Astronomy and Astrophysics (IUCAA), P.O., Post Bag 4,  Ganeshkhind, Pune 411 007, India\\
$^{14}$Center for Astro, Particle and Planetary Physics, New York University Abu Dhabi, PO Box 129188, Abu Dhabi, UAE\\
 $^{15}$School of Physics and Astronomy, University of Birmingham, Birmingham B15 2TT, UK\\
}
\maketitle 
\label{firstpage}
\begin{abstract}
  Multiwavelength studies indicate that nuclear activity and bulge
  properties are closely related, but the details remain unclear.  To
  study this further, we combine \mbox{{\it Hubble Space Telescope} }
  bulge structural and photometric properties with 1.5 GHz,
  \mbox{$e$-MERLIN} nuclear radio continuum data from the LeMMINGs
  survey for a large sample of 173 `active' galaxies (LINERs and
  Seyferts) and `inactive' galaxies (\mbox{H\,{\sc ii}}s and
  absorption line galaxies, ALGs).  Dividing our sample into active
  and inactive, they define distinct (radio core luminosity)--(bulge
  mass), $L_{\rm  R,core}$$-$$M_{*, \rm bulge}$, relations, with a mass turnover at
  $M_{*, \rm bulge}\sim 10^{9.8 \pm 0.3} \rm{ M_{\sun}}$ (supermassive
  black hole mass $M_{\rm BH} \sim 10^{6.8 \pm0.3} \rm M_{\sun}$),
  which marks the transition from AGN-dominated nuclear radio emission
  in more massive bulges to that mainly driven by stellar processes in
  low-mass bulges.  None of our 10/173 bulgeless galaxies host an
  AGN. The  AGN fraction increases with increasing $M_{*,\rm bulge}$ 
  such that  \mbox{$f_{\rm optical\_AGN}\propto M_{*,\rm bulge}^{0.24 \pm 0.06}$}
  and \mbox{$f_{\rm radio\_AGN}\propto M_{*,\rm bulge}^{0.24 \pm 0.05}$}. Between $M_{*,\rm bulge}\sim
10^{8.5}$ and $10^{11.3} \rm
M_{\sun}$,  \mbox{$f_{\rm optical\_AGN}$} steadily rises from $15 \pm 4$  to $80
\pm 5$  per cent.  We find that at fixed bulge mass, the radio loudness, nuclear radio
  activity and the (optical and radio) AGN fraction exhibit no
  dependence on environment.  Radio-loud hosts preferentially possess
  an early-type morphology than radio-quiet hosts, the two types are
  however indistinguishable in terms of bulge S\'ersic index and
  ellipticity, while results on the bulge inner logarithmic profile
  slope are inconclusive.  We finally discuss the importance of bulge
  mass in determining the AGN triggering processes, including
  potential implications for the nuclear radio emission in nearby
  galaxies.

  \end{abstract}


\section{Introduction}

All present-day massive galaxies host a supermassive black hole (SMBH)
at their centre
\citep{1998AJ....115.2285M,1998Natur.395A..14R,2005SSRv..116..523F}.
Theoretical models predict that feedback from an accreting SMBH (i.e.\
active galactic nucleus, AGN), through mechanisms such as radiation
pressure or radio jets (e.g.\ \citealt{2014ARA&A..52..589H}), injects
energy and momentum, which heat or expel the surrounding gas and
suppress star formation in the galaxy bulge\footnote{Traditionally the
  term `bulge' is associated with the spheroidal component of disc
  galaxies but it is used here to refer to the underlying host
  spheroid in case of elliptical galaxies and the spheroidal component
  for lenticular galaxies (S0s), spiral and irregular galaxies.}
component
\citep{1998A&A...331L...1S,2006MNRAS.365...11C,2006ApJS..163....1H,2015ARA&A..53..115K,2017MNRAS.465.3291W}.
This AGN self-regulated mechanism, where the gas removal by the
feedback shuts off the AGN itself, leads to the concurrent growth of
the SMBH and its host bulge and thus naturally explains the tight
correlations observed between SMBH mass ($M_{\rm BH}$) and the velocity dispersion of
bulges \citep{2000ApJ...539L...9F,2000ApJ...539L..13G} and the
luminosity and stellar and dynamical masses of bulges
\citep{1995ARA&A..33..581K,1998AJ....115.2285M,2002MNRAS.331..795M,2003ApJ...589L..21M,2004ApJ...604L..89H}.
AGN feedback is also believed to rapidly drive galaxy colour evolution
from blue to red (e.g.\
\citealt{2006MNRAS.370.1651C,2007ApJ...665..265F,2010ApJ...711..284S,2020ApJ...898...83D}). While
the evolution of the SMBH and its host bulge appear intertwined (e.g.\
\citealt{1998A&A...331L...1S,2000MNRAS.311..576K}), the details of the
regulatory role of the AGN   in building up the bulge stellar mass  and in shaping the stellar distributions of
local bulges are unclear.

Models of galaxy formation based on $\Lambda$ cold dark matter
($\Lambda$CDM) predict that massive local bulges have assembled
hierarchically through galaxy mergers (e.g.\
\citealt{1972ApJ...178..623T,1978MNRAS.183..341W,1984Natur.311..517B,1991ApJ...379...52W}).
Gravitational torques from the mergers convert discs into bulges and
funnel large amounts of gas into galaxy centres
\citep{1989Natur.340..687H,1991ApJ...370L..65B,1996ApJ...471..115B}. This
process is hypothesised to result in growth of the SMBH, AGN activity
and nuclear starburst events, and thus a centrally concentrated bulge.
On the other hand, low mass bulges, which are commonly associated with
spiral and irregular galaxies, may be consequences of dissipational
collapse of accreted gas within dark matter haloes
(\citealt{1980MNRAS.193..189F}). This initial  assembly  is accompanied by redistribution of
their disc material through secular evolution (e.g.\
\citealt{1982ApJ...257...75K,1993IAUS..153..209K}), 
which feeds the central SMBH
and triggers AGN activity (e.g.\
\citealt{1989Natur.338...45S,1995ApJ...454..623K,2011MNRAS.415.1027H,2015MNRAS.448.3442G,2020MNRAS.499.1406L,
  2021ApJ...917...53A}).   
  
Galaxy mergers, episodes of vigorous star
formation and SMBH growth are thought to be common at higher redshift
($z \sim 2.0 \pm 0.5$;
\citealt{2014ApJ...787...38F,2014ARA&A..52..415M}).  The scaled-down
analogies of these processes which are observed in nearby galaxies may
hold crucial clues into the build up of the bulge's stellar mass and
AGN triggering mechanisms: i.e.\ gas-rich or gas-poor mergers (e.g.\
\citealt{2008ApJ...674...80U,2020A&A...637A..94G}) and secular
evolution (e.g.\ \citealt{2011ApJ...726...57C,2013ARA&A..51..511K}).
Nearby galaxies commonly possess low-luminosity AGN (LLAGN), defined
to have H$\alpha$ luminosity $L_{\rm H\alpha} \la 10^{40}$ erg
s$^{-1}$
\citep{1997ApJS..112..315H,1997ApJ...487..568H,1997ApJS..112..391H}.
These are weaker than the powerful AGN which manifest themselves as
bright Seyfert galaxies \citep{1943ApJ....97...28S} and quasars
\citep{2008MNRAS.388.1011M}. LLAGN include low-luminosity Seyferts and
Low-Ionisation Nuclear Emission Line regions (LINERs,
\citealt{1980A&A....87..152H}), which generally display AGN-driven
nuclear activity \citep{1983ApJ...264..105F}, but the latter could be
powered by shocks \citep{1995ApJ...455..468D} and photoionisation from
hot stars \citep{2010MNRAS.402.2187S,2013A&A...558A..43S}. Aside from
LINERs and Seyferts, galaxies may either harbour \mbox{H\,{\sc ii}}
nuclei or be an absorption line galaxy (ALG). \mbox{H\,{\sc ii}}
nuclei are swamped with star formation events, which dominate the
nuclear ionisation processes, whereas the ALG nuclei lack distinctive
emission lines, nonetheless both spectral classes may contain LLAGN
nuclei in the form of a weakly accreting SMBH
\citep{2008ARA&A..46..475H,2014ARA&A..52..589H}.
While LLAGN represent the most common type of AGN in the local Universe
\citep{2006A&A...451...71F,2008ARA&A..46..475H}, they have
largely been overlooked in previous radio studies due in part to the
weak radio emission from the SMBHs, which is particularly true for
those hosted by late-type galaxies.  Radio continuum emission provides
a dust extinction-free constraint on the star formation and AGN
activity in LLAGNs (e.g.\
\citealt{2001ApJ...562L.133U,2005A&A...435..521N,2016MNRAS.458.2221N,2018A&A...616A.152S,2018MNRAS.476.3478B}).

In this work,  our primary objective is to utilise  multiwavelength observations, optimised for studying relations between
the AGN, host bulge properties and environment, for a comprehensive
view of the different evolution scenarios for the host galaxies. How
well we understand the feedback from the AGN and supernova explosions
and stellar winds (e.g.\ \citealt{2011MNRAS.415.3798K}) and the
low-level SMBH accretion activities in nearby galaxies (and by
extension LLAGNs) depends on how far down the radio and optical
luminosity function the data trace (e.g.\
\citealt{2014ARA&A..52..589H,2021MNRAS.508.2019B}).
Observations indicate that AGN feedback is a strong function of the
bulge mass and morphology for nearby galaxies (e.g.\
\citealt{2003MNRAS.341...54K,2005MNRAS.362...25B,2008MNRAS.384..953K,2014ARA&A..52..589H}). Radio
AGN are commonly associated with massive early-type galaxies, while
they are relatively rare in less massive late-type galaxies
\citep{1965ApJ...141.1560S,1995MNRAS.276.1373S,2002A&A...392...53N,2003MNRAS.340.1095D,2005A&A...435..521N,2017ApJ...847...72B,2018MNRAS.476.3478B,2021MNRAS.508.2019B,2021MNRAS.500.4749B}. \citet{2015MNRAS.450.2317S,2021MNRAS.508.2019B} constructed  
a scaling relation between the optical [\ion{O}{iii}] and radio
emission for local galaxies, and  revealed  distinct radio production
mechanisms for the different optical classes and SMBH
masses; AGN-driven sources dominate above
\mbox{$ M_{\rm BH} \sim 10^{6.5} \rm M_{\sun}$}, whereas below this
nominal $M_{\rm BH}$ threshold the nuclear radio emission is
predominantly from stellar processes associated with non-jetted
\mbox{H\,{\sc ii}} galaxies. 

\subsection{Bulge structure, environment and AGN activity   }

Past observational studies typically tie radio loudness to the
presence of two types of early-type bulges, namely `core-S\'ersic' and
`S\'ersic' (e.g.\
\citealt{1997AJ....114.1771F,2003AJ....125.2951G,2013ApJ...768...36D,2014MNRAS.444.2700D,2019ApJ...886...80D,2019ApJ...871....9D}),
with structurally distinct central regions (e.g.\
\citealt{2005A&A...440...73C,2007A&A...469...75C,2006A&A...447...97B,2005A&A...439..487D,2010ApJ...725.2426B,2009ApJS..182..216K,2011MNRAS.415.2158R,2018MNRAS.475.4670D}). Separating
core-S\'ersic and S\'ersic galaxies in the radio scaling relations is
important, as depleted cores are thought to be carved out by binary
SMBHs during the violent `dry' phase of the galaxy assembly
\citep{1980Natur.287..307B,1991Natur.354..212E,2018ApJ...864..113R,2020MNRAS.497..739N,2021MNRAS.502.4794N,2021arXiv210808317D}
and the AGN feedback is posited to play a role in the central
structural dichotomy of early-type galaxies (e.g.\
\citealt{2009ApJS..181..135H,2009ApJS..181..486H}).
Furthermore, observations on the link between environment  and AGN show  heating by episodic radio AGN which 
injects jets from central galaxies into the intra-cluster medium. This
can act to quench a cooling flow in galaxy groups and clusters
\citep{2009ApJ...698..594M,2011ApJ...727...39M,2012ARA&A..50..455F},
reconciling well with the ubiquity of depleted cores in massive
galaxies residing in such environments
\citep{2003AJ....125..478L,2019ApJ...886...80D}.

Whether AGN activity is environmentally driven is, however, currently
under debate.  
Tidal interactions and mergers
of galaxies can trigger the onset of nuclear
activity and/or enhance it by funnelling gas to innermost regions of
galaxies and subsequently onto the central SMBH (e.g.\
\citealt{1988ApJ...328L..35S,1989Natur.340..687H,1994mtia.conf.....S,1991ApJ...370L..65B}). 
In the
nearby universe, the incidence of AGN has been reported to increase
with the local density of the host galaxy \citep{2004MNRAS.353..713K},
while others reported a high incidence of AGN in lower density
environments \citep{2013MNRAS.430..638S,2020AJ....159...69M} or a lack
of significant environmental dependence for the AGN
\citep{2003ApJ...597..142M,2013MNRAS.429.1827P,2019ApJ...874..140A,2019MNRAS.488...89M}.
For further evidence in favour of the
`AGN-density' relation see
\citet{2007MNRAS.379..894B,2009MNRAS.393..377M,2013MNRAS.430..638S,2013MNRAS.430.3086G,2017MNRAS.466.4346M,2019A&A...622A..17S}.

In this work, we combine the results from an {\it HST} imaging
analysis with 1.5 GHz \emerlin\ radio data, allowing for a homogenous
study of the link between nuclear activity and host bulge properties
and environment in a representative sample of 173 active and inactive
nearby galaxies.  The sample covers a wide range in bulge mass,
nuclear activity, morphology and environment. The Legacy \emerlin\ Multi-band Imaging of Nearby
Galaxies Survey (LeMMINGs;
\citealt{2014evn..confE..10B,2021MNRAS.508.2019B,2021MNRAS.500.4749B,2022MNRAS.510.4909W})
is designed to exploit synergies from a large sample of
high-resolution, multiwavelength data (\emerlin\: radio, {\it HST}:
optical plus IR and {\it Chandra}: X-ray).  As part of LeMMINGs, we
have recently published \emerlin\ 1.5 GHz observations of all 280
galaxies above declination, $\delta > +20^{\circ}$ from the Palomar
bright spectroscopic sample of nearby galaxies
\citep{2018MNRAS.476.3478B,2021MNRAS.508.2019B,2021MNRAS.500.4749B}. In
\citet{2022MNRAS.510.4909W}, we presented the {\it Chandra} X-ray
properties for the nuclei of a statistically complete sample of 213
LeMMINGs galaxies. {\it Spitzer} and {\it Herschel} data for the full
sample of LeMMINGs galaxies are currently being analysed (Bendo et
al.\ in prep.). High-resolution optical and near-infrared observations
are desirable to derive accurate central and global galaxy structural
proprieties. In \citet{2023arXiv230311154D}, we
performed multicomponent decompositions of optical/near-IR surface
brightness profiles from {\it HST}, separating bulges, discs, bars,
spiral-arm and nuclear components.

The structure of this paper is as follows.  In Section~\ref{Sec02}, we
describe the LeMMINGs sample and the associated radio and optical
emission line data,  the bulge properties  characterised
and quantified using  {\it HST} imaging data and the 
ancillary ({\it GALEX} UV and {\it Spitzer} 3.6~$\upmu$m) data used in
the paper.  In Section~\ref{Secn3}, we examine how nuclear radio
emissions separated based on optical emission-line types relate to
bulge properties and environment. Section~\ref{Secn4} presents scaling
relations between the radio core luminosity and host bulge properties.
Section~\ref{Dissc} discusses our results in the context of models of
galaxy formation and evolution to provide insights into the AGN
triggering processes and their implications for the nuclear radio
activity.  Finally, in Section~\ref{ConV} we summarise our main
results and conclude.  There are two appendices. Appendix~\ref{EnvR}
gives details of our local density calculations, as a measure of the galaxy environment.
The global and central properties of the sample galaxies are presented
 in Appendix~\ref{DataTables}.

We use $H_{0}$ = 70 km s$^{-1}$ Mpc$^{-1}$, $\Omega_{m}$ = 0.3 and
$\Omega_{\Lambda}$ = 0.7 (e.g.\ \citealt{ 2019ApJ...882...34F}), which
is the average of the Planck 2018 Cosmology $H_{0}$ = 67.4 $\pm$ 0.5
km s$^{-1}$ Mpc$^{-1}$ \citep{2020A&A...641A...6P} and the LMC $H_{0}$
= 74.22 $\pm$ 1.82 km s$^{-1}$ Mpc$^{-1}$\citep{2019ApJ...876...85R}.
While {\it GALEX} UV and {\it Spitzer} 3.6~$\upmu$m data are in AB
magnitude system, other quoted magnitudes in the paper are in the Vega
system, unless specified otherwise.

\begin{table} 
\begin{center}
\setlength{\tabcolsep}{0.01859268281in}
\begin {minipage}{79mm}
\caption{Multiwavelength Data.}
\label{Tab1N}
\begin{tabular}{@{}llcc@{}}
\hline
Data&$N$\\
&(parent sample/this work)\\
 (1)&(2)&\\          
\hline       
 1.5 GHz \emerlin\ radio data$^{\rm [1r]}$   &280/173   \\
Optical spectral classification$^{\rm [1r]}$ &280/173 \\  
{\it HST} data$^{\rm [2r]}$ &173/173 \\                   
{\it GALEX} UV band 
and {\it Spitzer} 3.6~$\upmu$m data$^{\rm [3r]}$&1931/140  \\
\hline      
\end{tabular}     
{\it Notes.} Col (1): multiwavelength data used in this work.  Col (2): number of galaxies ($N$) in the parent sample and in the subsample used in this work. References.
 1r =  \citet{2018MNRAS.476.3478B,2021MNRAS.508.2019B,2021MNRAS.500.4749B}; 2r = \citet{2023arXiv230311154D}; 3r =  \citet{2018ApJS..234...18B}.
 \end{minipage}
\end{center}
\end{table}

\begin{table} 
\begin{center}
\setlength{\tabcolsep}{0.034881in}
\begin {minipage}{83mm}
\caption{LeMMINGs optical and radio properties.}
\label{Tab1}
\begin{tabular}{@{}llcccc@{}}
\hline
Galaxies&Number&Undetected&core-S\'ersic \\
&(our/full sample)&(our/full sample)&(our sample)\\
 (1)&(2)&(3)&(4)&\\          
\hline       
E    &23 (13.2\%)/26(9.3\%)& 52.2\%/50.0\% &65.2\%  \\
S0 & 42 (24.3\%)/55(19.6\%)&   40.5\%/41.8\%& 11.9\%\\                     
S  &102 (59.0\%)/189(67.1\%)&   54.0\%/60.3\% & 0\% \\
\vspace{0.2cm}
Irr  &6 (3.5\%)/10(3.6\%) & 66.7\%/70.0\%& 0\%  \\
Seyfert & 10 (5.8\%)/18 (6.4\%)& 20.0\%/27.8\% &  0\%    \\
ALG &  23 (13.3\%)/28 (10.0\%)   &78.3\%/75.0\%  &34.5\%  \\
LINER &71 (41\%)/94 (33.6\%)   &30.1\%/38.3\%& 15.5\%\\
\vspace{0.2cm}
{\sc h ii}  &69 (39.9\%)/140 (50.0\%) &     67.6\%/66.4\%& 1.7\%\\
Total&173 (100\%)/280 (100\%) &52.0\%/56.1\%&20/173 (11.6\%)\\ 
\hline
\end{tabular}     
{\it Notes.} The sample galaxies are first separated based on the galaxy
morphological and optical spectral classes (cols (1) and (2)) and then
further divided based on their radio non-detection and core-S\'ersic
type central structure (cols (3) and (4)). The term `full sample' refers to
the total LeMMINGs sample of 280 galaxies, whereas the term `our
sample' refers to the sub-sample of 173 LeMMINGs galaxies studied in
this paper.
 \end{minipage}
\end{center}
\end{table}

\section{Data}\label{Sec02}%

All the data used in this work are published  
elsewhere (\citealt{2018MNRAS.476.3478B,2021MNRAS.500.4749B,2021MNRAS.508.2019B,2018ApJS..234...18B,2023arXiv230311154D}; see Table~\ref{Tab1N}).

\subsection{The LeMMINGs }\label{Sec2}

In this work we use a sample drawn from the full LeMMINGs
\citep{2014evn..confE..10B,2018MNRAS.476.3478B,2021MNRAS.500.4749B,2021MNRAS.508.2019B}
sample of 280 nearby galaxies, which in turn is a subset of the
magnitude-limited ($B_{T} \le 12.5$ mag and declinations
$\delta > 0^{\circ}$) Palomar spectroscopic sample of 486 bright
galaxies \citep{1995ApJS...98..477H,1997ApJS..112..315H}.  By design,
all LeMMINGs galaxies have $\delta > +20^{\circ}$, ensuring reliable
radio visibility coverage for the \emerlin\ array.  The LeMMINGs
capitalises on the sub-mJy sensitivity ($\sigma$ $\sim$ 0.08~mJy
beam$^{-1}$) and high angular resolution ($\sim 0\farcs15$) radio
continuum observations of the full LeMMINGs sample taken with
\emerlin\ at 1.5 GHz for a total of 810~h
\citep{2018MNRAS.476.3478B,2021MNRAS.500.4749B}. This legacy survey
constitutes the deepest high-resolution radio study of the local
Universe and aims to investigate AGN accretion and star formation
events for a large sample of nearby galaxies. The full LeMMINGs
sample, which encompasses all the Palomar galaxies with
$\delta > 20^{\circ}$, is statistically complete akin to its parent
Palomar sample.

Full details of the goals, radio data reduction technique, radio
detection and flux measurements of the LeMMINGs can be found in
\citet{2018MNRAS.476.3478B,2021MNRAS.500.4749B}. Here, we use the
radio core properties of the LeMMINGs galaxies including radio
detection, sub-kpc radio structures and radio core luminosities which
span six orders of magnitude
(\mbox{$L_{\rm R,core} \sim 10^{34} - 10^{40}$ erg s$^{-1}$})
\citep{2018MNRAS.476.3478B,2021MNRAS.500.4749B}, as listed in
Tables~\ref{Tab1} and \ref{TableD}.  For the sub-kpc radio structures,
we adopt the classification scheme used in \citet{2018MNRAS.476.3478B}
namely class A = core/core jet, class B = one-sided jet, class C =
triple sources, class D = doubled-lobed and class E = jet + complex
shapes.

Our optical spectral classifications are taken from
\citet{2018MNRAS.476.3478B,2021MNRAS.500.4749B}.  In their
spectroscopic study of the nuclear ionisation mechanisms of the
Palomar galaxies, \citet[][their table~5]{1997ApJS..112..315H} applied
optical emission-line ratios along with spectral classification
criteria to divide the sample into Seyferts, LINERs, \mbox{H\,{\sc
    ii}} and Transition galaxies.  In order to revise the spectral
classification for the LeMMINGs galaxies,
\citet{2018MNRAS.476.3478B,2021MNRAS.500.4749B} used emission-line
ratios taken mainly from \citet[][see also
\citealt{1985ApJS...57..503F,1995ApJS...98..477H,1997ApJS..112..315H,1997ApJ...487..568H,1997ApJS..112..391H}]{1997ApJS..112..315H}
and applied the emission line diagnostic diagrams by
\citet{2006MNRAS.372..961K} and \citet{2010A&A...509A...6B}, see \citet{2023arXiv230311154D}.
 In this revised classification adopted here, the
LeMMINGs galaxies with emission lines were categorised as Seyfert,
LINER and \mbox{H\,{\sc ii}} galaxies, whereas those which lack
emission lines were dubbed  `absorption line galaxies (ALGs)', see Table~\ref{TableD}.

\subsection{Sample and detailed structural analysis with \it{HST}}

Our sample consists of 173 LeMMINGs galaxies (23 Es, 42 S0s, 102 Ss
and 6 Irrs), for which we were able to obtain decent {\it HST} imaging
in the Hubble Legacy Archive (HLA\footnote{\url
  {https://hla.stsci.edu.}}) at the start of this project
 \citep[][their table~1]{2023arXiv230311154D}.  The
aim of this paper is to investigate how the nuclear activity depends
on bulge structural properties utilising sub-arcsec resolution ($\sim$
$5-10$ pc at the mean distance for our sample of $\sim22$ Mpc) optical
and radio properties derived from {\it HST} and 1.5 GHz \emerlin\
radio observations. To achieve this, the bulge must first be isolated
from the rest of the galaxy through detailed photometric
decompositions. \citet{2023arXiv230311154D} used
{\it HST} (ACS, WFPC2, WFC3 and NICMOS) imaging to extract surface
brightness profiles which cover a large radial extent of
$R \ga 80-100\arcsec ( \ga 2R_{\rm e,bulge}$). This enabled our fitted
galaxy models to accurately constrain the shape of the stellar light
distributions associated with the bulges and the outer stellar galaxy
components including bars, discs, rings, haloes and spiral arms. 

We
performed accurate, multicomponent decompositions of the surface
brightness profiles and fitted up to six galaxy components (i.e.\
bulge, disc, partially depleted core, AGN, nuclear star cluster (NSC),
bar, spiral arm, and stellar halo and ring), simultaneously, using
S\'ersic and core-S\'ersic models. The galaxy decompositions
constitute the largest, most detailed structural analysis of nearby
galaxies with {\it HST} to date. Uncertainties on the fitted bulge and
other galaxy parameters were derived by decomposing simulated surface
brightness profiles generated via a Monte Carlo (MC) technique. The
bulge and galaxy structural data used here ($M_{\rm V,bulge}$,
$M_{\rm V,glxy}$, $M_{\rm *,bulge}$, $M_{\rm *,glxy}$,
$\epsilon_{\rm bulge}$, $B_{4,\rm bulge}$ and $\gamma$,
Tables~\ref{Tab1} and ~\ref{TableD}) are tabulated in \citet[][Tables~A1--A4]{2023arXiv230311154D}. We note that the
sample covers over six orders of magnitude in bulge stellar mass
($6 \la \log M_{*, \rm bulge} \la 12.5$) and contains all Hubble types
from Im to E (\citealt{1926ApJ....64..321H,1959HDP....53..275D}).

In order to study the scaling relations between the radio core
luminosity and the optical properties of galaxies, it is crucial to
examine the parameter space coverage afforded by our sample of 173
LeMMINGs galaxies. Motivated by this, in \citet[][section 2]{2023arXiv230311154D} we revealed that our sample, despite being
statistically incomplete, is representative of the statistically
complete, full LeMMINGs sample. We went on to show that the entire
ranges of radio core luminosity and galaxy luminosity probed by the
full LeMMINGs sample are also well traced by our sample.

\subsection{{\it GALEX} NUV, FUV band and {\it Spitzer} 3.6~$\upmu$m data}\label{Sec2.1}

AGN-driven feedback is postulated to be one of the main mechanisms for
the cessation of star formation in galaxies
\citep{1998A&A...331L...1S,2006MNRAS.365...11C,2015ARA&A..53..115K}. The
investigation of AGN activity therefore benefits from locating active
and inactive LeMMINGs hosts on colour$-$colour and colour$-$mass
diagrams. We make use of the \citet{2018ApJS..234...18B} UV and {\it
  Spitzer} 3.6~$\upmu$m magnitudes to construct colour-mass diagrams
for 140 LeMMINGs galaxies (see Section~\ref{Sec35}).
\citet{2018ApJS..234...18B} published total {\it GALEX} NUV, FUV and
{\it Spitzer} 3.6~$\upmu$m asymptotic, AB magnitudes for 1931 nearby
galaxies. There are 140 (85) galaxies in common between their sample
and the full (our {\it HST}) sample of 280 (173) LeMMINGs galaxies.  Following
\citet[][their eqs.\ 1-3]{2018ApJS..234...18B}, we separate these 140
overlapping LeMMINGs galaxies into `red sequence' (RS), `blue
sequence' (BS) and `green valley' (GV) based on the \mbox{(FUV --
  NUV)$-$(NUV $-$ [3.6])} colour $-$ colour diagram (see
Table~\ref{TableD}).  The FUV and {\it Spitzer} 3.6~$\upmu$m
magnitudes were corrected for Galactic extinction but no internal dust
attenuation correction was applied \citep{2018ApJS..234...18B}.

 For
the 55($=140-85$) galaxies which  are in  the full LeMMINGs sample but not in our {\it
  HST} sample, we compute
3.6~$\upmu$m bulge stellar masses ($M_{\rm bulge,3.6}$) to create the
colour--mass diagram (Section~\ref{Sec35}). To do so, we first
calculate total galaxy stellar masses using the 3.6~$\upmu$m
asymptotic AB magnitudes \citep{2018ApJS..234...18B}, distances from
NED, AB absolute magnitude for the Sun from
\citet[][]{2018ApJS..236...47W} of $6.0$ and a 3.6~$\upmu$m
mass-to-light ratio of 0.6 \citep{2014ApJ...788..144M}. The total
galaxy masses were then converted into bulge stellar masses using the
equations listed in \citet[][their table 5]{2023arXiv230311154D}. Because 
we did not separate the bulge component using photometric
decompositions the errors on $M_{\rm bulge,3.6}$ are large, typically
of $\sim$60 per cent.

\section{Nuclear radio emission, optical line emissions and the
  connection with bulge properties and environment
}\label{Secn3}%

In this section, we examine how the nuclear, radio and emission-line
properties for the LeMMINGs sample
\citep{2018MNRAS.476.3478B,2021MNRAS.500.4749B,2021MNRAS.508.2019B}
vary as a function of host bulge properties and environment.  We make
use of sets of bulge structural properties over large stellar mass and
morphology\footnote{To analyse the galaxies properties, we divided the
  sample into two morphological classes, early-type galaxies (Es and
  S0s) and late-type galaxies (Ss and Irrs).} ranges obtained from
accurate modelling of {\it HST} surface brightness profiles for a
sample of 173 LeMMINGs galaxies.  While the study of the dependency of
bulge structural properties on galaxy properties such as morphology is
not new (e.g.\
\citealt{2005MNRAS.362.1319L,2008MNRAS.388.1708G,2012ApJS..198....2K,2015ApJS..219....4S,2017A&A...598A..32M}),
a robust characterisation of galaxy structures using homogeneously
measured high-resolution optical and radio data for a large sample of
galaxies was not possible to date. In particular, our analysis
focusses on the bulge component as it is known to correlate better
with the mass of the SMBH and other central galaxy properties than the
galaxy disc at large radius.

\begin{figure}
\hspace{-3.591312cm}
\includegraphics[trim={-7.25197cm -0cm -5cm 1.69561cm},clip,angle=0,scale=0.48]{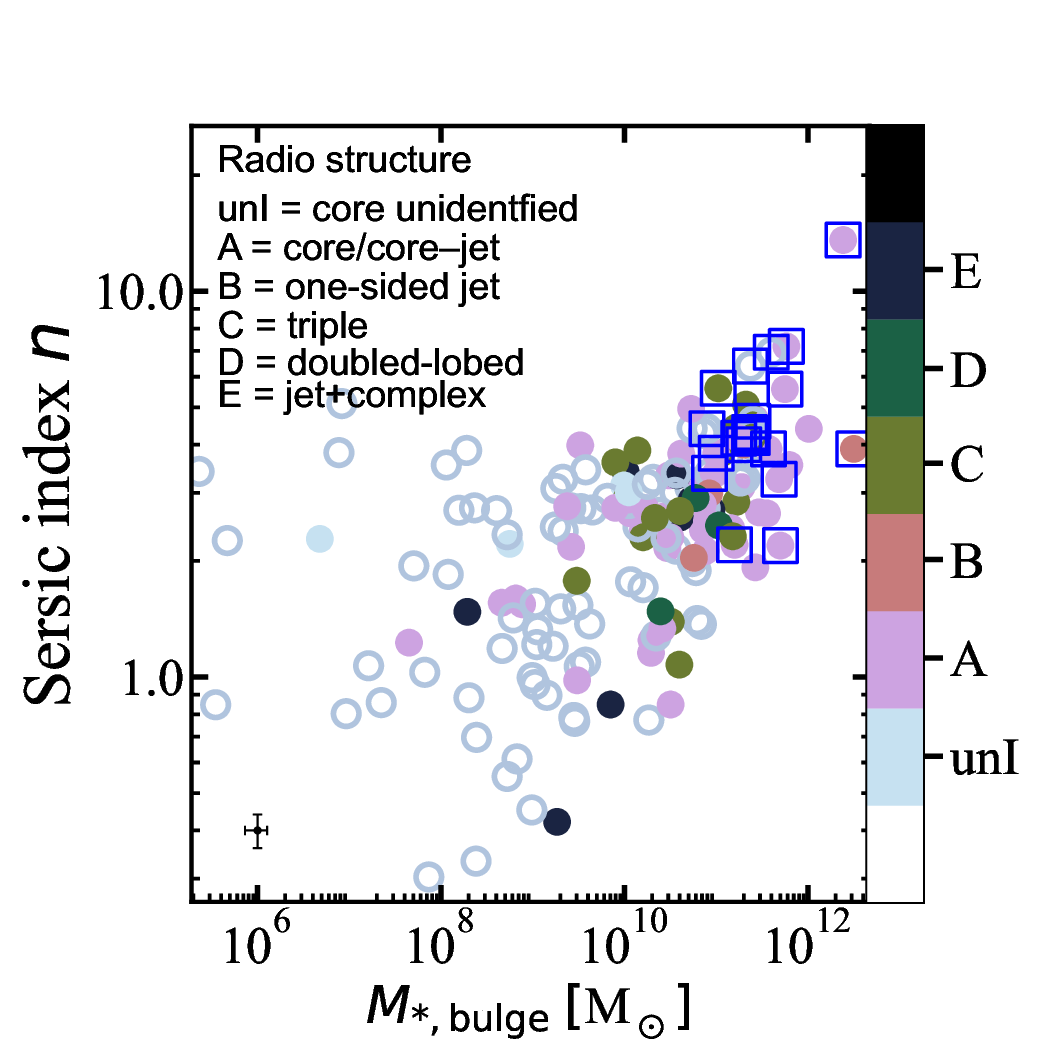}
\vspace{-.404990cm}
\caption{Dependence of radio morphological class \citep{2018MNRAS.476.3478B} on
  the S\'ersic index ($n$) and the stellar mass of the bulge
  ($M_{*,\rm bulge}$), see Table~\ref{TableD}. Filled circles show the
  galaxies in our sample that are radio detected with \emerlin\  at 1.5
  GHz, whereas open circles are for undetected galaxies. The radio
  morphological types of B (`one-sided jet'), C (`triple') and D
  (`doubled-lobed') are indicative of the presence of a jet (see the
  text for details). While radio detection and presence of jets
  strongly depend on  both $n$ and $M_{*,\rm bulge}$, the latter is a
  better predictor.  Core-S\'ersic galaxies (enclosed in blue squares)
  can assume morphologies A, B, C, D and unI, and 7/20 (35  per cent) of them
  are undetected with \emerlin\  at 1.5 GHz. }
  \label{RadA}
\end{figure}

\begin{figure*}
\hspace{-.180352363cm}
\includegraphics[angle=0,scale=0.269]{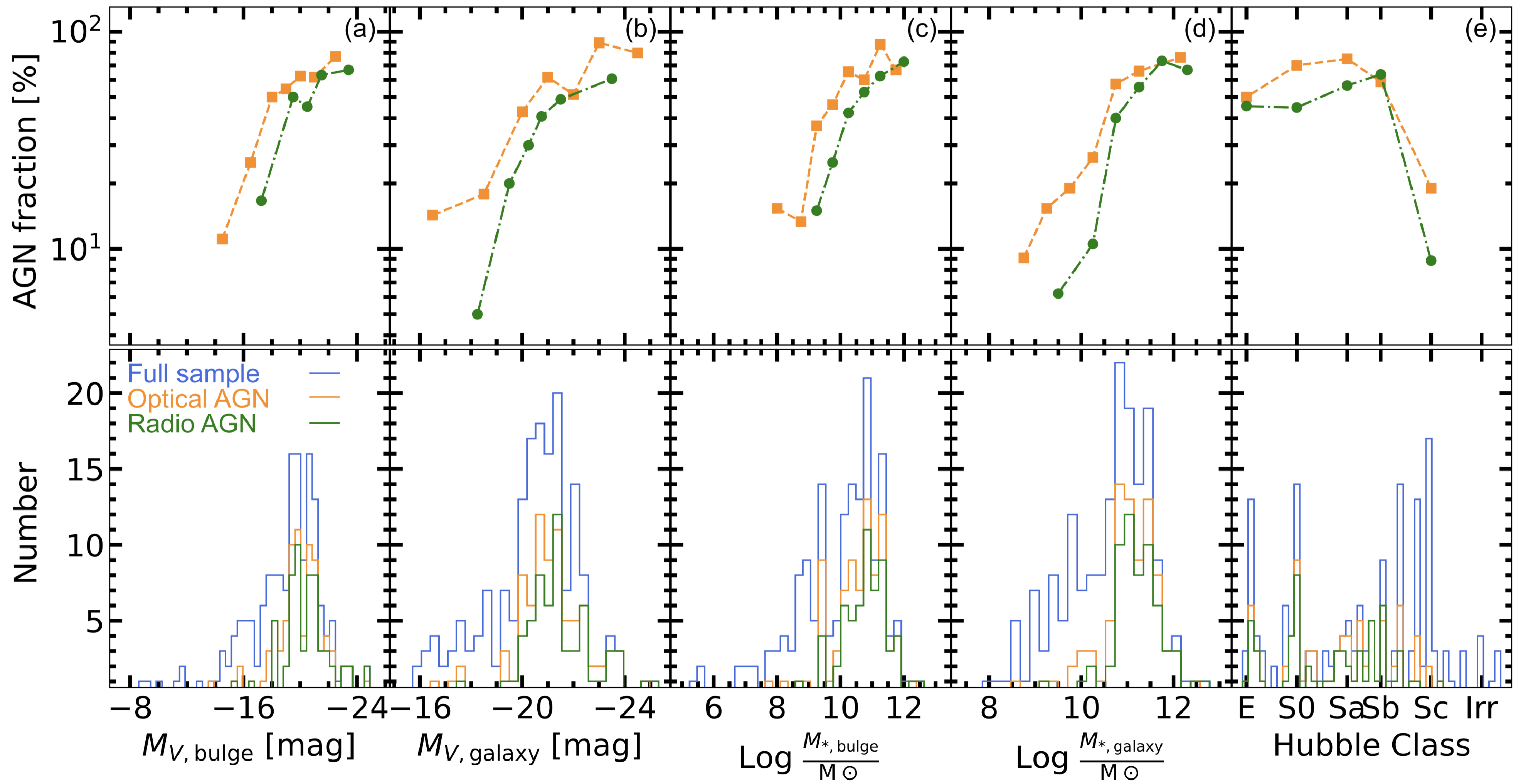}
\vspace{-.44950cm}
\caption{The relations between the AGN fraction and
  host galaxy properties. Top row:  (a) AGN fraction as a function of bulge
  luminosity $M_{V,\rm bulge}$, (b) total galaxy luminosity
  $M_{V,\rm glxy}$, (c) bulge stellar mass $M_{*,\rm bulge}$, 
(d)  galaxy stellar mass $M_{*,\rm glxy}$  and (e) morphology. Bottom
  row: distributions of $M_{V,\rm bulge}$, $M_{V,\rm glxy}$,
  $M_{*,\rm bulge}$, $M_{*,\rm glxy}$ and morphology for AGN and
  non-AGN hosts. The galaxies referred here to as `optical AGN' (orange curve) are
  LINERs and Seyferts and those referred to as `radio AGN' (green
  curve) are jetted objects with radio morphologies B (`one-sided
  jet'), C (`triple') and D (`doubled-lobed') and radio detected LINERs and Seyferts.}
  \label{Figure3}
\end{figure*}

\subsection{The relations between radio detection, radio
  morphology and bulge properties}\label{Sec3.1}

Fig.~\ref{RadA} shows the relation between the S\'ersic index and
bulge mass ($n-M_{*,\rm bulge}$) colour-coded by galaxy radio
morphology.  Of the 173 sample galaxies, we detect radio emission
$\gtrsim 0.2~\rm {mJy}$ from 83 galaxies (filled circles) with
\emerlin\ at 1.5 GHz. This gives a detection rate of 48 per cent, in
fair agreement with that from the full sample (44 per cent).  Of the
83 galaxies, five (NGC~3034, NGC~3838, NGC~4242, NGC~5273 and
NGC~5907) are radio-detected but core-unidentified; they have low
S\'ersic indices ($n \la 3$), faint bulge magnitudes
($M_{V,\rm bulge} \ga -18.9$ mag) and low stellar bulge masses
($M_{*,\rm bulge} \la 10^{10} \rm M_{\sun}$).  The remaining 90
undetected sources are denoted by open circles.  The radio detection
fraction increases with bulge mass, $n$ and bulge-to-total flux ratio
($B/T$). At $M_{*,\rm bulge} \ga 10^{11} \rm M_{\sun}$, this fraction is 77 per
cent, but it declines to 24 per cent for
$M_{*,\rm bulge} < 10^{10} \rm M_{\sun}$.  For  $M_{*,\rm bulge} \ga 10^{11}
 \rm M_{\sun}$ the mean values of $n$ and $B/T$ are $n = 4.00 \pm 1.50$,
  $B/T = 0.79 \pm 0.24$, while we measure   mean values of  $n = 1.52 \pm 1.06$,
$B/T = 0.10 \pm 0.10$  for 
$M_{*,\rm bulge} < 10^{10} \rm M_{\sun}$.  Large $M_{*,\rm bulge}$ and high $n$ values
however are not strictly associated with radio detection at \emerlin\
sensitivity. Most massive undetected sources are ALGs, the second most
common type being LINERs.

The radio morphological types of B (`one-sided jet'), C (`triple') and
D (`doubled-lobed') are indicative of the presence of a radio jet
\citep{2018MNRAS.476.3478B}. It is evident that radio jetted sources
are among the more massive galaxies wtih a mass range of
$M_{*,\rm bulge} \sim 3.0 \times10^{9} -3.2\times10^{12} \rm M_{\sun}$
and a median mass of
log $M_{*,\rm bulge}/\rm M_{\sun} = 10.8 \pm 0.6$. Radio
morphology type A (`core/core-jet') galaxies tend to be massive\footnote{Note that over the  mass range of
$M_{*,\rm bulge} \sim 3.0 \times10^{9} -3.2\times10^{12} \rm M_{\sun}$
  traced by our jetted galaxies, we find a median  mass for  type A galaxies  of log $M_{*,\rm bulge}/M_{\sun} = 10.9 \pm 0.7$).}
 (i.e.\ a median mass of
log $M_{*,\rm bulge}/M_{\sun} = 10.7 \pm 1.0$) but
they trace a wide range of $n$ ($\sim 1-13$) and mass
($M_{*,\rm bulge} \sim10^{7} -3\times10^{12} \rm M_{\sun}$). While
radio morphology types A and E \citep{2018MNRAS.476.3478B} do not
guarantee a jet with \emerlin\ at 1.5 GHz, the most massive
`core/core-jet' galaxies are probably jetted (see
Table~\ref{TableD}). Core-S\'ersic galaxies (enclosed in blue
squares) can assume all radio morphologies except E, and 7/20 (35 per
cent) of them are undetected with \emerlin\ at 1.5 GHz.

\subsection{The incidence of optical (emission-line) and Radio AGN
  as a function of stellar mass, luminosity and Hubble type
}\label{AGNFT}%

The motivation here is to see how the AGN fraction varies with the
host galaxy photometric properties.  Due to the high radio detection
and radio jet incidence for more massive bulges
(Section~\ref{Sec3.1}), a correlation is expected also between the AGN
fraction and the bulge mass, bulge magnitude and other related galaxy
properties.  Fig.~\ref{Figure3} shows the distributions of bulge
luminosity ($M_{V,\rm bulge}$), total galaxy luminosity
($M_{V,\rm glxy}$), bulge stellar mass ($M_{*,\rm bulge}$), galaxy
stellar mass ($M_{*,\rm glxy}$) and Hubble type for AGN and non-AGN
hosts. Galaxies with LINERs and Seyferts nuclei are active and simply
referred to as `optical AGN' (orange curve), while ALGs and \mbox{H\,{\sc ii}},
which are considered inactive, constitute the non-AGN subsample.
Galaxies which are dubbed as `radio AGN' (green curve) are limited to
those with radio morphologies B (`one-sided jet'), C (`triple') and D
(`doubled-lobed') and radio detected LINERs and Seyferts, see
Table~\ref{TableD}.  We note that the bulk (87 per cent) of the sample
galaxies with radio morphologies B, C and D are radio detected LINERs
and Seyferts. All the `Radio AGN' galaxies in the sample have AGN-like
spectral emission except for three inactive galaxies (NGC~3348,
NGC~3665 and NGC~4217).

Having defined the AGN fraction as the ratio between the number of
galaxies with an AGN and the total number of galaxies under
consideration, there is strong evidence for correlation between the
optical AGN fraction and bulge stellar mass and luminosity
(Fig.~\ref{Figure3}). Using the symmetrical ordinary least squares
(OLS) bisector regression \citep{1992ApJ...397...55F}, we find the
fraction of optical AGN galaxies is such that
$f_{\rm optical\_AGN}\propto M_{*,\rm bulge}^{0.24 \pm 0.06}$ and
$f_{\rm optical\_AGN}\propto M_{*,\rm glxy}^{0.30 \pm 0.05}$.  The majority 
of optical AGN (80  per cent) and radio AGN (90  per cent) hosts have bulges 
more massive than $M_{\rm bulge}$ $\sim 10^{10} \rm M_{\sun}$ ($M_{\rm
  glxy} \sim 10^{10.5} \rm M_{\sun}$) and brighter than $M_{V, \rm bulge}
\sim -18.2$ mag ($M_{V,\rm glxy} \sim
-20.0$ mag) (see Fig.~\ref{Figure3}).  From a bulge magnitude of $
-15.5$ to $
-18$ mag, the optical AGN fraction increases dramatically from
\mbox{$22 \pm 5$} to \mbox{$50 \pm 4$  per cent}, before rising gently to
$77 \pm 3$  per cent from $M_{V, \rm bulge} \sim
-18$  to $-22.5$ mag.  Between bulge mass of $M_{*,\rm bulge}\sim
10^{8.5}$ and $10^{11.3} \rm
M_{\sun}$, the optical AGN fraction steadily rises from $15 \pm 4$  to $80
\pm 5$  per cent. In general, the trend of the optical AGN fraction for the bulge 
mirrors that of the host galaxy.  We note that of the 32 galaxies in
the sample having $M_{*, \rm glxy} \la 5 \times 10^{9} \rm
M_{\sun}$ and thus defined to be low-mass by
\citet{2018MNRAS.476..979P} only three  (NGC~404, NGC~1058 and NGC~3982) host an
active AGN, yielding an optical AGN fraction for this low-mass domain of
9.4  per cent, which is comparable to the 10  per cent reported by \citet[][see also
\citealt{2019MNRAS.489L..12K,2019MNRAS.488..685M}]{2018MNRAS.476..979P}
at the same stellar mass range.

As with the optical AGN fraction, the `radio AGN' fraction
appears to be   a function of stellar mass and luminosity. The fraction of radio
AGN galaxies is such that
$f_{\rm radio\_AGN}\propto M_{*,\rm bulge}^{0.24 \pm 0.05}$ and
$f_{\rm radio\_AGN}\propto M_{*,\rm glxy}^{0.41 \pm 0.06}$. 
From a bulge magnitude of $-17$  to $
-19.5$ mag, the  `radio AGN' fraction increases  strongly from
\mbox{$17 \pm 5$} to \mbox{$50 \pm 4$  per cent} and then  rising gently to
$60 \pm 5$  per cent from $M_{V, \rm bulge} \sim
-19.5$  to $-23.0$ mag. From $M_{*,\rm bulge} \sim10^{10}$ to
$10^{11.7} \rm M_{\sun}$, the `radio AGN' fraction increases from of
$32 \pm 7$  to $68 \pm 5$  per cent, whereas when $M_{*, \rm glxy}$
is considered over a similar mass range the `radio AGN' fraction rises
from $3 \pm 3$   to $25 \pm 5$  per cent. 

\begin{figure*}
\hspace{2.1528cm}
\includegraphics[trim={.05538050087cm -8cm -2.965591cm 1.099443cm},clip,angle=0,scale=0.71]{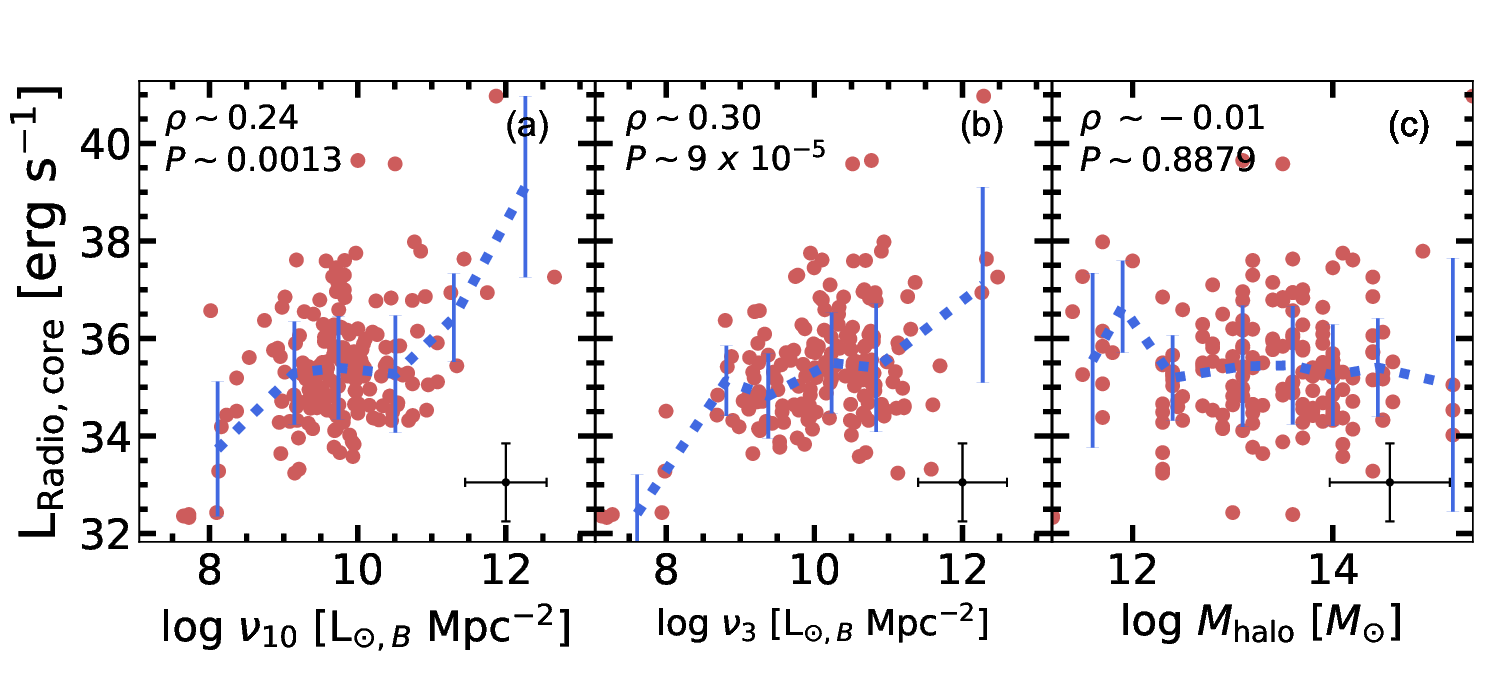}
\vspace{-6.41740cm}
\caption{Radio core luminosity for our sample of 173 galaxies
  ($L_{\rm R,core}$) plotted as a function of $B$-band luminosity
  surface density inside a cylinder enclosing the 10th nearest galaxy
  neighbour $\nu_{10}$ (a), $B$-band luminosity surface density for a
  cylinder containing 3 nearest neighbour $\nu_{3}$ (b) and halo mass
  $M_{\rm halo}$ (c).  The blue curve indicates the median
  $L_{\rm R,core}$ values for the corresponding local density and halo
  mass bins.  The error bars show the 1$\sigma$ error on the
  measurements.  Overall there is a mild tendency of the median
  $L_{\rm R,core}$ values to increase with $\nu_{10}$ and $\nu_{3}$
  but no relation of $L_{\rm R,core}$ with halo mass. However,
  removing the six outlying sample data points with high \mbox{
    $\nu_{3}$ (> $10^{12}$ ${\rm L}_{\sun,B}$ Mpc$^{-2}$}) and low
  \mbox{$\nu_{3}$ (< $10^{8}$${\rm L}_{\sun,B}$
    Mpc$^{-2}$}), the mild tendency disappears and the correlations
  between $L_{\rm R,core}$ and
  $\nu$ become very weak and no correlation is seen between $L_{\rm
    R,core}$ and $M_{\rm
    halo}$: Spearman's correlation coefficients and probabilities for
  the full smaple are (a) $\rho \sim 0.13, P \sim 0.0924$, (b) $\rho
  \sim 0.20, P \sim 0.01231$ and (c) $\rho \sim -0.07, P \sim
  0.3811$. }
  \label{LcoreENV}
\end{figure*}

We also checked the `radio AGN' fraction by restricting the `radio
AGN' subsample to contain only jetted galaxies with radio morphologies
B, C and D.  We find that the `jetted' radio AGN fraction is a
function of stellar mass and luminosity such that $f_{\rm
  jetted\_radio\_AGN}\propto M_{*,\rm bulge}^{0.17 \pm
  0.05}$ and $f_{\rm jetted\_radio\_AGN}\propto M_{*,\rm glxy}^{0.18
  \pm 0.06}$.  From $M_{*,\rm bulge} \sim10^{10}$ to $10^{11.3} \rm
M_{\sun}$, the jetted radio AGN fraction increases from of $10 \pm
5$ to $30 \pm 5$ per cent, while for $M_{*, \rm
  glxy}$ considered over a similar mass range the jetted radio AGN
fraction rises from $3 \pm 3$ to $25 \pm
5$ per cent.  Owing to the low number statistics associated with the
jetted radio AGN galaxies, we remind the need for caution when
interpreting the aforementioned results. Furthermore, the quoted
jetted radio AGN fractions are probably a lower limit, since massive
galaxies with radio morphologies A and E might host a radio jet and
inactive galaxies might have a weak AGN cloaked at their centres.

Although the trend of increasing radio AGN activity with increasing
stellar mass has been previously reported (e.g.\
\citealt{2003MNRAS.341...54K,2005MNRAS.362...25B,2008MNRAS.384..953K,2014ARA&A..52..589H}),
our relation
($f_{\rm radio\_AGN}\propto M_{*,\rm glxy}^{0.41 \pm 0.06}$) is
significantly shallower than the (radio-loud AGN fraction)-(galaxy
stellar mass) relation found by
\citet[][$f_{\rm Radio-loud\_AGN}\propto M_{*,\rm
  glxy}^{2.5}$]{2005MNRAS.362...25B}. The discrepancy between the
slopes from our work and \citet{2005MNRAS.362...25B} may be due to the
higher sensitivity and higher angular resolution of our $e$-MERLIN
radio data which allow for the  identification of jetted, radio-quiet
AGN. Note that the bulk of the outflows of ionised matter in a  jet can 
 be either relativistic, giving  rise to a radio-loud AGN, or sub-relativistic,
  resulting  in a radio-quiet AGN. Also shown in Fig.~\ref{Figure3}  are 
  trends with optical morphological classes; we
notice a variation in the AGN fraction across the Hubble sequence:
47.8  per cent of elliptical galaxies, 64.3  per cent of S0s and 42.2  per cent of spiral
galaxies (75  per cent of Sa, 54.6  per cent of Sb, and 20.5  per cent Sc) are optical AGN.

\begin{figure*}
\hspace{-.029886cm}
\includegraphics[trim={.35965591cm -3.3cm -2.09991cm .040043cm},clip,angle=0,scale=0.382599081]{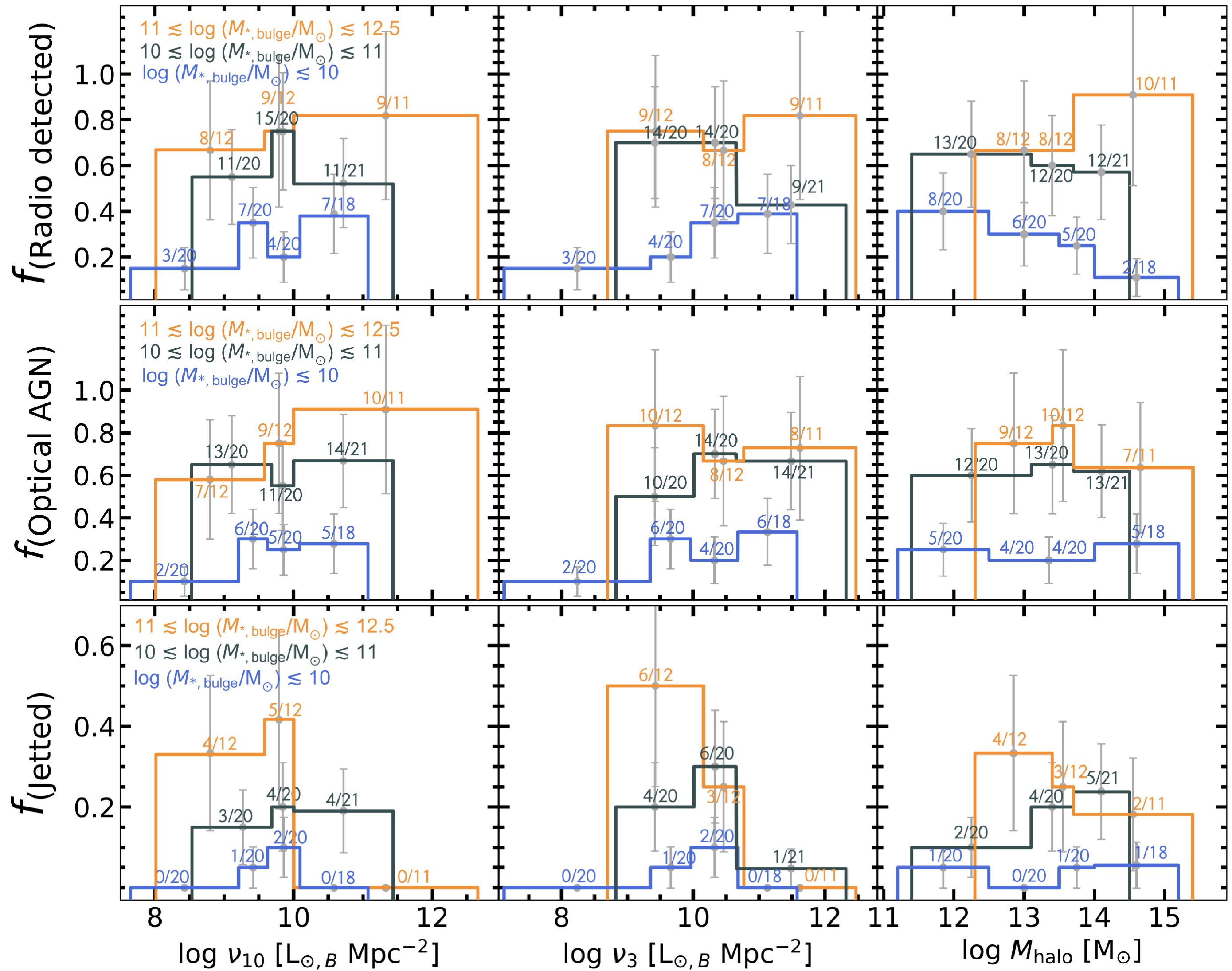}
\vspace{-1.5899cm}
\caption{Radio detection, optical emission-line AGN and radio AGN
  fractions ($f_{(\rm Radio~detection)}$, $f_{(\rm Optical~AGN)}$ and
  $f_{(\rm Radio~AGN (Jetted))}$, respectively) are plotted as a
  function of luminosity surface densities ($\nu_{10}$ and $\nu_{3}$)
  and halo mass $M_{\rm halo}$ colour-coded by three (high-,
  intermediate- and low-)bulge mass ($M_{*,\rm bulge}$) bins. The
  error bars denote the Poisson errors. Within each $M_{*,\rm bulge}$
  bin considered, the radio detection and optical AGN fractions vary only
  mildly with $\nu_{10}$, $\nu_{3}$ and $M_{\rm halo}$. The only
  notable exception is the decrease in $f_{(\rm Jetted)}$ with
  increasing $\nu_{3}$ for the massive bin (bottom row, middle
  panel). }
 \label{AGNENV}
 \end{figure*}

 \subsection{Dependence of nuclear radio emission and AGN fraction on
   environment} %
   
There are conflicting reports in the literature about the role of
galaxy environment in triggering AGN activities and nuclear radio
emissions (e.g.\ \citealt{2019MNRAS.488...89M}). This motivates us to
investigate the environmental dependence of nuclear activities for the
LeMMINGs galaxies.  We apply two methods to measure the galaxy
environments: local densities based on the nearest neighbour method
and dark matter halo mass.  Measures of the environment based on the
nearest neighbours method are shown to approximate well the internal
properties of the dark matter halo \citep{2012MNRAS.419.2670M}.  For
each galaxy in the sample, we derive local surface densities ($\rho$)
and luminosity surface densities ($\nu$) inside a cylinder containing
the 3rd and 10th nearest neighbours with $M_{B} \la -18.0$ mag and a
velocity cut of \mbox{$\upDelta V_{\rm hel} <$ 300 km s$^{-1}$}
(denoted by $\rho_{3}$, $\nu_{3}$ and $\rho_{10}$, $\nu_{10}$,
respectively, see  Table~\ref{TableD}, also
\citealt{2011MNRAS.416.1680C}). A detailed discussion of the
derivations of these environmental measures and their comparison with
those from \citet{2011MNRAS.416.1680C} are available online. Given the strong correlations we find  between $\rho$ and
$\nu$ (Spearman's correlation coefficient
$r_{s} \sim 0.93-0.96, P \sim$ 0),
throughout this paper we only consider the luminosity surface
densities $\nu_{10}$ and $\nu_{3}$ (see Table~\ref{TableD}). We
determine that the typical uncertainty on $\log \nu_{10}$ and
$\log \nu_{3}$ is 0.60 dex.

For the halo mass ($M_{\rm halo}$), we follow
\citet{2017MNRAS.471.1428V} and use the halo mass based on the
projected mass estimator \citep{1985ApJ...298....8H} from the
\citet{2007ApJ...655..790C} high density contrast (HDC) group
catalogue which provides halo mass estimates quantified based on the
two different methods for groups with at least three members (see
Table~\ref{TableD}).

\subsubsection{Radio core luminosity against   
local environment density and halo mass }\label{RcRD}%

We examine here whether the radio core luminosities ($L_{\rm R,core}$)
for our sample galaxies correlate with the local environment measures
and halo masses. Fig.~\ref{LcoreENV} shows $L_{\rm R,core}$ plotted
against (a) the $B$-band luminosity surface density inside a cylinder
enclosing the 10th nearest neighbour $\nu_{10}$, (b) the same density
but now inside a cylinder including the 3rd nearest neighbour
$\nu_{3}$ and (c) the halo mass $M_{\rm halo}$. Note that a galaxy
that inhabits a dense region, i.e.\ a galaxy located close to its
neighbours, has high values of $\nu_{3}$ and$\nu_{10}$.  Overall, we
find the median radio core luminosities for our sample appear to
increase slightly as a function of $\nu$, but our galaxies reside in
all environments, regardless of their radio core luminosities. 

There
is a weak correlation between $L_{\rm R,core}$ and $\nu_{10}$
($r_{s} \sim 0.24$, $P \sim 0.0013$), a marginally significant
correlation between $L_{\rm R,core}$ and $\nu_{3}$ ($r_{s} \sim 0.30$,
$P \sim 9\times10^{-5}$) and no correlation between $L_{\rm R,core}$
and $M_{\rm halo}$ ($r_{s} \sim -0.01$, $P \sim 0.8879$). Excluding
the six outlying sample data points, i.e.\ 3 with high \mbox{
  $\nu_{3}$ (> $10^{12}$ ${\rm L}_{\sun,B}$ Mpc$^{-2}$}) and 3 low
\mbox{$\nu_{3}$ (< $10^{8}$${\rm L}_{\sun,B}$
  Mpc$^{-2}$}), see the middle panel, which seem to be driving the
relations in Fig.~\ref{LcoreENV}a, b, we rerun the analysis, finding
only weak correlations between $L_{\rm R,core}$ and $\nu$ ($r_{s} \sim
0.13-0.20, P \sim 0.01231-0.0924$). For $L_{\rm R,core}$ and $M_{\rm
  halo}$ the relation is consistent with the null hypothesis of no
correlation ($r_{s} \sim -0.07$, $P \sim 0.3811$).

\begin{figure*}
\hspace{-.57299019993cm}
\includegraphics[trim={.36851856730560994871cm -3.05cm -.9709cm 01.403508043639943cm},clip,angle=0,scale=0.5100]{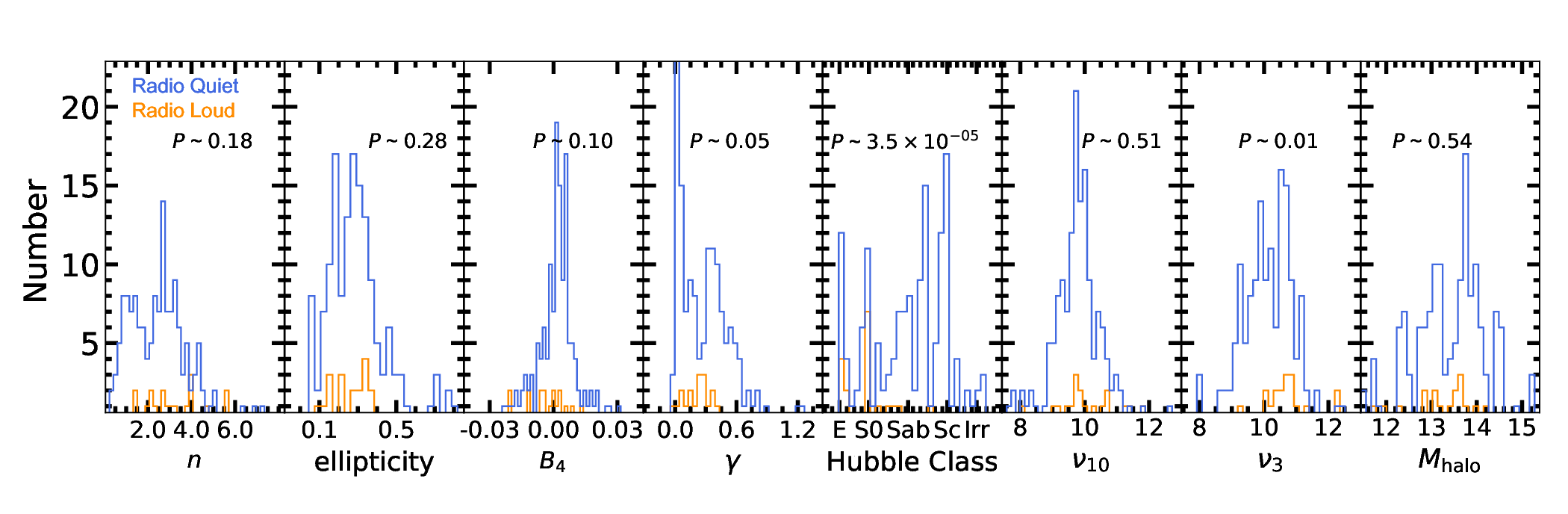}
\vspace{-2.43720cm}
\caption{Properties of radio-quiet (RQ, blue curve) and radio-loud
  (RL, orange curve) sample galaxies. We show the distributions of the
  S\'ersic index ($n$), ellipticity, isophote shape parameter
  $B_{4}$, negative, inner logarithmic slope
  ($\gamma$), Hubble type, luminosity surface densities within
  cylinders enclosing the 10th and 3rd nearest neighbours
  ($\nu_{10}$ and $\nu_{3}$) and halo mass ($M_{\rm
    halo}$). We run two-sample Kolmogorov-Smirnov (KS) tests on the
  cumulative distributions of the RQ and RL datasets and show the
  corresponding
  $P$-values. For the galaxies' Hubble type and
  $\nu_{3}$, we can reject the null hypothesis that the RQ and RL
  datasets are drawn from the same distributions as the
  $P$-values are $< 0.05$. While the comparison of the
  $\gamma$ distributions for RQ and RL galaxies yields a borderline
  case with $P \sim 0.05$, the null hypothesis cannot be rejected ($P
  > 0.05$) when KS tests are performed on the remaining properties of RQ
  and RL galaxies ($n$, ellipticity, $B_{4}$, $\nu_{3}$ and $M_{\rm
    halo}$).  }
 \label{RQRL}
\end{figure*}

\subsubsection{(Radio detection)-density and AGN-density relations in
  bins of fixed bulge stellar
  mass}\label{RDAGND}%

The incidence of radio activity and AGN are strong functions of bulge
mass $M_{\rm bulge}$ (Figs.~\ref{RadA} and \ref{Figure3}). Studies
have also shown that bulge stellar mass depends on the host galaxy
environment, with large numbers of massive, early-type (non-star
forming) galaxies preferentially in higher density local environments
\citep{1984ARA&A..22..185D,2003ApJ...584..210G,2003MNRAS.346.1055K,2004MNRAS.353..713K,2006MNRAS.373..469B,2010ApJ...721..193P},
see also Appendix A1.  As such (radio
detection)-density and AGN-density relations may be driven by
$M_{\rm bulge}$. Therefore in Fig.~\ref{AGNENV} we separate our sample
into three (high-, intermediate- and low-bulge) mass bins and examine
how the radio detection, optical and radio AGN
fractions vary as a function of environment as measured by $\nu_{10}$,
$\nu_{3}$ and $M_{\rm halo}$. Within each fixed $M_{\rm bulge}$ bin,
we find no obvious dependency of the radio detection and AGN fraction
on environment across the sample, regardless of the method we apply to
define galaxy environment. For most of the cases in Fig.~\ref{AGNENV},
the radio detection and (both optical and radio) AGN fraction remain
unchanged over the five (four) orders of magnitude in $\nu_{10}$ and
$\nu_{3}$ ($M_{\rm halo}$) probed by our sample.  In general, the
environments of the active galaxies in the sample are
indistinguishable from those of inactive ones. Our results are in
agreement with some past work
\citep{2003ApJ...597..142M,2013MNRAS.429.1827P,2019ApJ...874..140A,2019MNRAS.488...89M},
albeit see conflicting claims in the literature (e.g.\
\citealt{2004MNRAS.353..713K}).

\subsection{Radio loudness }\label{Sec735}

In order to gain further insight into the nuclear radio emissions in
our galaxies, we divide them into radio-loud (RL) and radio-quiet (RQ)
galaxies.  Following \citet{2021MNRAS.508.2019B}, in  \citet{2023arXiv230311154D}
 we classified galaxies as RL when log
($L_{\rm core}/L_{\rm {[\ion{O}{iii}]}}$) > $-$2 and log
($M_{\rm BH}/\rm M_{\sun}$) > 7.7. This yielded 155 RQ and 18 RL galaxies
in our sample; 17 of the 18 RL galaxies are LINERs and one is a
(jetted) \mbox{H\,{\sc ii}} galaxy NGC~3665 (Table~\ref{TableD}).

In Fig.~\ref{RQRL}, we present comparisons of the distributions of the properties of RQ and RL galaxies: S\'ersic index ($n$), ellipticity,
isophote shape parameter $B_{4}$, negative, inner logarithmic slope
($\gamma$), Hubble type, $\nu_{10}$, $\nu_{3}$ and $M_{\rm halo}$. We
run several two-sample \mbox{Kolmogorov--Smirnov} (KS) tests on the cumulative
distributions of the RQ and RL datasets. When compared to RQ galaxies, we find that RL galaxies tend
to posses an early-type morphology  and inhabit a denser environment as defined by $\nu_{3}$. 
For the Hubble type and $\nu_{3}$, we can reject the null
hypothesis that the RQ and RL datasets are drawn from the same
distributions at  $P$-values of $< 0.05$.  The evaluation of the
$\gamma$ distributions for RQ and RL galaxies with $P \sim 0.05$ is a
borderline case, although there is a mild tendency for RL galaxies to
have shallower inner slopes than RQ galaxies.  Having performed KS
tests on the remaining galaxy properties, we cannot reject the null
hypothesis that the RQ and RL galaxies are drawn from identical
distributions in $n$, ellipticity, $B_{4}$, $\nu_{10}$ and
$M_{\rm halo}$ ($P$-values $\sim 0.10-0.54$).
  
Note that the only significant trend between nuclear activity and
environment from our work is the tendency for RL galaxies to have high
values of $\nu_{3}$, thus they preferentially reside in the highest
local densities, however other environmental measures ($\nu_{10}$ and
$M_{\rm halo}$) do not corroborate the trend (see
Fig.~\ref{RQRL}). Also, we have not ruled out whether the bulge mass
is driving the trend between radio loudness and $\nu_{3}$.  In
summary, our results again suggest AGN and radio loudness  are not
environmentally driven and thus agree with what we found in Sections~\ref{RcRD} and \ref{RDAGND}.
 
\subsection{Active and inactive  galaxies across the   \mbox{(FUV--[$3.6$]})-$M_{*,\rm bulge}$ plane}\label{Sec35}

\begin{figure*}
\hspace{-.5cm}
\includegraphics[trim={-.0296cm -1.3cm -1.9cm 4.25 cm},clip,angle=0,scale=0.53]{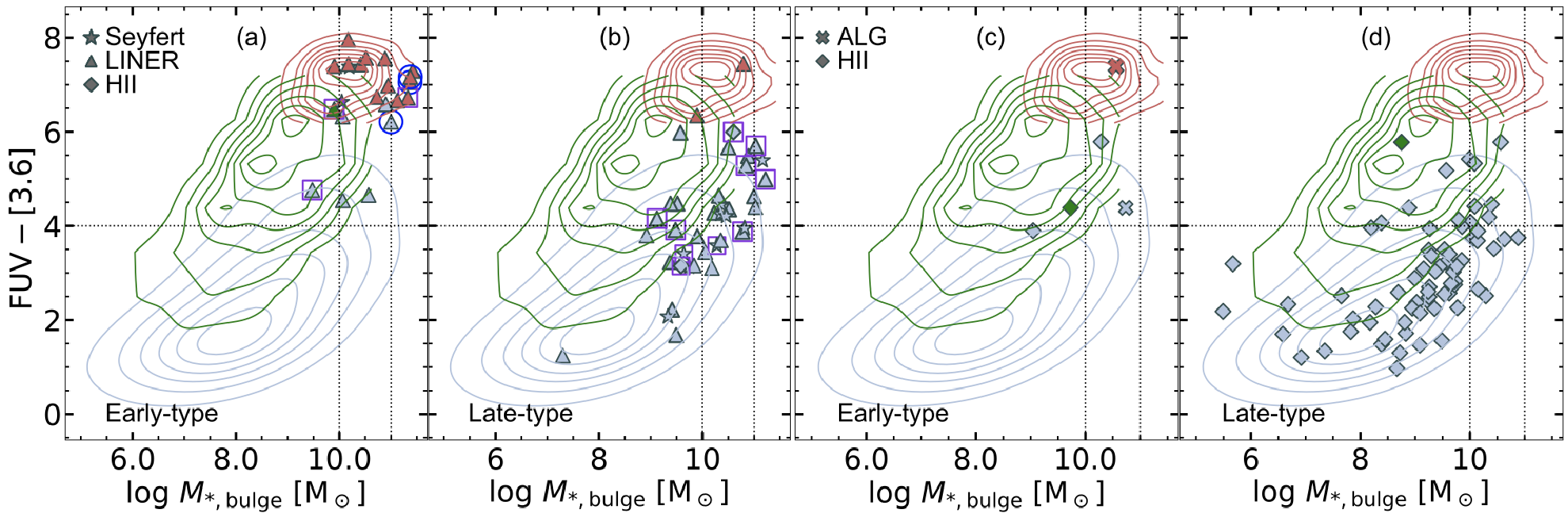}
\vspace{-1.2830cm}
\caption{(FUV--[3.6]) galaxy colour versus bulge mass relation for 140
  LeMMINGs galaxies that are in common with the sample of 1931 nearby
  galaxies from \citet{2018ApJS..234...18B}, who published {\it GALEX}
  NUV, FUV and {\it Spitzer} 3.6~$\upmu$m asymptotic galaxy
  magnitudes.  We divide the sample based on their optical spectral
  class (active: 12 Seyferts + 28 LINERs + 2 jetted \mbox{H\,{\sc ii}}
  galaxies, panels (a) and (b) and inactive: 2 ALGs plus 76
  \mbox{H\,{\sc ii}} galaxies, panels (c) and (d)) and on morphology
  (27 early-type and 113 late-type).  Of the 140 galaxies, 85 are in
  our sample and have accurate bulge mass measurements used here
  (Table~\ref{TableD}). The bulge masses for the remaining 55 galaxies
  have been computed from the 3.6~$\upmu$m magnitudes
  \citep{2018ApJS..234...18B} and are not well constrained, see the
  text for further detail (Section~\ref{Sec2.1}). Red-sequence (RS),
  blue-sequence (BS) and green valley (GV) galaxies are shown in red,
  light-blue and green contours which trace the {\it GALEX}/S$^{4}$G
  results from the sample of 1931 (early- and late-type) galaxies,
  after converting the galaxies' 3.6~$\upmu$m magnitudes
  \citep{2018ApJS..234...18B} into bulge masses using the equations
   in \citet[][their table 5]{2023arXiv230311154D}, Section~\ref{Sec2.1}. The
  vertical lines delineate the low-, intermediate- and high-(bulge
  mass) boundaries that we adopted in this work (see
  Section~\ref{Sec5.3}).  Filled stars, triangles and diamonds denote
  the Seyfert, LINER and \mbox{H\,{\sc ii}} galaxies, while the filled
  `x' symbols represent ALGs.  `Jetted' radio AGN hosts and
  core-S\'ersic galaxies are enclosed in magenta squares and open blue
  circles, respectively. In general, the hosts of active galaxies (a
  and b) typically have \mbox{FUV--[3.6]} colours $ >$ 4.0, and lie
  above the blue, star-forming main sequence defined by inactive, star
  forming galaxies (d).  }
\label{RedBlue}
\end{figure*}

In a continued effort to better examine the relationship between AGN
activity and bulge growth, in Fig.~\ref{RedBlue} we show the
\mbox{(FUV--[$3.6$]}) galaxy colour, which is a good proxy for the
global star-formation per unit galaxy stellar mass
\citep{2018ApJS..234...18B,2020ApJ...898...83D}, against bulge stellar
mass ($M_{\rm bulge}$) (e.g.\
\citealt{2004ApJ...600..681B,2014MNRAS.440..889S}) for 140 LeMMINGs
galaxies, 85 of which are in our sample (Table~\ref{TableD}, see
Section~\ref{Sec2.1}). We divide the sample based on their optical
spectral class and morphology (27 early-type and 113 late-type). Also
shown in Fig.~\ref{RedBlue} are red, light-blue and green contours
which trace the {\it GALEX}/S$^{4}$G red-sequence (RS), blue-sequence
(BS) and green valley (GV) defined by the sample of 1931 galaxies of
\citet{2018ApJS..234...18B}.  We find that the vast majority of active
early-type galaxies lie on the RS, while the remaining ones, with
intermediate-mass bulges, appear on the GV
(Fig.~\ref{RedBlue}a). These results suggest that the star formation
in massive bulges was quenched by the AGN feedback on short timescales
($\upDelta \tau \la 1$ Gyr;
\citealt{1992MNRAS.254..601B,1999MNRAS.302..537T,2006MNRAS.372..537W}). Active
late-type galaxies reside preferentially in the red end of the blue
sequence and they appear to define a transitioning sequence, where the
bulges become more massive and redder as they march gradually on the
RS (see also
\citealt{2007ApJ...665..265F,2007MNRAS.382.1541W,2010ApJ...711..284S,
  2015ApJ...798...52V, 2018MNRAS.477.3014B}). There is a caveat
here. Fig.~\ref{RedBlue} plots integrated galaxy colours, while we
actually seek to examine the bulge-SMBH co-evolution. In the case of
disc galaxies, the use of galaxy colours likely bias the bulge colours
blue-wards, therefore most active late-type bulges would actually lie
on the GV (Fig.~\ref{RedBlue}).

Overall, the hosts of active galaxies (Fig.~\ref{RedBlue}a and b),
typically have \mbox{FUV--[$3.6$]} colours $ >$ 4.0, and sit above the
blue, star-forming main sequence defined by inactive, star forming
galaxies (Fig.~\ref{RedBlue}d). We also find that radio-loud LINERs
lie on the RS. The `jetted' radio AGN hosts (Fig.~\ref{RedBlue}a and
b) are predominantly (85 per cent) BS galaxies, co-spatial with the
massive GV hosts or lie on the red end of the BS.  These support the
notion that AGN feedback shuts off or suppresses star formation in
galaxies.  We also find that all the three core-S\'ersic galaxies in
the subsample (Fig.~\ref{RedBlue}, blue circles) are located on the
red contour, this is expected since massive bulges are thought to form
via gas-poor major merger events involving massive SMBHs.

The key question as to whether the AGN feedback and suppression of
star formation in bulges are strictly coincidental is still being
debated (e.g.\ \citealt{2007ApJ...671.1388D}). Tackling this issue and
addressing the time-scale of the AGN and its duty cycle (e.g.\
\citealt{2008MNRAS.388..625S,2015MNRAS.451.2517S}) is beyond the scope
of this paper.  For low-mass galaxies in clusters
and groups the effects of the (weak) AGN and/or stellar feedback
events may be limited to the central regions and thus less efficient
in quenching the galaxies globally. Instead, for such galaxies, 
quenching is likely and maybe largely driven by environmental
processes such as ram pressure stripping
\citep{1972ApJ...176....1G},  galaxy-galaxy harassment 
\citep{1996Natur.379..613M} and/or strangulation 
\citep{1980ApJ...237..692L,2015Natur.521..192P}.

\subsection{Nuclear radio activity and kinematic morphology }\label{Sec3.2}%

In \citet{2023arXiv230311154D} we use a sample of 30 early-type
galaxies in common with the ATLAS$^{\rm 3D}$ sample
\citep{2011MNRAS.414..888E} consisting of 4 slow rotators (SRs) and 26
fast rotators (FRs) to investigate the link between kinematics,
central structure and nuclear radio detection.  The low number
statistics in that work prohibited us from drawing conclusions,
therefore we expand our sample here to 48 early-type galaxies (11 SRs
+ 37 FRs) by adding 8 galaxies that are in common between us and
\citet{2017MNRAS.471.1428V} plus 10 galaxies from the full LeMMINGs
sample with no {\it HST} data (thus excluded in  \citealt{2023arXiv230311154D})
that are in common with \citet{2011MNRAS.414..888E}.  Of the 48
early-type galaxies, 38 (10 SRs + 28 FRs) have {\it HST} data and
these are thus classified as S\'ersic and core-S\'ersic galaxies; we
find 9 out of 10 (90 per cent) SRs are core-S\'ersic galaxies, whereas
23/28 (82 per cent) FRs are S\'ersic galaxies. It is apparent that the
vast majority of SRs and FRs are core-S\'ersic and S\'ersic galaxies,
respectively, consistent with the work of
\citet{2013MNRAS.433.2812K,2017MNRAS.464..356V}.  The results can be
naturally reconciled under the scenario that SRs are largely a
consequence of experiencing predominantly gas-poor (`dry') mergers,
whereas FRs are associated with wet mergers and gas-rich processes.
For the 48 overlapping early-type galaxies
\citep{2011MNRAS.414..888E,2017MNRAS.471.1428V}, we find that, while
SRs have a slightly higher radio detection rate than FRs, the radio
detection fraction for the two kinematic classes are broadly
consistent within the errors: 19/37 FRs (0.51 $\pm$ 0.15) and 7/11 SRs
(0.64 $\pm$ 0.31) are radio detected with $e$-MERLIN at 1.5 GHz.

\section{Scaling Relations between radio core luminosity and bulge
  properties }\label{Secn4}%

\begin{figure*}
\hspace{-1.5cm}
\includegraphics[trim={.33cm -3cm -7.6cm .0827cm},clip,angle=0,scale=0.371722306]{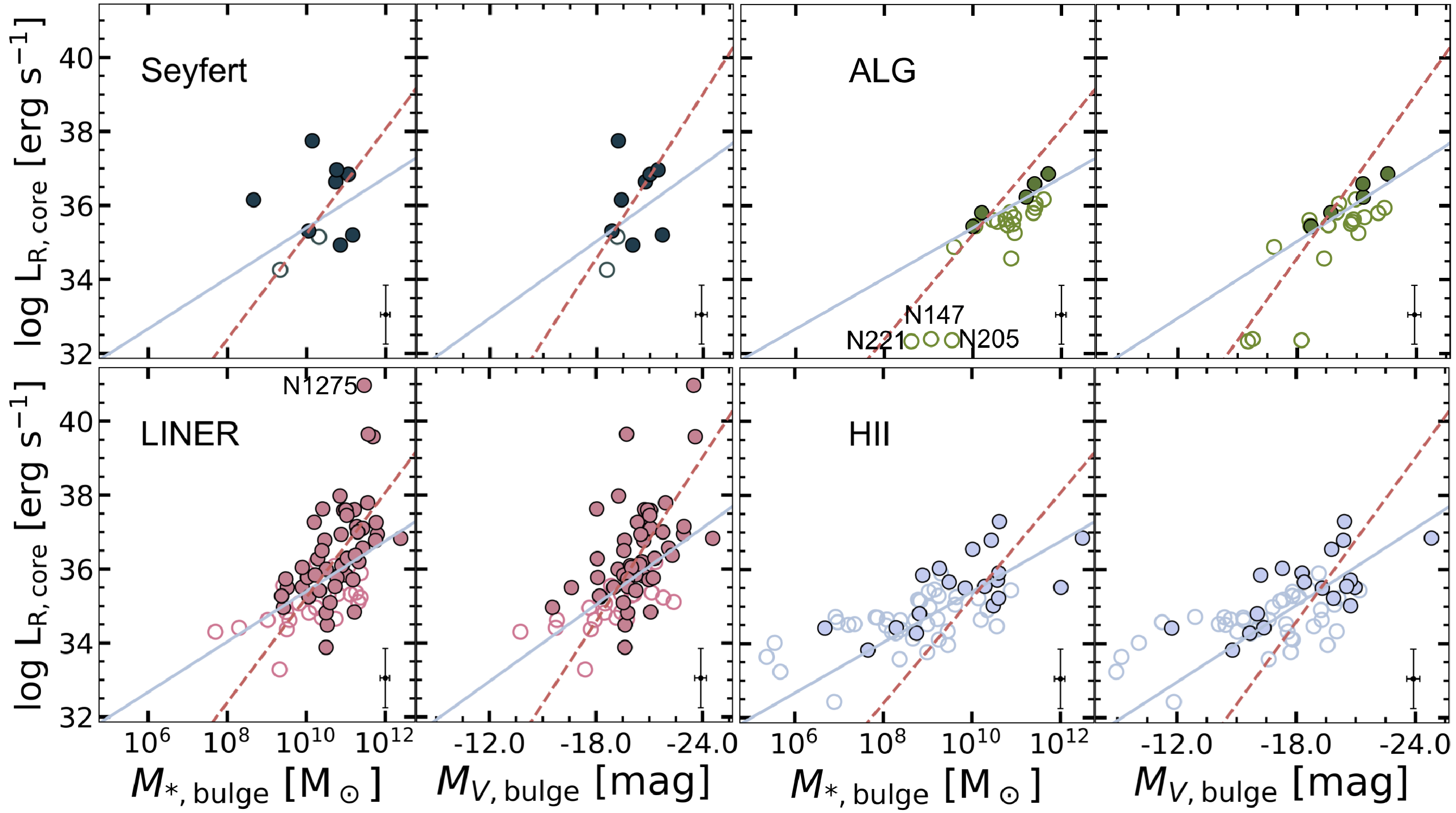}
\vspace{-1.630cm}
\caption{Correlations between the \emerlin\ 1.5 GHz radio core
  luminosity ($L_{\rm R,core}$) and bulge stellar mass
  ($M_{*,\rm bulge}$) and absolute {\it V}-band bulge magnitude
  ($M_{V,{\rm bulge}}$) for our sample of 173 galaxies, separated by
  spectral classes. Filled circles show the galaxies in our sample
  that are radio-detected with \emerlin\ at 1.5 GHz, whereas open
  circles are for the undetected sources.  Our $L_{\rm R,core}$ values
  for the undetected sources are $3\sigma$ upper limits.  The dashed
  and solid lines are OLS bisector regressions for the active galaxies
  (Seyferts (dark-green circles) and LINERs (pink circles)) and
  inactive galaxies (ALGs (green circles) and \mbox{H\,{\sc ii}}s
  (light-blue circles)), respectively. NGC~147, NGC~205 and NGC~221,
  which are all ALGs and dwarf satellites of the Andromeda Galaxy,
  deviate significantly from the relations defined by inactive
  galaxies.  A typical error bar associated with the data points is
  shown at the bottom of each panel. }
   \label{Fig6}
\end{figure*}

\begin{figure}
\hspace{-.5992cm}
\includegraphics[trim={-1.8077cm -7cm -2.6cm 01.65620443cm},clip,angle=0,scale=0.3300]{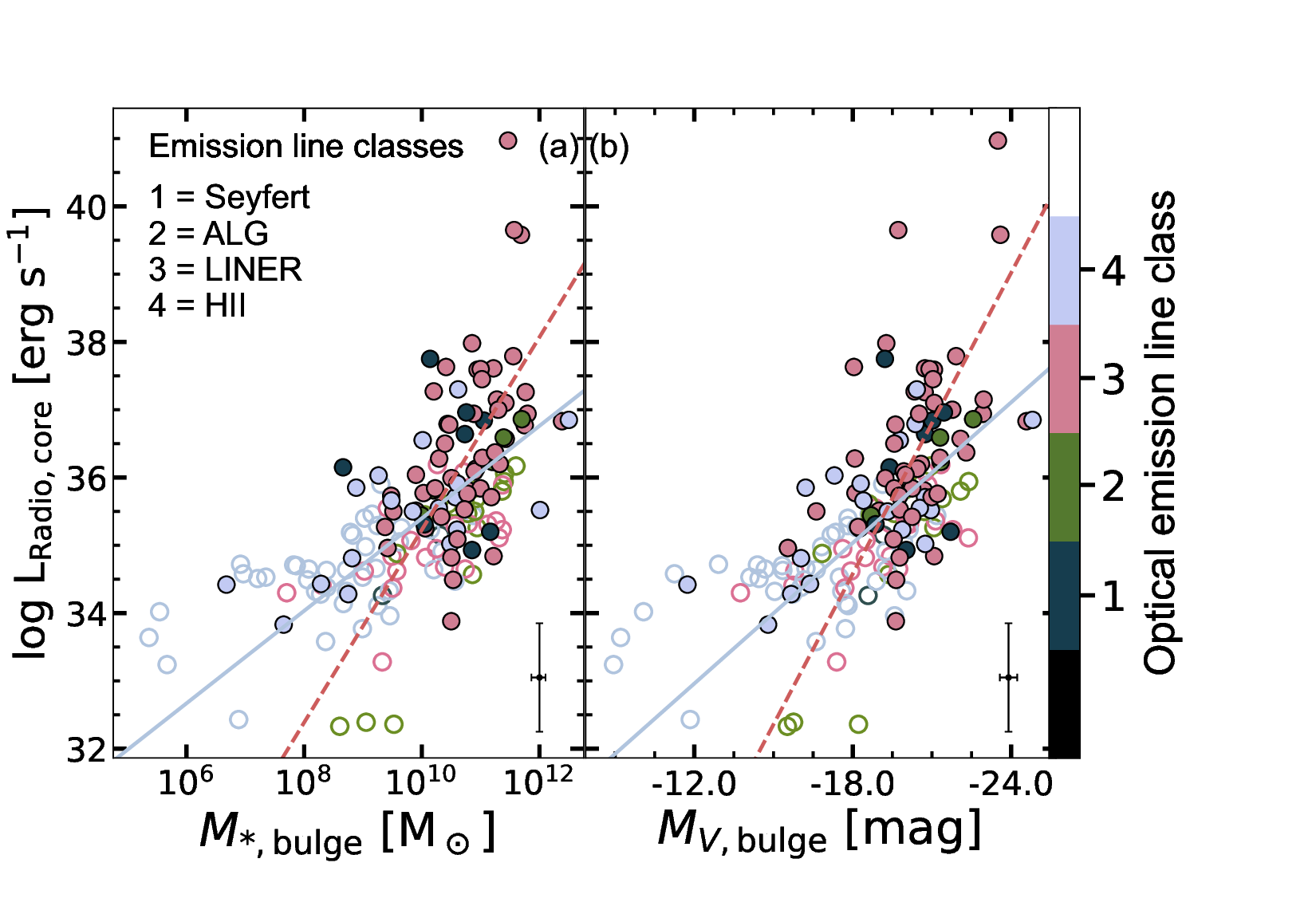}
\vspace{-3.0480cm}
\caption{Similar to Fig.~\ref{Fig6} but now showing the spectral
  classes together.  }
 \label{Fig7}
\end{figure}

\begin{figure}
\hspace{.08883cm}
\includegraphics[trim={-.27cm -6.cm 1cm 1.588113573cm},clip,angle=0,scale=0.61]{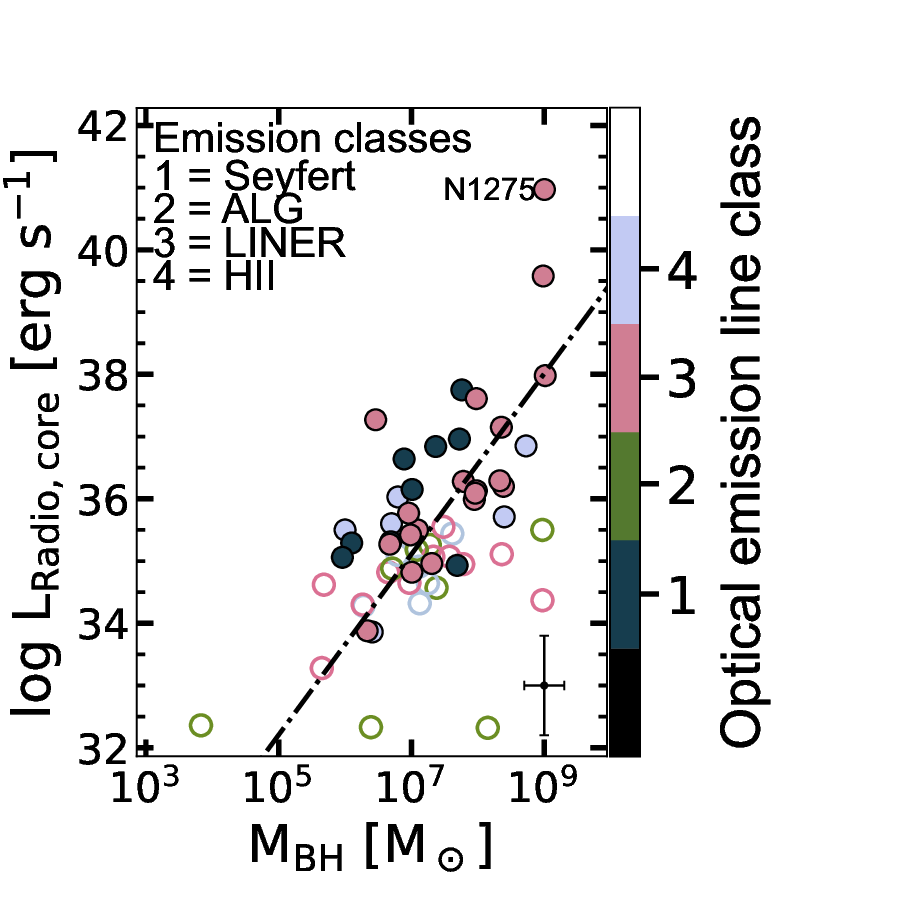}
\vspace{-4.30530cm}
\caption{The radio core luminosity ($L_{\rm R,core}$)   as a function of the SMBH mass
  ($M_{\rm BH}$) for all 59 LeMMINGs galaxies with measured
  $M_{\rm BH}$.  The dashed-dotted line is the OLS
bisector regression fit to the data. A typical error bar associated with the data points
  is shown at the bottom.}
\label{Fig8}
\end{figure}

Bulge prominence, a high bulge S\'ersic index and a large bulge mass
are preferentially associated with nuclear radio activity (see
Section~ \ref{Sec3.1},  \citealt{2023arXiv230311154D}). Scaling
relations between the central galaxy properties (such as bulge mass,
bulge luminosity, SMBH mass and velocity dispersion) and the radio
luminosity provide insight into the origin of nuclear radio emissions
and they can help constrain galaxy formation models
\citep{2001ApJ...558..561U,2002A&A...392...53N,2004A&A...418..429F,2005A&A...435..521N,2014ARA&A..52..589H,2018MNRAS.476.3478B,2019NatAs...3..387P,2021MNRAS.500.4749B,2021MNRAS.508.2019B}.
Taking advantage of the bulge masses and luminosities of LeMMINGs
galaxies, accurately measured through our multicomponent
decompositions, we investigate several radio core luminosity
$L_{\rm R,core}$ scaling relations. These are displayed in
Figs.~\ref{Fig6}, \ref{Fig7}, ~\ref{Fig8}, \ref{Fig9} and \ref{Fig10}.

\subsection{$L_{\rm R,core}-M_{*, \rm bulge}$ and $L_{\rm R,core}-M_{V,\rm bulge}$}\label{Scal1}
Figs.~\ref{Fig6} and \ref{Fig7} show the relation between the radio
core luminosity ($L_{\rm R,core}$) and the bulge's stellar mass
$M_{*, \rm bulge}$ and absolute {\it V}-band magnitude
$M_{V,\rm bulge}$, colour-coded based on the optical emission line
class. Separating the sample into active (Seyferts, LINERs and two
jetted \mbox{H\,{\sc ii}}) and inactive (ALGs and \mbox{H\,{\sc ii}})
galaxies, we fit each type separately, and find strong correlations
between $L_{\rm R,core}$ and $M_{*, \rm bulge}$ and $M_{V,\rm
  bulge}$. Note that the sample spans more than six orders of
magnitude in stellar mass and over a dozen in absolute magnitude. The
OLS bisector regressions that we fit to the
($L_{\rm R,core},M_{*,\rm bulge}$) and
($L_{\rm R,core},M_{V,\rm bulge}$) data set reveal that the
$L_{\rm R,core}-M_{*,\rm bulge}$ and $L_{\rm R,core}-L_{V,\rm bulge}$
relations are not log--linear but instead active and inactive galaxies
follow two distinct relations with markedly different slopes. The OLS
bisector regression of our active galaxies yields
$L_{\rm R,core} \propto M_{*, \rm bulge}^{1.42\pm0.18}$ and
$L_{\rm R,core} \propto L_{V,\rm bulge}^{1.85\pm0.23}$, while we see
substantial flattening of the trend for inactive galaxies
$L_{\rm R,core} \propto M_{*, \rm bulge}^{0.68\pm0.08}$ and
$L_{\rm R,core} \propto L_{V,\rm bulge}^{0.88\pm0.13}$. Note that  our 
regression analyses  did not take into account  (accurately measured) 
errors on $L_{\rm R,core}$ and bulge proprieties, therefore we opted
 for a  symmetrical bisector regression, rather than an OLS(X|Y) or 
 OLS(Y|X)  regression. The results of
the regression fits are presented in Table~\ref{Tab04}.  The
transition from an inactive (i.e.\ star-formation driven) to an active
optical AGN seems to occur at
$M_{*, \rm bulge}\sim 10^{9.8 \pm 0.3} \rm M_{\sun}$ and
$M_{V, \rm bulge}\sim -18.5 \pm 0.3$ mag.
To assess the strength
of the correlations, we use the Spearman and Pearson correlation
coefficients ($r_{\rm s}, r_{\rm p}$, see Table~\ref{Tab04}).  For the
$L_{\rm R,core}-M_{*, \rm bulge}$ relations we find
$r_{\rm s} \sim 0.56-0.73$ and $r_{\rm p} \sim 0.54-0.64$ and the
pertaining $P$-values for a null hypothesis that two sets of data are
uncorrelated are very low ($P \sim 10^{-14}-10^{-7}$). The tight
$L_{\rm R,core}-M_{V,\rm bulge}$ relations are such that
-$r_{\rm s} \sim$ $0.53-0.70$ and -$r_{\rm p} \sim 0.53-64$ and
$P \sim 10^{-13}-10^{-7}$.

Given that 52 per cent of our sample do not have radio emission
detected with \emerlin\ at 1.5 GHz and that our $L_{\rm R,core}$
values for such undetected sources are upper limits, we performed a
statistical censored analysis of the $L_{\rm R,core}-M_{*, \rm bulge}$
and $L_{\rm R,core}-M_{V,\rm bulge}$ relations using the {\sc asurv}
package \citep{1985ApJ...293..192F,1986ApJ...306..490I} accounting for
upper limits. The slopes from the censored and uncensored analyses are
consistent within the errors
($L_{\rm R,core} \propto M_{*, \rm bulge}^{1.19\pm0.22}$ and
$L_{\rm R,core} \propto L_{\rm V,bulge}^{1.33\pm0.30}$, for active
galaxies, censored;
$L_{\rm R,core} \propto M_{*, \rm bulge}^{0.74\pm0.17}$ and
$L_{\rm R,core} \propto L_{\rm V,bulge}^{0.85\pm0.20}$, for inactive
galaxies, censored).  The slopes of the
$L_{\rm R,core}-M_{*, \rm bulge}$ and
$L_{\rm R,core}-L_{V, \rm bulge}$ relations for active and inactive
galaxies are different, irrespective of the applied statistical
method. There are however marked discrepancies between intercepts of
the relations from the censored and uncensored analyses, in which the
former intercepts have large errors and are markedly offset towards lower values. Despite
the high level scatter at the low radio core luminosity end, the
undetected radio sources which are mostly \mbox{H\,{\sc ii}} galaxies
appear to be simply luminosity scaled-down versions of the radio
detected inactive galaxies. However, three dwarf satellites of M31, the
Andromeda Galaxy (NGC~147, NGC~205 and NGC~221, e.g.\
\citealt{2013Natur.493...62I}), which are all ALGs at a distance of
0.8 Mpc with $L_{\rm R,core}$ upper limits
\citep{2018MNRAS.476.3478B,2021MNRAS.500.4749B}, deviate notably from
the inactive $L_{\rm R,core}-M_{*, \rm bulge}$ and
$L_{\rm R,core}-L_{V, \rm bulge}$ relations.  With the assumption that
all ALGs follow the inactive $L_{\rm R,core}-M_{*, \rm bulge}$ and
$L_{\rm R,core}-L_{V, \rm bulge}$ relations, a plausible
interpretation is that the very low $L_{\rm R,core}$ upper limit
values of these low mass galaxies arise from an observational bias associated 
with their close proximity to us, which allowed very  low $L_{\rm R,core}$ values to
 be measured. An alternative interpretation for the lack of radio emission may be that
  the gravitational interaction of these three satellites with M31 has 
stripped them of their  gas (see e.g.\ \citealt{2021ApJ...913...53P}),  subsequently  quenching  
  star formation in the galaxies.

\setlength{\tabcolsep}{0.11790in}
\begin{table*}
\begin {minipage}{182mm}
\caption{Scaling relations between radio core luminosity and galaxy properties.  } 
\label{Tab04}
\begin{tabular}{@{}lllccccccccccccccccc@{}}
\hline
Relation &OLS bisector fit &$r_{\rm
                                              s}/P$-${\rm value}$&$r_{\rm
                                                            p}/P$-${\rm value}$&Sample&\\
\hline
 \multicolumn{5}{c}{\bf Active galaxies (LINERs+ Seyferts) }\\

$L_{\rm R,core}-M_{\rm *,bulge} $
        &$\mbox{$\log$}\left(\frac{L_{\rm R,core}}{\rm erg~ s^{-1}}\right)= (1.42\pm
           0.18) \mbox{log}\left(\frac{M_{\rm *,bulge}}{\mbox{$5\times10^{10}$ $\rm M_{\sun}$}}\right)$ +~($36.22~ \pm  0.15$)
 &0.56/$4.5\times10^{-8}$&0.54/$ 2.5\times10^{-7}$&81 \\

$L_{\rm R,core}-M_{V,\rm bulge}$
        &$\mbox{log}\left(\frac{L_{\rm R,core}}{\rm erg ~s^{-1}}\right)= (-0.74   \pm
            0.09)\left( M_{V,\rm bulge}+20.5 \right)$ +~($ 36.43 ~ \pm  0.16$)
             &$-$0.53/$3.06\times10^{-7}$&$-$0.53/$ 2.95\times10^{-7}$&81 \\
 \vspace{-.185cm}
$L_{\rm R,core}-\sigma $
        &$\mbox{log}\left(\frac{L_{\rm R,core}}{\rm erg ~s^{-1}}\right)= (4.85\pm
             0.58) \mbox{log}\left(\frac{\sigma}{\mbox{170 {\rm km s$^{-1}$}}}\right)$ +~($ 36.27~ \pm  0.11$)
 &0.72/$7.9\times10^{-19}$&$0.66/2.8\times10^{-15}$&112\\
 \\[4pt]
 \multicolumn{5}{c}{\bf Inactive galaxies (ALG+ \mbox{H\,{\sc ii}}) }\\
$L_{\rm R,core}-M_{\rm *,bulge} $
        &$\mbox{log}\left(\frac{L_{\rm R,core}}{\rm erg ~s^{-1}}\right)= (0.68\pm
           0.08) \mbox{log}\left(\frac{M_{\rm *,bulge}}{\mbox{$2\times10^{9}$ $\rm M_{\sun}$}}\right)$ +~($ 34.92 ~ \pm  0.10$)
 & 0.73/$7.1 \times 10^{-15}$&0.64/$ 6.4\times10^{-11}$&82\\

 $L_{\rm R,core}-M_{\rm V,bulge}$
        &$\mbox{log}\left(\frac{L_{\rm R,core}}{\rm erg ~s^{-1}}\right)= (-0.35   \pm
            0.05)\left( M_{V,\rm bulge}+18.5 \right)$ +~($ 35.21 ~ \pm  0.08$)
             &$-$0.70/$3.3\times10^{-13}$&$-$0.64/$1.1\times10^{-10}$&82 \\          
             
 \vspace{-.185cm}
$L_{\rm R,core}-\sigma $
        &$\mbox{log}\left(\frac{L_{\rm R,core}}{\rm erg ~s^{-1}}\right)= (3.06\pm
              0.23) \mbox{log}\left(\frac{\sigma}{\mbox{65 {\rm km s$^{-1}$}}}\right)$ +~($  34.91~ \pm  0.07$)
 &0.58/$7.9 \times10^{-16}$&0.56/$ 1.2 \times10^{-14}$&164 \\
 \\[4pt]

 \multicolumn{4}{c}{\bf All galaxies}\\

$L_{\rm R,core}-M_{\rm BH}$  
        &$\mbox{log}\left(\frac{L_{\rm R,core}}{\rm erg ~s^{-1}}\right)= ( 1.45\pm
             0.20) \mbox{log}\left(\frac{M_{\rm BH}}{\mbox{$3\times10^{7} \rm M_{\sun}$ }    }\right)$    +~($ 35.80 ~\pm  0.19$)
 &0.52/$2.6\times10^{-5}$&0.57/2.3$\times10^{-6}$&59 \\

$L_{\rm R,core}-\sigma $
        &$\mbox{log}\left(\frac{L_{\rm R,core}}{\rm erg ~s^{-1}}\right)= (3.61\pm
              0.24) \mbox{log}\left(\frac{\sigma}{\mbox{100 {\rm km s$^{-1}$}}}\right)$ +~($ 35.46~ \pm  0.06$)
 &0.65/$7.2 \times10^{-35}$&0.64/$2.9\times10^{-33}$&276 \\

 \\[4pt]
 \multicolumn{5}{c}{\bf  Censored analysis}\\
\hline
Relation &Linear regression from {\sc asurv}&$r_{\rm
                                              s}/P$-${\rm value}$&$r_{\rm
                                                            p}/P$-${\rm value}$&Sample&\\
\hline
 \multicolumn{5}{c}{\bf Active galaxies (LINERs+ Seyferts) }\\
$L_{\rm R,core}-M_{\rm *,bulge} $
        &$\mbox{$\log$}\left(\frac{L_{\rm R,core}}{\rm erg~ s^{-1}}\right)= (1.19\pm
           0.22) \mbox{log}\left(\frac{M_{\rm *,bulge}}{\mbox{$5\times10^{10}$ $\rm M_{\sun}$}}\right)$ +~($35.9~ \pm  1.8$)
 &---&---&81 \\

$L_{\rm R,core}-M_{V,\rm bulge}$
        &$\mbox{log}\left(\frac{L_{\rm R,core}}{\rm erg ~s^{-1}}\right)= (-0.53   \pm
            0.12)\left( M_{V,\rm bulge}+20.5 \right)$ +~($ 36.0 ~ \pm  1.7$)
             &---&---&81 \\
 \vspace{-.185cm}
$L_{\rm R,core}-\sigma $
        &$\mbox{log}\left(\frac{L_{\rm R,core}}{\rm erg ~s^{-1}}\right)= (5.10\pm
             0.60) \mbox{log}\left(\frac{\sigma}{\mbox{170 {\rm km s$^{-1}$}}}\right)$ +~($ 35.9~ \pm  0.9$)
 &---&---&112\\
 \\[4pt]
 \multicolumn{5}{c}{\bf Inactive galaxies (ALG+ \mbox{H\,{\sc ii}}) }\\
$L_{\rm R,core}-M_{\rm *,bulge} $
        &$\mbox{log}\left(\frac{L_{\rm R,core}}{\rm erg ~s^{-1}}\right)= (0.74\pm
           0.16) \mbox{log}\left(\frac{M_{\rm *,bulge}}{\mbox{$2\times10^{9}$ $\rm M_{\sun}$}}\right)$ +~($ 33.7 ~ \pm  1.6$)
 &---&---&82\\

 $L_{\rm R,core}-M_{V,\rm bulge}$
        &$\mbox{log}\left(\frac{L_{\rm R,core}}{\rm erg ~s^{-1}}\right)= (-0.36   \pm
            0.08)\left( M_{V,\rm bulge}+18.5 \right)$ +~($ 34.0 ~ \pm  1.6$)
             &---&---&82 \\          
             
 \vspace{-.185cm}
$L_{\rm R,core}-\sigma $
        &$\mbox{log}\left(\frac{L_{\rm R,core}}{\rm erg ~s^{-1}}\right)= (2.70\pm
              0.48) \mbox{log}\left(\frac{\sigma}{\mbox{65 {\rm km s$^{-1}$}}}\right)$ +~($  33.9~ \pm  0.8$)
 &---&---&164 \\
 \\[4pt]
 \multicolumn{4}{c}{\bf All galaxies}\\
 \vspace{0.09cm}
$L_{\rm R,core}-\sigma $
        &$\mbox{log}\left(\frac{L_{\rm R,core}}{\rm erg ~s^{-1}}\right)= (3.97\pm
              0.34) \mbox{log}\left(\frac{\sigma}{\mbox{100 {\rm km s$^{-1}$}}}\right)$ +~($ 34.7~ \pm  0.6$)&---&---&276 \\
 \hline
\end{tabular} 
{\it Notes.}  Radio core luminosity ($L_{\rm R_{core}}$) as a function of
bulge stellar mass ($M_{\rm *,bulge}$), absolute magnitude
($M_{\rm V,bulge}$), central velocity dispersion ($\sigma$) and
measured black hole mass ($M_{\rm BH}$) for active and inactive
galaxies. The different columns represent: the OLS bisector fits to
the data, the Spearman and Pearson correlation coefficients
($r_{\rm s}$ and $r_{\rm p}$, respectively) and the associated
probabilities for a serendipitous correlation. We also show censored
linear regressions using {\sc asurv}, accounting for upper limits, see
Section~\ref{Scal1} for details.
\end{minipage}
\end{table*}

\subsection{$L_{\rm R,core}-M_{\rm BH}$ \mbox{and} $L_{\rm R,core}-\sigma$ } \label{Scal2}%

\begin{figure*}
\hspace{-.0883cm}
\includegraphics[trim={.337cm -1.9cm -1.5cm .04cm},clip,angle=0,scale=0.38563]{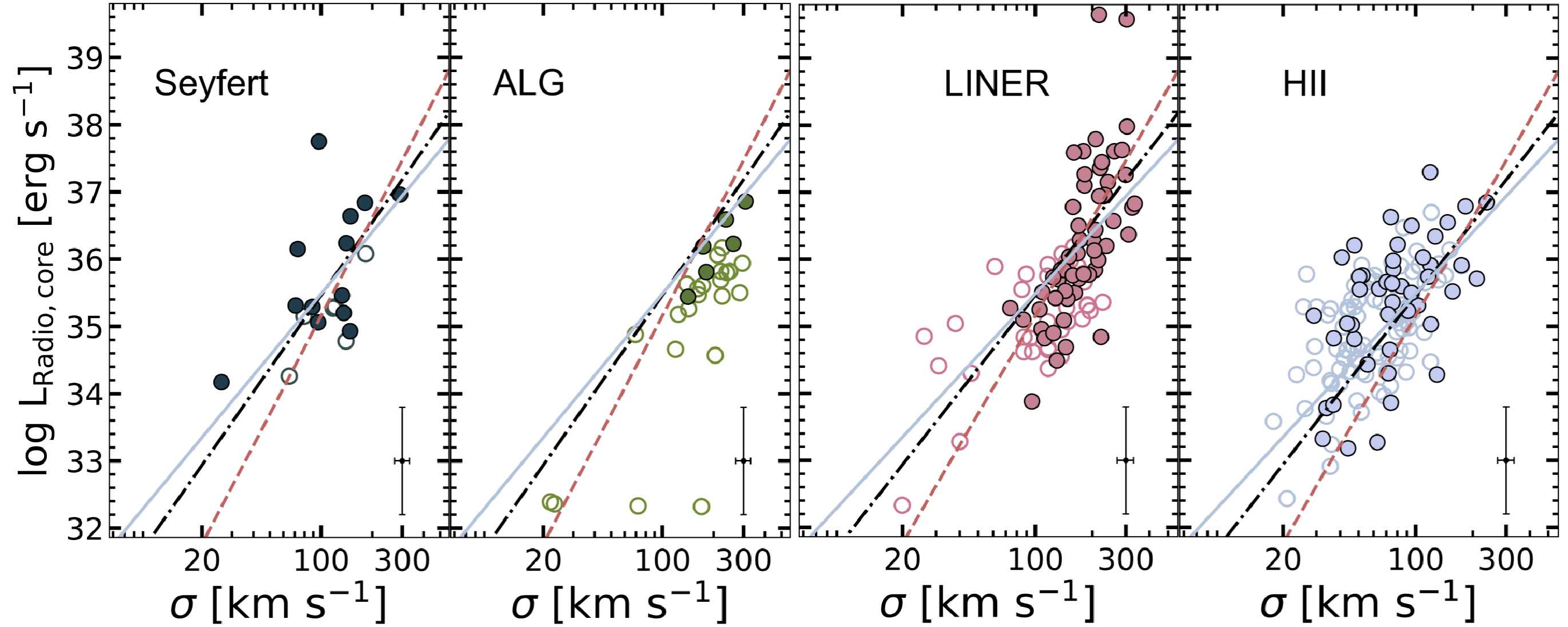}
\vspace{-1.18950cm}
\caption{The radio core luminosity   ($L_{\rm R,core}$)  as a function of the central stellar velocity
  dispersion ($\sigma$) for all 276 LeMMINGs galaxies with $\sigma$
  available from \citet{2009ApJS..183....1H}. Symbolic representations are as in Fig.~\ref{Fig6}. The dashed, solid and dashed-dotted lines are  OLS
bisector fits for the
  active, inactive and full sample of galaxies under
  consideration. NGC~1275 falls outside the range  radio core luminosity shown here. A typical error bar associated with the data points
  is shown at the bottom of each panel.}
\label{Fig9}
\end{figure*}

We next examine the correlations between the radio core luminosity and
the SMBH mass ($M_{\rm BH}$) for those 59 LeMMINGs galaxies with
measured $M_{\rm BH}$ (\citealt{2016ApJ...831..134V,
  2019ApJ...872..104N}, see Table~\ref{TableD}), and the central
stellar velocity dispersion ($\sigma$) for all 276 LeMMINGs galaxies
with $\sigma$ available from \citet[][]{2009ApJS..183....1H}; see
Figs.~\ref{Fig8}, \ref{Fig9}, see Table~\ref{TableD}.  To measure the 
central velocity dispersions, \citet[][]{2009ApJS..183....1H} used
  long-slit optical spectra  placed across the nucleus of their galaxies 
  with a rectangular aperture of size $2\arcsec \times4\arcsec$  \citep{1995ApJS...98..477H}. 
The measured SMBH masses are based on
the methods of gas dynamics, stellar dynamics and megamasers, except
for three galaxies (NGC~3516, NGC~5273 and NGC~5548) with
reverberation mapping SMBH masses.
 
The \mbox{$L_{\rm R,core}-M_{\rm BH}$} relation is regarded as a good
diagnostic to separate star-forming and AGN galaxies \citep[e.g.\
][]{2002A&A...392...53N,2018A&A...616A.152S,2018MNRAS.476.3478B}. Brighter
and more massive galaxies which host massive BHs are associated with
radio emission from the AGN as reflected by the trend of increasing
AGN fraction with SMBH mass, whereas at low luminosities/masses the
radio emission is primarily due to star formation.  The 59 SMBH
masses, although securely measured, are fairly high. With only 12/59
SMBHs less massive than $M_{\rm BH} \sim10^{6.5} \rm M_{\sun}$, the
presence of two distinct relations for active and inactive galaxies is
not evident from the \mbox{$L_{\rm R,core}-M_{\rm BH}$} diagram
(Fig.~\ref{Fig8}), therefore we have not sought to separate the
galaxies into these two groups. In
\citet{2018MNRAS.476.3478B,2021MNRAS.500.4749B} we reported `broken'
scaling relations involving $L_{\rm core}$ and $M_{\rm BH}$ with
distinct slopes which correspond to non-jetted star-forming
($\sim L_{\rm core} \propto M_{\rm BH}^{0.61}$) and active galaxies
($\sim L_{\rm core} \propto M_{\rm BH}^{1.65}$), the break occurring
at $M_{\rm BH} \sim 10^{6.5} \rm M_{\sun}$; the bulk (221/280) of the
\citet{2018MNRAS.476.3478B,2021MNRAS.500.4749B} galaxies had their
SMBH masses predicted using the $M-\sigma$ relation. In
Fig.~\ref{Fig8} we refrain from using SMBH masses predicted with
well-known SMBH scaling rations (e.g.\ $M_{\rm BH}-\sigma$,
$M_{\rm BH}-L_{\rm bulge}$ and $M_{\rm BH}-M_{*,\rm bulge}$) because
departures from a single power-law relation between $M_{\rm BH}$ and
host galaxy properties have been reported to exist particularly at the
low- and high-mass ends (e.g.\
\citealt{2008ApJ...688..159G,2013ApJ...764..184M,2017MNRAS.465...32B,2018ApJ...855L..20M,2021ApJ...908..134D}). We
find that the OLS bisector fit gives the relation between
$L_{\rm R,core}$ and $M_{\rm BH}$ as
$L_{\rm R,core} \propto M_{\rm BH}^{1.45 \pm0.20}$
($r_{\rm s}/P \sim 0.52/2.6 \times 10^{-5}$ and
$r_{\rm p} \sim 0.57/2.3 \times 10^{-6}$), Table~\ref{Tab04}.

\begin{figure}
\hspace{-.8583cm}
\includegraphics[trim={-1.12998727038527cm -8.cm -1.6cm 1.5507384cm},clip,angle=0,scale=0.7015]{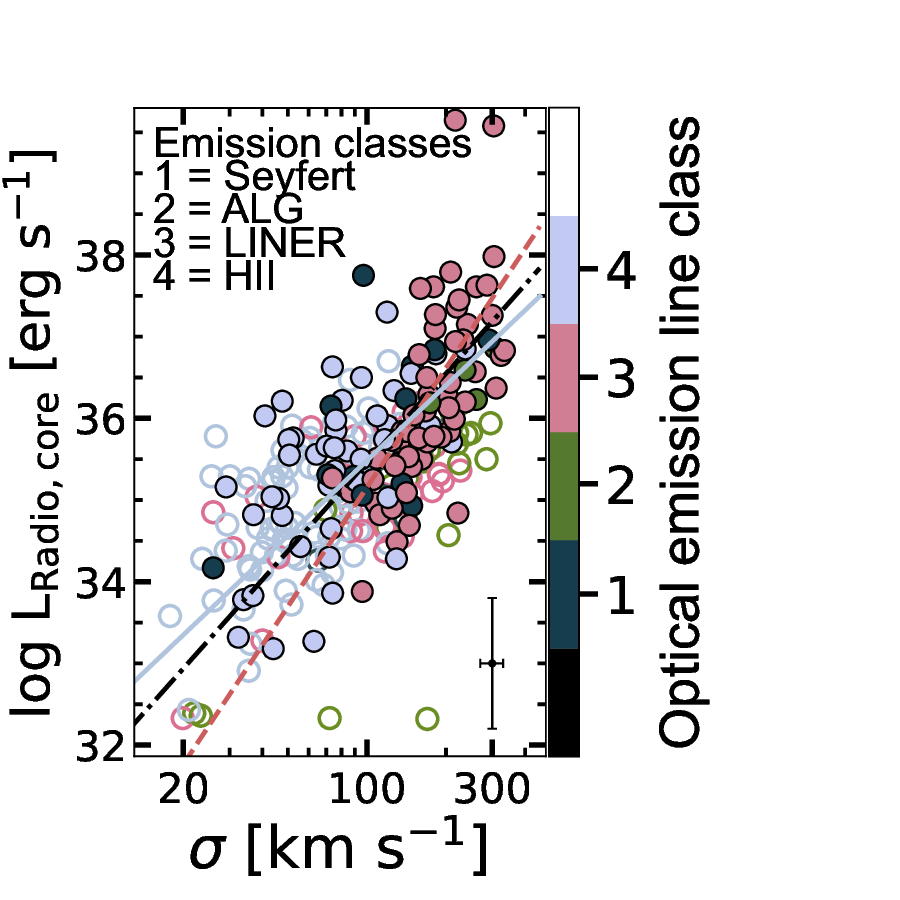}
\vspace{-6.35620cm}
\caption{Similar to Fig.~\ref{Fig9} but now showing the spectral classes together for ease of comparison. A typical error bar 
  is shown at the bottom right.}
\label{Fig10}
\end{figure}

In the \mbox{$L_{\rm R,core}-\sigma$} diagram (Figs.~\ref{Fig9}, \ref{Fig10}),
\mbox{H\,{\sc ii}} galaxies, galaxies with LINER, Seyferts nuclei and
ALGs cleanly separate into two sequences. For a given radio core
luminosity, \mbox{H\,{\sc ii}} galaxies seem to possess lower central velocity
dispersions than LINERs, Seyferts, and ALGs. We note that  \mbox{H\,{\sc ii}} galaxies are
 likely to contain a low-dispersion disc component which may reduce  the measured 
  $\sigma$ values. However,   \citet[][]{2009ApJS..183....1H} measured $\sigma$ 
  using  the spectra of the central regions of the galaxies typically dominated by
   the bulge.  Furthermore,   \cite{2022MNRAS.510.5639M} reported weak 
   dependence of  $\sigma$ on  disc component.  
On the other hand, we  find ALGs do not
deviate from the best-fitting inactive \mbox{$L_{\rm R,core}-\sigma$}
relation defined by the \mbox{H\,{\sc ii}}+ALG subsample, but they
reside almost exclusively under this relation.

 We therefore fit OLS
bisector regressions first by splitting the sample into active and
inactive galaxies and then using the sample of 276 LeMMINGs
galaxies. The OLS fits yield the relations between $L_{\rm R,core}$
and $\sigma$ such that $L_{\rm R,core} \propto \sigma^{4.85 \pm0.58}$
($r_{\rm s}/P \sim 0.72/7.9 \times 10^{-19}$ and
$r_{\rm p} \sim 0.66/2.8 \times 10^{-15}$),
$L_{\rm R,core} \propto \sigma^{3.06 \pm0.23}$
($r_{\rm s}/P \sim 0.59/7.9 \times 10^{-16}$ and
$r_{\rm p} \sim 0.56/1.2 \times 10^{-14}$) and
$L_{\rm R,core} \propto \sigma^{3.61 \pm0.24}$
($r_{\rm s}/P \sim 0.65/7.2 \times 10^{-35}$ and
$r_{\rm p} \sim 0.64/2.9\times 10^{-33}$) for active, inactive and
combined sample galaxies, respectively (Table~\ref{Tab04}). Our
censored $L_{\rm R,core}-\sigma$ relations are
$L_{\rm R,core} \propto \sigma^{5.10 \pm0.60}$,
$L_{\rm R,core} \propto \sigma^{2.75 \pm0.48}$ and
$L_{\rm R,core} \propto \sigma^{3.97 \pm0.34}$ for active, inactive
and combined sample galaxies, respectively (Table~\ref{Tab04}). As
noted above, while the slopes from the censored and uncensored
analyses agree within the errors, the intercepts of the relations are
different.  The transition from a steep \mbox{$L_{\rm R,core}-\sigma$}
relation for optical AGN galaxies to the less steep, inactive
\mbox{$L_{\rm R,core}-\sigma$} relation happens at
$\sigma \sim 85 \pm 5$ km s$^{-1}$.

Figs.~\ref{Fig6},  \ref{Fig8} and \ref{Fig9} reveal an outstanding outlier
NGC~1275. This  LINER galaxy  is $2.5-3.8$
dex (in the $\log L_{\rm R,core}$ direction) above the
$L_{\rm R,core}-M_{*, \rm bulge}$, $L_{\rm R,core}-L_{V, \rm bulge}$,
\mbox{$L_{\rm R,core}-\sigma$} and \mbox{$L_{\rm R,core}-M_{\rm BH}$}
best-fit lines. Given that NGC~1275 is offset upward in several
$L_{\rm R,core}$ scaling diagrams, the most favoured interpretation is
that the radio core luminosity is higher than expected.  We note  the level of
scatter pertaining to the $L_{\rm R,core}$ scaling relations of
\mbox{$0.80-1.0$ dex} (in the $\log L_{\rm R,core}$ direction). For
comparison, the typical uncertainly on $\log L_{\rm R,core}$ is 
$\sim$1 dex.

\section{Discussion}\label{Dissc}%

\subsection{Is the $L_{\rm R,core}-M_{\rm BH}$ relation broken? }

We have shown previously unreported breaks in the slopes of the
observed \mbox{$L_{\rm R,core}-M_{*, \rm bulge}$},
\mbox{$L_{\rm R,core}-M_{V, \rm bulge}$} and
\mbox{$L_{\rm R,core}-\sigma$} relations when the sample includes
active and inactive galaxies (Sections ~\ref{Scal1}, \ref{Scal2} and
Table~\ref{Tab04}).  The slopes of these $L_{\rm R,core}$ scaling
relations for galaxies whose radio emission is dominated by AGN are
steeper than those for star formation dominated galaxies, with the
breaks occurring at
$M_{*, \rm bulge}\sim 10^{9.8 \pm 0.3} \rm M_{\sun}$,
$M_{V, \rm bulge}\sim -18.5 \pm 0.3$ mag and $\sigma \sim 85 \pm 5$ km
s$^{-1}$. We have interpreted the broken $L_{\rm R,core}$ scaling
relations as being due to differences in the origin of nuclear radio
emissions for AGN and star forming galaxies
\citep{2017A&A...601A.143F,2018MNRAS.476.3478B,2021MNRAS.500.4749B,2021MNRAS.508.2019B}. Our
motivation here is twofold: (i) examine if the single (log--linear)
\mbox{$L_{\rm R,core}-M_{\rm BH}$} relation (Fig.~\ref{Fig8}) is
internally consistent with other $L_{\rm R,core}$ scaling relations
and (ii) examine if it holds for all galaxies.  As noted above,
\citet{2021MNRAS.508.2019B} reported a break in the
\mbox{$L_{\rm R,core}-M_{\rm BH}$} relation at
$M_{\rm BH} \sim 10^{6.5} \rm M_{\sun}$ for the full sample of 280
LeMMINGs galaxies, analogous to the aforementioned breaks (Sections
~\ref{Scal1} and \ref{Scal2}); in that work the SMBH mass for 221
LeMMINGs galaxies were predicted from the $M_{\rm BH}-\sigma$ relation
\citep{2002ApJ...574..740T}. Combining our single (log--linear)
\mbox{$L_{\rm R,core}-M_{\rm BH}$} relation constructed using 59
measured $M_{\rm BH}$ with the
\mbox{$L_{\rm R,core}-M_{*, \rm bulge}$},
\mbox{$L_{\rm R,core}-L_{V, \rm bulge}$} and
\mbox{$L_{\rm R,core}-\sigma$} relations (Table~\ref{Tab04}), we find
the mass, luminosity and velocity dispersion turnovers given above
(that marks the transition from star formation to AGN) correspond to a
SMBH mass of $M_{\rm BH} \sim 10^{6.8 \pm0.3} \rm M_{\sun}$. This
suggests that breaks in the different $L_{\rm R,core}$ scaling
relations are related and that the break we identified in the
\mbox{$L_{\rm R,core}-M_{\rm BH}$} relation
\citep{2021MNRAS.508.2019B} may be a consequence of a broken
\mbox{$L_{\rm R,core}-\sigma$} relation. As such a single
\mbox{$L_{\rm R,core}-M_{\rm BH}$} relation can hold over the entire
radio core luminosity range, although our results do not preclude a
broken $L_{\rm R,core}-M_{\rm BH}$ relation. To ascertain the
universality of the $L_{\rm R,core}-M_{\rm BH}$ relation more directly
measured SMBHs for our sample with
$M_{\rm BH} \la 10^{6.5} \rm M_{\sun}$ are needed.

\subsection{The influence of environment on AGN activity and the
  importance of bulge mass }%

Our results have revealed no evidence for environmental effects on the
AGN activity.  Across a small range in bulge mass, we find no
significant dependence of the radio core luminosity, radio loudness,
nuclear radio activity or AGN fraction (both optical and radio) on
environmental metrics and halo mass (Fig.~\ref{AGNENV}). Our work is
consistent with a growing body of literature
\citep{1995A&A...293..683L,2001ApJ...559..606C,2003ApJ...597..142M,2004ASSL..319..189K,2010MNRAS.404.1231V,2013MNRAS.429.1827P,2019ApJ...874..140A,2019MNRAS.488...89M,2021ApJ...922L..17M},
and the results favour a scenario in which the AGN activity is largely
bulge mass dependent (see Section~\ref{Sec5.3}; also
\citealt{2003ApJ...597..142M,2014MNRAS.441..599B,2020MNRAS.493.3838M}).
We recall that radio loudness, the radio detection fraction and the
fraction of galaxies hosting optical (emission-line) AGN and/or radio
AGN are strong functions of the bulge stellar mass and luminosity,
$M_{*,\rm bulge}$, $M_{V,\rm bulge}$ (see
Sections~\ref{Sec3.2}$-$\ref{AGNFT}; also
\citealt{2003MNRAS.340.1095D,2005MNRAS.362...25B,2013MNRAS.429.1827P}).
We have also shown that the radio core luminosity scales strongly with
$M_{*,\rm bulge}$ and $M_{V,\rm bulge}$. These observed relations are
ascribed here to a formation scenario in which bulges and their SMBHs
evolve together in lockstep (e.g.\
\citealt{2002ApJ...564..120H,2002A&A...392...53N,2005A&A...435..521N,2014ARA&A..52..589H}). Galaxies
with more massive bulges harbour massive SMBHs efficient at generating
strong AGN-driven outflows and radio emissions, which consequently
blow away gas and produce lower star formation efficiencies (e.g.\
\citealt{1998A&A...331L...1S,2006MNRAS.365...11C,2006ApJS..163....1H,2008MNRAS.388..587L}).
The very deep central potential in the bulges of massive galaxies also
help retain the hot X-ray-emitting interstellar medium (ISM) that is
being kept hot by the energy from the radio jet (e.g.\
\citealt{1993MNRAS.263..323T,2007ARA&A..45..117M,2009Natur.460..213C,2009ApJS..182..216K,2012ARA&A..50..455F,2014ARA&A..52..589H}).
In contrast, low-mass bulges with a shallow potential well likely
undergo a weak AGN activity and stellar feedback
\citep{2005MNRAS.364.1337S}.  Finally, we note that the so-called
morphology--density relation \citep{1980ApJ...236..351D} is such that
elliptical and early-type disc galaxies, apart from hosting more
powerful AGN, live in denser environments. Therefore this relation
could give rise to an apparent link between the AGN and host galaxy
environment.

\subsection{Implications for host galaxy formation mechanisms}\label{Sec5.3}%

We now discuss the observed connections between the stellar mass,
optical spectral line and radio continuum emission, and the potential
implications for the formation of core-S\'ersic, S\'ersic, Seyfert,
LINER and {H\,{\sc ii}} galaxies, and ALGs.  We focus on the bulge
mass $M_{*, \rm bulge}$ which is a powerful predictor of galaxy
properties such as Hubble type, $n$, $B/T$, $\sigma$ and $M_{\rm BH}$.

  \subsubsection{Core-S\'ersic galaxies }\label{CoreSD}%

  Core-S\'ersic galaxies have depleted cores
  \citep{2003AJ....125.2951G,2004AJ....127.1917T,
    2006ApJS..164..334F,2009ApJS..182..216K,2013AJ....146..160R,2012ApJ...755..163D,2013ApJ...768...36D,2014MNRAS.444.2700D,2017MNRAS.471.2321D,2019ApJ...886...80D,2020A&A...635A.129K} and  are thought to be built up  through  one or more successive `dry' (gas-poor) mergers involving SMBHs \citep{1980Natur.287..307B,1991Natur.354..212E,1997AJ....114.1771F,2001ApJ...563...34M,2006ApJ...648..976M}.
They have high radio luminosities ($L_{R,\rm core} \ga 10^{35}$ erg
  s$^{-1}$) and represent 11.6 per cent (20/173) of our sample. They
  are massive ($M_{*,\rm bulge} \ga 10^{11} \rm M_{\sun}$) and
  bulge-dominated early-type galaxies hosting either LINER- or
  ALG-type nuclear emission, although one exception which has an
  \mbox{H\,{\sc ii}} nucleus is the S0 NGC~3665. Most (65 per cent) of
  the core-S\'ersic galaxies in the sample are radio-detected. In  \citet{2023arXiv230311154D}
   we revealed that core-S\'ersic galaxies tend to
  be systematically radio-loud and round
  (\mbox{$\epsilon = 0.21 \pm 0.08$}). They also show a tendency to
  have `boxy' distorted or pure elliptical (i.e.\ neutral) isophotes
  (\mbox{$B_{4} = (-0.007 \pm 0.08)\times10^{-2}$}), confirming past
  work (e.g.\
  \citealt{1983A&A...127..205H,1997AJ....114.1771F,1984MNRAS.206..899D,1991AJ....101..148W,1965ApJ...141.1560S,1995MNRAS.276.1373S,2005A&A...440...73C,2006A&A...447...97B,2011MNRAS.415.2158R,2015ApJ...798...55D,2018MNRAS.475.4670D}). Our
  results here also confirm that most (90 per cent) slow-rotators are
  core-S\'ersic galaxies
  \citep[see][]{2013MNRAS.433.2812K,2017MNRAS.464..356V}.

  Despite both core-S\'ersic LINERs and ALGs having large bulge
  masses, high bulge luminosities, high $n$ values ($\ga 4$) and
  similar inner profile slopes ($\gamma \la 0.3$) and $B_{4}$ values,
  there is a significant difference in the radio detection fraction
  for the two types, i.e.\ 9/11 (82 per cent) and 3/8 (38 per cent),
  respectively.  We attribute these results as being associated with
  the level of gas accretion onto the SMBH (or lack thereof).  In
  light of the {\it HST} structural analysis and radio emission
  analysis, we postulate that core-S\'ersic galaxies with LINER nuclei
  likely involve fuelling gas for the AGN (in hot accretion mode,
  e.g.\ \citealt{2007MNRAS.379..894B,2009Natur.460..213C,2013MNRAS.430.3086G}) and some
  low-level, residual star formation (see Fig.~\ref{RedBlue}), whereas the growth of core-S\'ersic ALGs (most of which are ellipticals and thus do not have a supply of cold gas) may largely involve a simple addition
  of stars and little gas (e.g.\ \citealt{1997AJ....114.1771F,2009ApJS..181..486H}).  The cooling of hot gas and/or accretion of
  a gas-rich satellite are mechanisms through which massive early-type
  galaxies and brightest cluster galaxies (BCGs) can acquire cold gas,
  allowing for episodes of low-level star formation (e.g.\
  \citealt{2003A&A...412..657S,2008ApJ...681.1035O,2010MNRAS.406..382K,2011MNRAS.414..940Y,2014ApJ...784...78R,2017MNRAS.471L..66S}). An alternative interpretation compatible with the lack of nuclear
  activity in massive ALGs is an intermittent gas accretion of the AGN
  that leads to short lived AGN episodes, namely the recurrence
  scenario (e.g.\ \citealt{2015MNRAS.451.2517S,2017NatAs...1..596M}).

For core-S\'ersic LINER galaxies, the stronger AGN feedback from their
massive SMBHs can generate high nuclear radio emission, explaining why
they trace the high-mass end of the steeper $L_{R,\rm core}$ scaling
relations, in contrast to core-S\'ersic ALGs which lie below their
LINER counterparts on the inactive relations.  In
\citet{2020ApJ...898...83D} we have demonstrated such core-S\'ersic
galaxies are red but not strictly dead (see
\citealt{2011MNRAS.418L..74D,2019MNRAS.486.1404D}). With UV fluxes
being detectable by GALEX (Fig.~\ref{RedBlue}), core-S\'ersic LINER
galaxies fall on the high-mass end of the tight (SMBH
mass)-(UV$-$[3.6] colour) red sequence which is populated by older
bulges with more massive SMBHs. Within the core-S\'ersic LINER galaxy
formation scenario, any residual star formation has to be kept at low
level by the AGN feedback, otherwise a substantial late-time gas
inflow and subsequent starburst events would regenerate a S\'ersic
galaxy after having replenished any preexisting depleted core scoured
by binary SMBHs.  Radio jets are invoked to inject energy as an
additional source of heating to prevent the hot X-ray gas from cooling
and to maintain radio-mode feedback episodes (e.g.\
\citealt{2006MNRAS.365...11C,2006MNRAS.368L..67B,2007ARA&A..45..117M,2007MNRAS.380..877S,2008MNRAS.391..481S,2009ApJ...698..594M,2011ApJ...727...39M,2009Natur.460..213C,2012ARA&A..50..455F,2014ARA&A..52..589H,2021A&A...648A..17V})
in the most massive early-type galaxies and BCGs which reside in
groups and clusters (e.g.\ \citealt{2008ApJ...681.1035O,
  2014ApJ...784...78R,2017MNRAS.471L..66S,2019MNRAS.490.3025R}). Of
the 20 core-S\'ersic galaxies in our sample, four (20 per cent) indeed
display clear evidence for such radio jets.
  
\subsubsection{S\'ersic galaxies}%

S\'ersic galaxies make up $\sim$88 per cent (153/173) of our
sample. They are coreless and span a wide range in radio luminosities
($L_{R,\rm core} \sim 10^{32}-10^{40}$ erg s$^{-1}$).  S\'ersic bulges
exhibit a large $\gamma$ range, from 0.01 to 0.70, and the vast
majority ($\sim 87$ per cent) of them have
$M_{*,\rm bulge} \la10^{11} \rm M_{\sun}$.  They are associated with
all optical emission classes and have a radio detection fraction (46
per cent) lower than their core-S\'ersic counterparts (65 per cent),
unsurprising given that radio detection increases with increasing
bulge stellar mass. The fraction of S\'ersic galaxies that are jetted
is $\sim$12 per cent, lower than that for core-S\'ersic galaxies (20
per cent). All but one of the Seyferts and \mbox{H\,{\sc ii}} hosts
are S\'ersic galaxies; the only exception, NGC~3665, is mentioned
above.  S\'ersic bulges grow primarily via gas-rich processes
(gas-rich mergers and secular evolution), accompanied by cold gas
dissipation and starbursts in the nuclear region
\citep{1997AJ....114.1771F,2009ApJS..181..135H}.  Their stellar
populations are formed over a protracted period of time
\citep{2005ApJ...621..673T,2009ARA&A..47..371T,2010MNRAS.404.1775T,2011MNRAS.418L..74D,2015MNRAS.448.3484M}.

\subsubsection{Seyferts} %

All ten Seyferts in the sample are S\'ersic galaxies and overall have
intermediate-mass bulges ($10^{10}-10^{11} \rm M_{\sun}$) which are
similar to those for the S\'ersic ALGs and intermediate-mass S\'ersic
LINERs.  Over this mass range, 88 per cent of Seyferts and 75 per cent
of S\'ersic LINER galaxies have radio detections, in contrast with the
low detection rate found for ALGs (9 per cent).  Having their
$M_{*,\rm bulge}, M_{V, \rm bulge}$ and $\sigma$ correlated well with
$L_{\rm R,core}$, Seyfert galaxies unite with LINERs to jointly define
the active $L_{\rm R,core}-M_{*,\rm bulge}$,
$L_{\rm R,core}-M_{V, \rm bulge}$ and $L_{\rm R,core}-\sigma$
sequences.  These results suggest an AGN origin for the radio emission
in Seyfert galaxies.  The key difference between Seyferts and
intermediate-mass S\'ersic LINERs is their Hubble type; most (70 per
cent) Seyferts are spiral galaxies, while the latter bifurcate into 57
per cent spirals and 43 per cent S0s. Our observations suggest that,
in their late evolution, Seyfert bulges probably have undergone major,
gas-rich mergers and possibly gas-rich accretion events (e.g.\
\citealt{2009ApJ...694..599H,2009ApJS..181..135H,2012A&A...540A..23S}),
which are gentler than the very violent major merger events that
core-S\'ersic and massive S\'ersic galaxies are formed (e.g.\
\citealt{2006ApJ...636L..81N,2009ApJS..181..486H}).  \citet{2006ApJS..166....1H,2009ApJ...694..599H} suggested  the accretion of cold gas stochastically onto a central SMBH for fuelling low mass ($M_{\rm *,bulge} \la 10^{10} \rm M_{\sun}$ and $M_{\rm BH} \la 10^{7} \rm M_{\sun}$) Seyferts. The same cold  accretion mode may account for our intermediate-mass Seyfert galaxies \citep{2006ApJS..166....1H}. In
\citet{2020ApJ...898...83D} we highlighted that early- and late-type
galaxies define a red and blue (SMBH mass)-color sequence, attributed
to two distinct channels of SMBH growth for the two Hubble types (e.g.\ \citealt{2020ApJ...897..102C}).
Within this picture, the formation and SMBH growth of Seyfert
galaxies, most of which are late-type galaxies, are likely to involve
a higher level of nuclear activity than their LINER counterparts due
to the increased gas availability for enhanced feeding of their
SMBHs. This is also evident from their high radio detection rate.
  
 \subsubsection{Absorption Line Galaxies }
 
ALGs are commonly regarded to lack nuclear activity \citep[][and references therein]{2018MNRAS.476.3478B}. While a full understanding of their nature is lacking,  the inactivity in some ALGs can be due to a strong, internal dust  obscuration that diminishes  a weakly active SMBH   \citep{2009MNRAS.398.1165G,2015A&A...584A..42A}. Of the 23 ALGs
 in the sample, eight are core-S\'ersic galaxies and the remaining 15
 are S\'ersic galaxies.  As noted in Section~\ref{CoreSD}, our
 core-S\'ersic ALGs have large bulge masses
 ($M_{*,\rm bulge} \ga10^{11} \rm M_{\sun}$). This implies massive
 SMBHs ($M_{\rm BH} \ga 5 \times 10^{8} \rm M_{\sun}$), considering
 the mutual growth of SMBHs and galaxies, i.e.\ the
 $M_{\rm BH}-M_{*, \rm bulge}$ correlation (e.g.\
 \citealt{2002MNRAS.331..795M}) but core-S\'ersic ALGs appear to have
 low radio detection rates (38 per cent) for their large bulge and
 SMBH masses. S\'ersic ALGs consist of low mass, local dwarfs
 ($M_{*,\rm bulge} \la 10^{10} \rm M_{\sun}$), and intermediate mass
 ($M_{*,\rm bulge} \sim 10^{10}-10^{11} \rm M_{\sun}$) lenticular and
 elliptical galaxies. Together with \mbox{H\,{\sc ii}} bulges, not
 only do the ALG bulges define the inactive
 $L_{R,\rm core}-M_{*,\rm bulge}$, $L_{R,\rm core}-M_{V, \rm bulge}$
 and $L_{R,\rm core}-\sigma$ sequences, but they also typically
 populate the high mass end of the relations.  Note that all 23 ALGs
 are early-type galaxies, the bulk (13/23) of which are elliptical
 galaxies. They appear to have lower radio emission than Seyferts and
 LINERs over the same $M_{*,\rm bulge}$, $M_{V,\rm bulge}$ and
 $\sigma$ ranges. Therefore, it seems reasonable to posit a gas-poor
 formation origin for ALGs, gas dissipation being less important for
 them than for any other emission line class. One ALG (NGC~3348)
 clearly exhibits a radio jet suggestive of an active AGN, conflicting
 with the optical emission line classification.  ALG bulges, when
 treated as a whole, their early-type morphology and the position on
 the $L_{R,\rm core}$ scaling relations points towards growth being
 dominated by gas-poor processes. For the massive and
 intermediate-mass ALGs, our observations favour gas-poor mergers as
 the formation path, which starve the massive SMBHs and give rise to
 very weak AGN accretion activity
 \citep{2018MNRAS.476.3478B,2021MNRAS.500.4749B,2021MNRAS.508.2019B},
 whereas for the low mass ALGs, environmental processes such as tidal stripping
  \citep{1972ApJ...176....1G} could account for the removal
 of gas from the galaxy \citep{2005ApJ...629..259R} explaining the low
 nuclear radio emissions from the low mass BHs. As noted above,
 massive ALGs may be due to an intermittent gas accretion of the AGN.
 
\subsubsection{LINERs}

LINERs are among the most massive galaxies in the sample. They have
high incidence of radio detection and brighter radio core
luminosities. They split nearly evenly between early-type (51 per
cent) and late-type (49 per cent) morphologies and show a large range
in mass. As noted above, they unite with Seyfert galaxies and jointly
follow the active $L_{R,\rm core}$ scaling relations.  Of the 71
LINERs in our sample, 12 are core-S\'ersic galaxies
($M_{*,\rm bulge} \ga 10^{11} \rm M_{\sun}$), 14 massive S\'ersic
galaxies ($M_{*,\rm bulge} \ga10^{11} M_{\sun}$), 31 intermediate-mass
S\'ersic galaxies
($M_{*,\rm bulge} \sim 10^{10}-10^{11} \rm M_{\sun}$), and 15 low-mass
S\'ersic galaxies ($M_{*,\rm bulge} \la 10^{10} \rm M_{\sun}$). As
discussed above, the properties of the core-S\'ersic LINER galaxies
are reproduced by gas-poor major merger events that involve low-level,
residual star formation and gas fuelling for the AGN. We find
radio-loud LINERs, which are known to behave like radio galaxies
(e.g.\
\citealt{2018MNRAS.476.3478B,2021MNRAS.500.4749B,2021MNRAS.508.2019B}),
typically sit on the red-sequence in the
(FUV--[$3.6$])-$M_{*,\rm bulge}$ plane.  Massive S\'ersic LINERs and
core-S\'ersic LINERs show similar $M_{*,\rm bulge} $,
$L_{\rm R,core}$, $\sigma$ and radio detection fractions, and the
S\'ersic index values for the two types agree within the errors
($n_{\rm Massive\_Ser\_LINER} = 3.22 \pm 0.84$,
$n_{\rm coreSer\_LINER} = 4.23 \pm 1.22$). Furthermore, the former are
hosted mainly by S0s (40 per cent) and spiral galaxies (43 per cent),
while the latter are largely associated with elliptical galaxies (82
per cent).  Although our analysis supports a dissipative major merger
scenario as a formation path for S\'ersic LINERs with massive bulges
\citep{1997AJ....114.1771F,2009ApJS..181..135H}, an alternative
mechanism is the rejuvenation of a core-S\'ersic galaxy by recent
nuclear star formation which erases the depleted core (e.g.\
\citealt{2010MNRAS.404.1775T,2019ApJ...877...48C}).

Our results indicate that intermediate-mass S\'ersic LINERs which
exhibit similarities with Seyfert galaxies are a lower-mass extension
of the S\'ersic LINERs populating the massive end.  Collectively, our
observations are well reconciled with predominantly AGN-driven radio
emission for LINER galaxies with intermediate-to-massive
bulges. Low-mass LINER and {\mbox{H\,{\sc ii}} galaxies appear
  indistinguishable in terms of $L_{\rm R,core}$, $n$ and Hubble type,
  although the former show slightly higher radio detection fraction
  (40$\pm$19 per cent) than the latter (21$\pm$7 per cent) within the
  errors.  Consequently, the formation of low-mass, LINER bulges, akin
  to their \mbox{H\,{\sc ii}} analogues, likely involves non-merger
  mechanisms, e.g.\ secular processes driven by non-axisymmetric
  stellar structures (e.g.\
  \citealt{2011ApJ...727L..31S,2013MNRAS.429.2199S}).

 \subsubsection{\mbox{H\,{\sc ii}} galaxies}

 Of the 69 \mbox{H\,{\sc ii}} galaxies in the sample, 68 are S\'ersic,
 the only core-S\'ersic \mbox{H\,{\sc ii}} galaxy is NGC~3665.  The
 majority (80 per cent) of the \mbox{H\,{\sc ii}} galaxies have
 low-mass bulges ($M_{*,\rm bulge} \la10^{10} \rm M_{\sun}$).  The
 remaining 20 per cent possess intermediate-mass bulges
 ($M_{*,\rm bulge} \sim 10^{10}-10^{11} \rm M_{\sun}$), except for one
 galaxy (NGC~3665) whose bulge is extremely massive
 ($ M_{*,\rm bulge} \sim 10^{12.5} \rm M_{\sun}$). In fact, NGC~3665 and
 NGC~4217 are the two \mbox{H\,{\sc ii}} galaxies in our sample having
 a clear `jet-like' radio structure, while for another two
 \mbox{H\,{\sc ii}} galaxies with intermediate masses (NGC 2782, and
 NGC 3504) their jet morphology is less secure.  As expected, all our
 bulgeless galaxies harbour \mbox{H\,{\sc ii}} nuclei.  The low-mass
 {\mbox{H\,{\sc ii}} galaxies (98 per cent of which are late-type
   galaxies) exhibit a low radio detection fraction of 12/55 (21.8 per
   cent), compared with a higher detection fraction of 10/15 (64 per
   cent) for the higher mass bin (which consists of 79 per cent
   late-types and 21 per cent early-types). \mbox{H\,{\sc ii}}
   galaxies and ALGs are described jointly by the inactive
   $L_{R,\rm core}-M_{*,\rm bulge}$, $L_{R,\rm core}-M_{V, \rm bulge}$
   and $L_{R,\rm core}-\sigma$ correlations, although upper limit
   $L_{R,\rm core}$ values are used for 67.6 per cent of the
   \mbox{H\,{\sc ii}} galaxies. Despite both being categorised as
   inactive, ALGs and \mbox{H\,{\sc ii}} galaxies have a crucial
   distinction: ALGs are associated with gas-poor processes and weakly
   active (or fully switched off) AGN, while the \mbox{H\,{\sc ii}}
   galaxies, particularly at the low-mass end, which likely contain
   high molecular gas fractions, are consistent with their nuclear
   radio emission being generated mainly by star formation
   \citep{2018MNRAS.476.3478B,2021MNRAS.500.4749B,2021MNRAS.508.2019B}.
   
   Intermediate-mass \mbox{H\,{\sc ii}} bulges appear
   indistinguishable from similar-mass Seyfert and LINER bulges in
   terms of radio detection fraction, $L_{R,\rm core}$,
   $M_{V, \rm bulge}$, $n$ and location on the $L_{\rm R,core}$
   scaling diagrams, but the former contain slightly more late-type
   galaxies (85$\pm$35 per cent) than the latter two (70$\pm$40 per
   cent and 57$\pm$17 per cent, respectively). At intermediate masses,
   the growth of \mbox{H\,{\sc ii}} bulges might be dominated by
   gas-rich major merger events, but it is unclear whether the nuclear
   radio emission has an AGN or a star formation origin. For the
   low-mass \mbox{H\,{\sc ii}} galaxies, the high-level of star
   formation is generally the main source of energy for the nuclear
   radio emission, dominating over the low-mass BH activity (e.g.\
   \citealt{2017MNRAS.472L.109A}). For such galaxies, the bulge growth
   scenario is through secular evolution (e.g.\
   \citealt{1993IAUS..153..209K,2013ARA&A..51..511K,2004ARA&A..42..603K}),
   where gravitational torques induced by non-axisymmetric stellar
   structures such as bars and spirals drive the rearrangement of disc
   material by channeling it from the large-scale disc into the
   nuclear region
   \citep{1989Natur.338...45S,1995ApJ...454..623K,2000ApJ...528..677E,
     1992MNRAS.259..345A,2005A&A...441.1011G,2021ApJ...917...53A},
   fuelling star formation at the centre and slowly feeding the low
   mass BH, $M_{\rm BH} \sim 10^{4}-10^{6} \rm M_{\sun}$
   \citep{2010ApJ...721...26G,2011ApJ...737L..45J,2011ApJ...727L..31S,2012ApJ...744..148K,2013MNRAS.429.2199S}. Indeed,
   41 (59 per cent) of our \mbox{H\,{\sc ii}} galaxies are classified
   as barred in the Third Reference Catalogue, RC3
   \citep{1991rc3..book.....D}.
 
\section{Summary and conclusions}\label{ConV} 

The main goal of this work has been to investigate the nuclear
activity in 173 LeMMINGs galaxies (23 Es, 42 S0s, 102 Ss and 6 Irrs)
both in the optical and radio at sub-arcsec resolutions, and in so
doing highlight the role of the bulge in dictating the AGN activity
and star formation events in nearby galaxies.  Given that most
LeMMINGs galaxies are low-luminosity AGN, we use our 1.5 GHz,
high-sensitivity and high angular resolution \mbox{$e$-MERLIN} radio
observations
\citep{2018MNRAS.476.3478B,2021MNRAS.500.4749B,2021MNRAS.508.2019B}. These
allowed for accurate identification of the AGN, singling out any
potential contamination from star formation.  This is coupled with
accurate photometric and structural parameters, luminosities and
stellar masses for the bulge components and the full galaxies, which
we determined from detailed, multicomponent decompositions of
high-resolution {\it HST} surface brightness profiles, fitting
simultaneously up to six galaxy components, e.g.\ bulges, discs,
depleted core, AGN, nuclear star clusters, bars, spiral arms, rings
and stellar haloes  \citep{2023arXiv230311154D}.  These
complementary LeMMINGs studies provided bulge stellar masses and radio
core luminosities that span wide ranges of
$M_{*, \rm bulge}\sim 10^{6}-10^{12.5} \rm M_{\sun}$ and
$L_{\rm R, \rm core}\sim 10^{32}-10^{40}$ erg s$^{-1}$. They
represent, to date, the most comprehensive radio and optical view of
the connection between galactic nuclei and bulges in the nearby
Universe.  We have also quantified the galaxy environment using two
approaches: halo mass and local luminosity surface densities measured
within the 3rd and 10th nearest neighbour of the target galaxy
($\nu_{3}$ and $\nu_{10}$). There are 140 galaxies in common between
the full LeMMINGs and {\it GALEX}/S$^{4}$G \citep{2018ApJS..234...18B}
samples. We first separate these overlapping galaxies into red
sequence, blue sequence and green valley objects using the {\it
  GALEX}/S$^{4}$G colour-colour diagrams \citep{2018ApJS..234...18B}
and then located them on the \mbox{(FUV--[$3.6$]})-$M_{*,\rm bulge}$
diagrams.

We find the following:

(1) The radio detection fraction increases with bulge mass
$M_{*,\rm bulge}$ and S\'ersic index $n$. At
$M_{*,\rm bulge} \ga 10^{11} \rm M_{\sun}$, the radio detection fraction
is 77 per cent, declining to 24 per cent for $M_{*,\rm bulge} < 10^{10} \rm M_{\sun}$.
Radio-jetted sources are hosted by the more massive galaxies over a
mass range of $M_{*,\rm bulge} \sim3.0\times10^{9} -3.2\times10^{12} \rm M_{\sun}$
and a median mass of
$M_{*,\rm bulge} \sim 6.0\times 10^{10} \rm M_{\sun}$.

 (2) We confirm that the fraction of galaxies harbouring emission-line
AGN and/or radio AGN is a strong function of $M_{*,\rm bulge}$ and
$M_{V,\rm bulge}$, although the correlations are less constrained for
the radio AGN.  The majority of AGN (80 per cent) and radio AGN (90 per cent) hosts
have $M_{\rm bulge}$ $\ga 10^{10} \rm M_{\sun}$
($M_{\rm glxy} \ga 10^{10.5} \rm M_{\sun}$). None of our 10 bulgeless galaxies host an AGN. The fraction of AGN galaxies
is such that $f_{\rm optical,AGN}\propto M_{*,\rm bulge}^{0.24 \pm 0.06}$ and
$f_{\rm optical,AGN}\propto M_{*,\rm glxy}^{0.30 \pm 0.05}$. The radio AGN
fraction scales as
$f_{\rm radio\_AGN}\propto M_{*,\rm bulge}^{0.24 \pm 0.05}$ and
$f_{\rm radio\_AGN} \propto M_{*,\rm glxy}^{0.41 \pm 0.06}$, markedly
different from that reported for the host of radio-loud galaxies by
\citet[][$f_{\rm radio\_loud} \propto M_{*,\rm
  glxy}^{2.5}$]{2005MNRAS.362...25B}.  For low-mass galaxies
($M_{*, \rm glxy} \la 5 \times 10^{9} \rm M_{\sun}$), we find an AGN
fraction of 9.4 per cent in agreement with the 10 per cent reported by \citet{2018MNRAS.476..979P}. 

(3)  Overall, there are   only weak correlations between  the radio core luminosity ($L_{\rm
  R,core}$) and   luminosity surface density ($\nu$) ($r_{s} \sim 0.13-0.20, P \sim
0.01231-0.0924$). The relation between $L_{\rm R,core}$ and halo mass ($M_{\rm
  halo}$) is consistent with the null hypothesis of no
correlation ($r_{s} \sim -0.07$, $P \sim
0.3811$). While the median radio core luminosities for our sample
appear to increase slightly as a function of
$\nu$, our galaxies reside in all environments, regardless of
their $L_{\rm
  R,core}$. 

(4) At fixed bulge mass, our results are compatible with no
significant dependence of the radio core luminosity, radio loudness,
nuclear radio activity and AGN (both optical and radio) fraction on
environmental metrics (i.e.\ including halo mass).

(5) Compared to RQ hosts, RL hosts preferentially possess an
early-type morphology and inhabit a denser environment as defined by
$\nu_{3}$.  While there is a mild tendency for RL galaxies to have
shallower inner logarithmic slopes ($\gamma$) than RQ galaxies, the KS
test on the $\gamma$ distributions for RQ and RL galaxies with
$P \sim 0.05$ is a borderline case.  Having performed KS tests on the
data sets for galaxy properties $n$, ellipticity, $B_{4}$, $\nu_{10}$
and $M_{\rm halo}$, the null hypothesis that the RQ and RL galaxies
are drawn from identical distributions cannot be rejected ($P$-values
$\sim 0.10-0.54$).

(6) $L_{\rm R,core}$ scales with  $M_{*, \rm bulge}$,
\mbox{$L_{V, \rm bulge}$}, \mbox{$\sigma$} and  $M_{\rm BH}$.  We find  hitherto unreported breaks in the
\mbox{$L_{\rm R,core} -M_{*, \rm bulge}$}, 
\mbox{$L_{\rm R,core}-L_{V, \rm bulge}$} and
\mbox{$L_{\rm R,core}-\sigma$} relations for datasets including upper
limits.  These turnovers, which separate AGN and  star formation dominated radio emissions 
occur at $M_{*, \rm bulge}\sim 10^{9.8 \pm 0.3} \rm M_{\sun}$ and
$M_{V, \rm bulge} = -18.5 \pm 0.3$ mag, $\sigma = 85 \pm 5$ km
s$^{-1}$ and $M_{\rm BH} = 10^{6.8 \pm0.3} \rm M_{\sun}$. Separating
the sample into active (LINERs+Seyferts) and inactive
(ALG+\mbox{H\,{\sc ii}}) galaxies, an OLS regression analysis yields a
relation $L_{\rm R,core} \propto M_{*, \rm bulge}^{1.42 \pm 0.18 }$
for active galaxies with a slope that is double that for the inactive
galaxies $L_{\rm R,core} \propto M_{*, \rm bulge}^{0.68 \pm 0.08
}$. This trend of steeping slope for the AGN galaxies is echoed in the
\mbox{$L_{\rm R,core}-L_{V, \rm bulge}$} and
\mbox{$L_{\rm R,core}-\sigma$} relations. Whether a single
log--linear  $L_{\rm R,core}-M_{\rm BH}$ relation holds over the entire
radio core luminosity range is unclear. 

(7) The hosts of active galaxies in general sit above the blue
sequence defined by inactive, star-forming galaxies.  Similarly, the
`jetted' radio AGN hosts are predominantly (85 per cent) blue sequence
galaxies; co-spatial with the massive green valley hosts or lie on the
red end of the blue sequence. These results reinforce the notion that
AGN feedback shuts off or suppresses star formation in galaxies.

(8) Core-S\'ersic galaxies, which represent $\sim$12 per cent (20/173)
of our sample, have high radio luminosities
($L_{\rm R,\rm core} \ga 10^{35}$ erg s$^{-1}$) and are, in general, 
bulge-dominated early-type galaxies hosting either a LINER- or an
ALG-type nuclear emission. We conclude that
core-S\'ersic galaxies with LINER nuclei likely involve AGN fuelling
in hot gas accretion mode and some low level, residual star formation,
whereas the growth of core-S\'ersic ALGs may largely involve simple
addition of stars and little gas. On the other hand, S\'ersic galaxies, which have 
$L_{\rm R,\rm core} \sim 10^{32}-10^{40}$ erg s$^{-1}$, are associated
with all optical emission classes.

(9) All Seyferts are S\'ersic galaxies, typically having intermediate mass
bulges ($10^{10}-10^{11} \rm M_{\sun}$).  Our observations suggest
that Seyfert bulge formation and SMBH growth are likely to
involve higher level of nuclear activity than their LINER analogues
due to  increased gas availability. LINERs are among the most massive galaxies in the
sample having the highest incidence of radio detection and brightest
radio core luminosities.  ALGs have lower radio emissions than Seyferts and LINERs over
the same $M_{*,\rm bulge}$, $M_{V,\rm bulge}$ and $\sigma$
ranges. Our observations suggest gas-poor formation origin for
ALGs. The vast majority (80 per cent) of the
\mbox{H\,{\sc ii}} galaxies have low mass bulges
($M_{*,\rm bulge} \la10^{10} \rm M_{\sun}$) and a plausible scenario is that such
 bulges grow in a secular fashion  where non-axisymmetric stellar structures fuel star
formation at the centre, slowly feeding their SMBHs. All 10 bulgeless
galaxies in the sample harbour \mbox{H\,{\sc ii}} nuclei.

\section{ACKNOWLEDGMENTS}
We thank the referee for valuable comments which improved the presentation of this paper. 
B.T.D. acknowledges support from grant `Ayudas para la realizaci\'on
de proyectos de I+D para j\'ovenes doctores 2019.' for the HiMAGC
(High-resolution, Multi-band Analysis of Galaxy Centres) project
funded by Comunidad de Madrid and Universidad Complutense de Madrid
under grant number PR65/19-22417. B.T.D and A.G.d.P acknowledge
financial support from the Spanish Ministry of Science, Innovation and
Universities (MCIUN) under grant No. RTI2018-096188-B-I00.
J.H.K. acknowledges financial support from the State Research Agency
(AEI-MCINN) of the Spanish Ministry of Science and Innovation under
the grant `The structure and evolution of galaxies and their central
regions' with reference PID2019-105602GB-I00/10.13039/501100011033,
from the ACIISI, Consejer\'{i}a de Econom\'{i}a, Conocimiento y Empleo
del Gobierno de Canarias and the European Regional Development Fund
(ERDF) under grant with reference PROID2021010044, and from IAC
project P/300724, financed by the Ministry of Science and Innovation,
through the State Budget and by the Canary Islands Department of
Economy, Knowledge and Employment, through the Regional Budget of the
Autonomous Community. AA thanks the Spanish Ministerio de Ciencia e 
Innovación (grant
PID2020-117404GB-C21) and the State Agency for Research of the Spanish 
MCIN through the “Center of Excellence Severo Ochoa” award for the 
Instituto de Astrofísica de Andalucía (SEV-2017-0709). We would like to 
acknowledge the support the
\emerlin\ Legacy project `LeMMINGs', upon which this study is based.
\emerlin\, and formerly, MERLIN, is a National Facility operated by
the University of Manchester at Jodrell Bank Observatory on behalf of
the STFC. We acknowledge Jodrell Bank Centre for Astrophysics, which
is funded by the STFC. This research has made use of the NASA/IPAC
Extragalactic Database (NED), which is operated by the Jet Propulsion
Laboratory, California Institute of Technology, under contract with
the National Aeronautics and Space Administration.

This work has made use of {\sc numpy} \citep{2011CSE....13b..22V},
{\sc matplotlib} \citep{Hunter:2007} and {\sc corner}
\citep{2016JOSS....1...24F} and {\sc astropy}, a community-developed
core {\sc python} package for Astronomy
\citep{2013A&A...558A..33A,2018AJ....156..123A}, {\sc cubehelix }
\citep{2011BASI...39..289G}, {\sc jupyter}
\citep{2016ppap.book...87K}, {\sc scipy} \citep{2020NatMe..17..261V},
and of {\sc topcat} (i.e.\ `Tool for Operations on Catalogues And
Tables', \citealt{2005ASPC..347...29T}).

\section{Data availability}

The data underlying this article are available in the article.

\bibliographystyle{mn2e}
\bibliography{Bil_Paps_biblo.bib}

\setcounter{section}{0}
\renewcommand{\thesection}{A\arabic{section}}

\section{Measures of galaxy  environments}\label{EnvR}

\setcounter{equation}{0}
\renewcommand{\theequation}{A\arabic{equation}}

We employ four local density estimators based on the nearest neighbour
method as measures of the galaxy environment ($\rho_{10}$, $\rho_{3}$,
$\nu_{10}$ and $\nu_{3}$, see e.g.\
\citealt{2011MNRAS.416.1680C,2012MNRAS.419.2670M,2017MNRAS.471.1428V}). We
choose the number of nearest neighbours to be $N_{\rm gal} =$ 3 and 10
which generally are suitable to probe different scales of the galaxy
local environment at the distances of our sample
\citep{2012MNRAS.419.2670M}.

The mathematical expression for the local surface density of a galaxy
$\rho_{10}$ is
 \begin{equation}
 \rho_{10}=  \frac{N_{\rm gal}}{\pi R_{10}^{2}}, 
  \label{EqA1}
 \end{equation}
 where $R_{10}$ is the radius, which is centred on the target galaxy
 and defines a cylinder that encloses the nearest $N_{\rm gal}$ (=10)
 neighbours. We use a $B$-band magnitude cut of $M_{B} \la -18.0$ mag
 for the neighbours, which are selected to be within a velocity
cylinder of $\pm$300 km s$^{-1}$ around the target galaxy (i.e.\ a
relative heliocentric velocity
\mbox{$|V_{\rm hel,target} -V_{\rm hel,neighbour}|<$ 300 km s$^{-1}$},
see \citealt[][their section~3.1]{2011MNRAS.416.1680C}). We note that
$N_{\rm gal}$ does not include the target galaxy.
  
The surface density $\rho_{3}$ is similar to $\rho_{10}$ but the
former is defined by a cylinder containing 3 nearest neighbours with
$M_{B} \la -18.0$ mag and a velocity cut of
\mbox{$\upDelta V_{\rm hel} <$ 300 km s$^{-1}$} and it can be written
as \renewcommand{\theequation}{A\arabic{equation}}
\begin{equation}
 \rho_{3}=  \frac{N_{\rm gal}}{\pi R_{3}^{2}}, 
  \label{EqA2}
 \end{equation}
where $R_{3}$ is the radius centred on the target galaxy enclosing
 the third nearest neighbour.
  
The quantity $\nu_{10}$ is the luminosity surface density of a galaxy
 inside a cylinder containing its 10 nearest neighbours and it can be
 written as
\begin{equation}
 \nu_{10}=  \frac{\sum^{10}_{i=1} 10^{-0.4(M_{i,B}-M_{\sun,B})}}{\pi R_{10}^{2}}, 
  \label{EqA3}
 \end{equation}
 where  the $B$-band solar  luminosity  is $M_{\sun,B} = 5.44$ mag
 \citep{2018ApJS..236...47W}. The index $i$ runs from 1 to 10,
 indicating the 10 nearest neighbours of the target galaxy and
 $R_{10}$ is defined in the same manner as in  $\rho_{10}$
 (Eq.~\ref{EqA1}). Similarly,  the luminosity surface density  $\nu_{3}$ is defined as 
\begin{equation}
 \nu_{3}=  \frac{\sum^{3}_{i=1} 10^{-0.4(M_{i,B}-M_{\sun,B})}}{\pi R_{3}^{2}}, 
  \label{EqA4}
 \end{equation}
 where $R_{3}$ is defined as in  $\rho_{3}$ (Eq.~\ref{EqA2}).

 The $B$-band absolute magnitudes of the galaxies are from Hyperleda
 \citep{2014A&A...570A..13M}.  We use NED for an automated search of
 the nearest neighbours around a target galaxy and for obtaining their
 radial velocities. NED represents the most current and comprehensive
 assimilation of extragalactic objects and their radial
 velocities. That is, it offers the most complete available data to
 define the environment of a galaxy using its radial velocity and the
 radial velocities of other galaxies typically observed along and
 around that same line of sight.  The maximum search radius for a
 nearest neighbour allowed by NED is 10 Mpc. The luminosity surface
 densities $\nu_{10}$ and $\nu_{3}$ are listed in see
 Table~\ref{TableD}.

 Fig.~\ref{Distribution_D3and10} plots the distribution of the values
 of $\rho_{10}$, $\rho_{3}$, $\nu_{10}$ and $\nu_{3}$ (see
 Table~\ref{TableD}) for our sample of 173 galaxies. We note that for
 the sample galaxies at distances $D \la 5$ Mpc (23/173) our values of
 $\rho_{10}$, $\rho_{3}$, $\nu_{10}$ and $\nu_{3}$ are upper
 limits. These local galaxy density estimates span roughly five orders
 of magnitude, suitable to test the influence of the local environment
 of galaxies on their evolution.
       
 Fig.~\ref{Env_Density_3vs_10} explores which local density method
 results in a better measurement of the galaxy environment by
 comparing our values of the surface density $\rho$ and luminosity
 surface density $\nu$ for a given number of neighbours. We find
 strong correlations between the two local density estimators $\rho$
 and $\nu$ (Spearman's correlation coefficient
 $r_{s} \sim 0.93-0.96, P \sim 10^{-99}-10^{-78}$), revealing that
 they provide a comparable measure of the galaxy environment.
 
 We next compare our $\rho_{10}$, $\rho_{3}$, $\nu_{10}$ and $\nu_{3}$
 values with those from \citet[][their tables 2 and
 3]{2011MNRAS.416.1680C} who tabulated local galaxy density estimates
 for the ATLAS$^{\rm 3D}$ parent sample of 871 nearby galaxies in the
 ATLAS$^{\rm 3D}$ volume with $M_{K} \le -21.5$ mag drawn from the Two
 Micron All Sky Survey (2MASS, \citealt{2006AJ....131.1163S}).
 Comparing the data, we find that their $K$-band magnitude cut of
 $M_{K} \le -21.5$ mag approximately corresponds to our $B$-band
 magnitude cut of $M_{B} \la -18.0$ mag. There are 91 galaxies in
 common between our sample and the ATLAS$^{\rm 3D}$ parent
 sample. Fig.~\ref{Env_AT3D_Ours} (a) shows how their values of
 $\rho_{10}$ compare with ours. Although the agreement is good for the
 bulk (75 per cent) of the overlapping galaxies, our $\rho_{10}$
 values are larger by more than a factor of 2 for the remaining 25 per
 cent. We attribute this discrepancy to differences in the database
 and filter (used to identify the nearest neighbours and measure their
 magnitudes) adopted by us and \citet{2011MNRAS.416.1680C}.  On
 average we have identified more neighbours over a given projected
 radius than \citet{2011MNRAS.416.1680C}.  As such, their median value
 of $R_{10}$ = 3.8 Mpc is larger than ours \mbox{$R_{10} \sim 2.9$
   Mpc}. While we do not find a one-to-one relation between the
 $\rho_{10}$ values from the two studies, Fig.~\ref{Env_AT3D_Ours} (b)
 reveals a tight correlation between the $K$-band luminosity surface
 density $\nu_{10}$ \citep{2011MNRAS.416.1680C} and our $B$-band
 $\nu_{10}$ for the 91 overlapping galaxies
 ($r_{s} \sim 0.83, P \sim 10^{-25}$).

 Finally, because galaxy morphology is well known to depend on the
 local environmental density through the so-called morphology-density
 relation \citep{1980ApJ...236..351D}, we go on and check if we can
 recover this relation. Having expanded our sample of galaxies by
 adding is 32 massive elliptical galaxies from
 \citet{2014MNRAS.444.2700D,2019ApJ...886...80D}, in
 Fig.~\ref{Morph_density_reln} we display the fraction of elliptical,
 S0 and spiral+Irr galaxies as a function of $\nu_{10}$ for 205
 (=173+32) nearby galaxies. We reproduce the traditional
 morphology-density relation reasonably well, see \citet[][their
 fig.~9]{2011MNRAS.416.1680C}, although we caution that our sample,
 which is representative of the statistically complete full LeMMINGs
 sample, suffers from incompleteness  \citep[see][]{2023arXiv230311154D}.  The
 trend in Fig.~\ref{Morph_density_reln} is that the fraction of
 late-type (spiral+Irr) galaxies declines with increasing local
 environmental density, and correspondingly the early-type (E+S0)
 fraction increases.

 \setcounter{figure}{0}
\renewcommand{\thefigure}{A\arabic{figure}} 

\begin{figure}
\hspace{-.94cm}
\includegraphics[trim={-1.99150915cm -8cm -3.cm 1.20cm},clip,angle=0,scale=0.42]{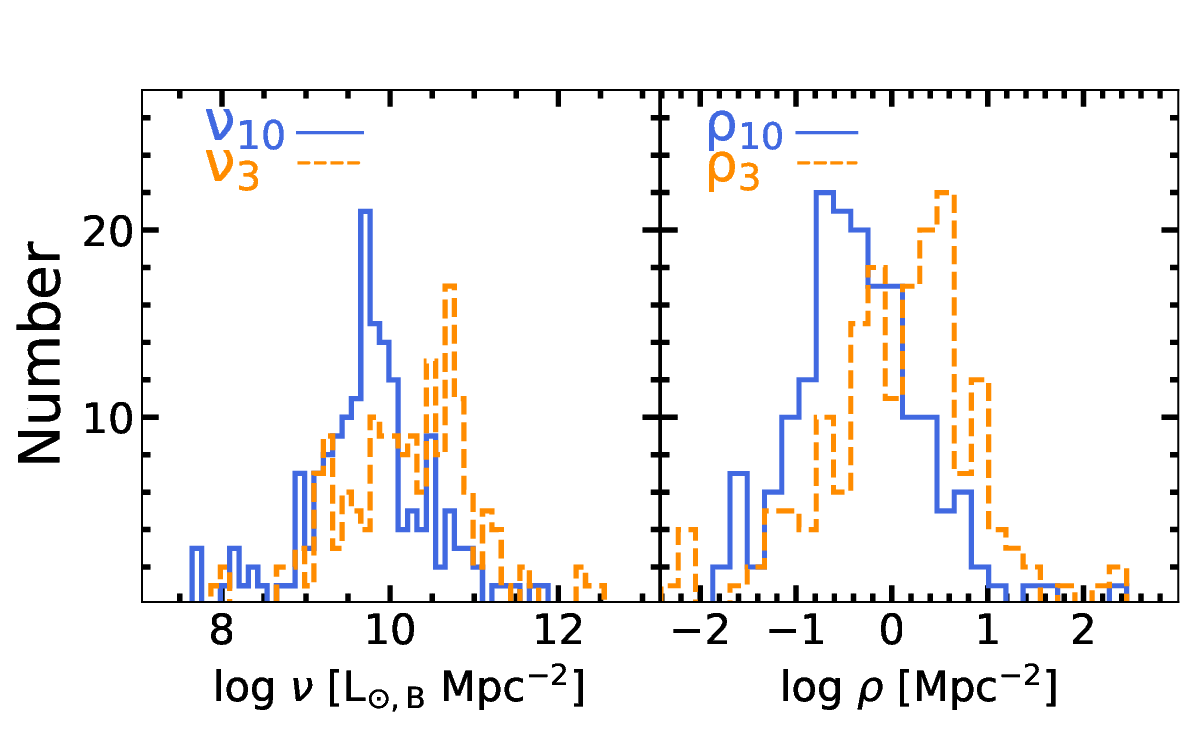}
\vspace{-4.093cm}
\caption{Histogram of the luminosity surface densities ($\nu_{10}$ and
  $\nu_{3}$) and surface densities ($\rho_{10}$ and $\rho_{3}$) for
  our sample of 173 LeMMINGs galaxies.}
   \label{Distribution_D3and10}
\label{Fig2}
\end{figure}

\begin{figure}
\hspace{-.7709208883cm}
\includegraphics[trim={-1.554215cm -11cm -1.6cm 1.380073cm},clip,angle=0,scale=0.401]{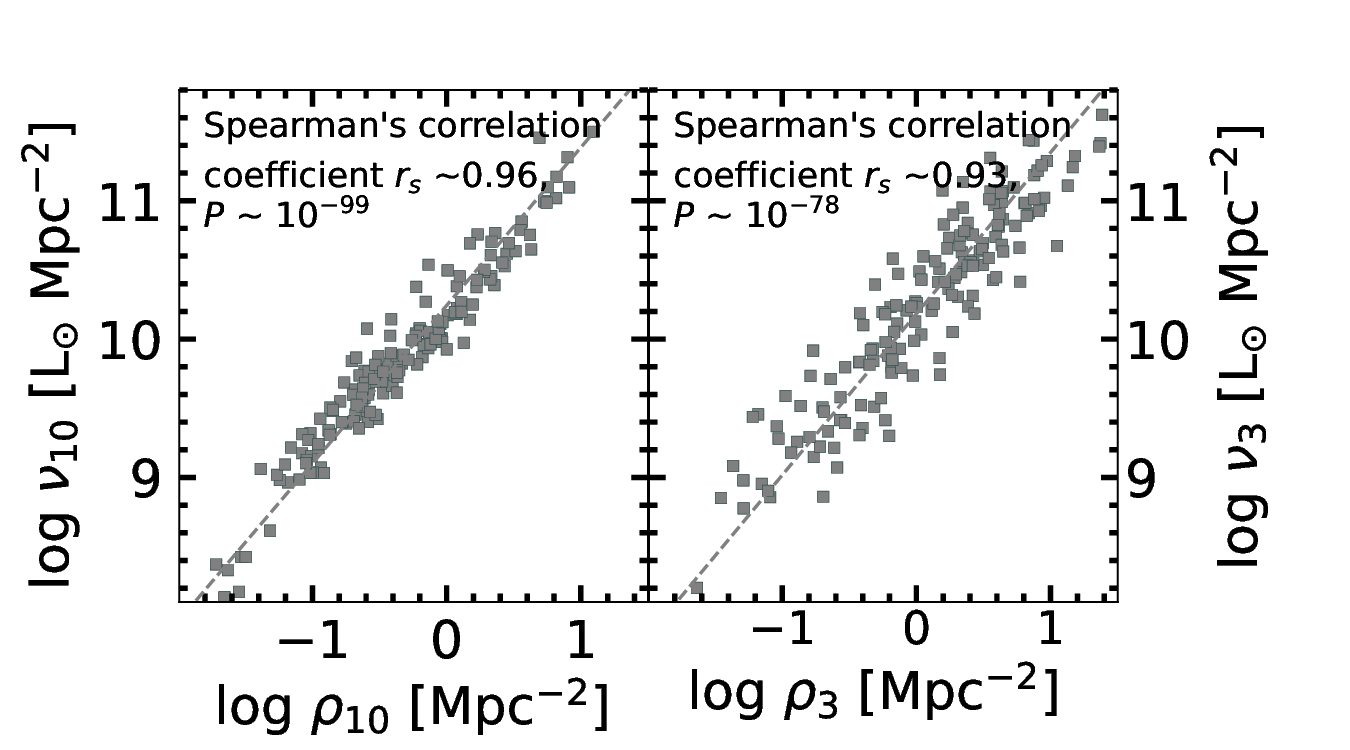}
\vspace{-4.9150cm}
\caption{ Correlations between luminosity surface density ($\nu$) and
  surface density ($\rho$) for our sample.}
\label{Env_Density_3vs_10}
\end{figure}

\begin{figure}
\hspace{-1.209208883cm}
\includegraphics[trim={-2.8cm -11cm -1.6cm 1.390073cm},clip,angle=0,scale=0.401]{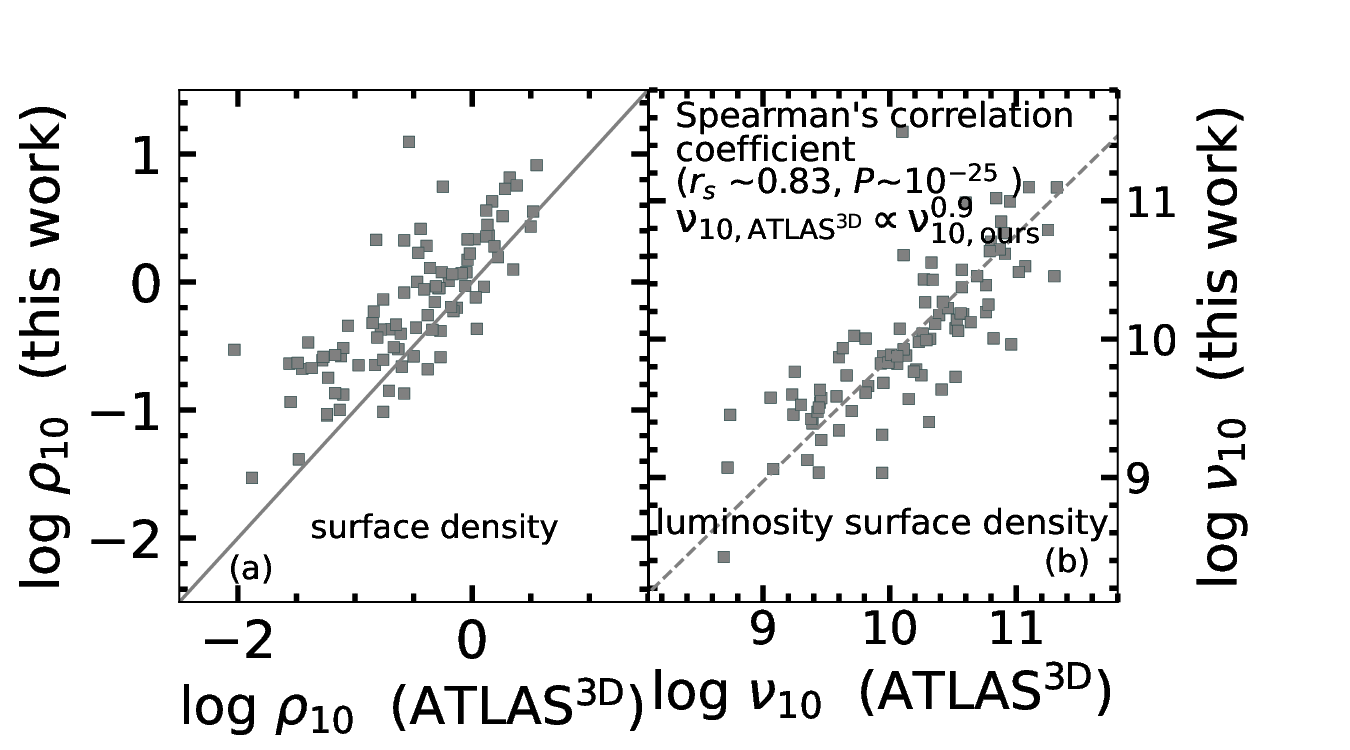}
\vspace{-4.92150cm}
\caption{(a) Comparison of our local surface density values with those
  from the ATLAS$^{\rm 3D}$ \citep{2011MNRAS.416.1680C} for 91
  overlapping galaxies.  The solid line shows a one-to-one
  relation. (b) Our $B$-band luminosity surface density values are plotted as a
  function of the ATLAS$^{\rm 3D}$ $K$-band luminosity surface density
  values. The dashed line is the least-squares fit to the luminosity
  surface density data ($\nu_{\rm 10,ATLAS^{3D}} \propto \nu_{\rm 10,ours}^{0.9}$). }
\label{Env_AT3D_Ours}
\end{figure}

\begin{figure}
\hspace{-.2654cm}
\includegraphics[trim={-.51834cm -6cm -3.cm 1.9940cm},clip,angle=0,scale=0.5483]{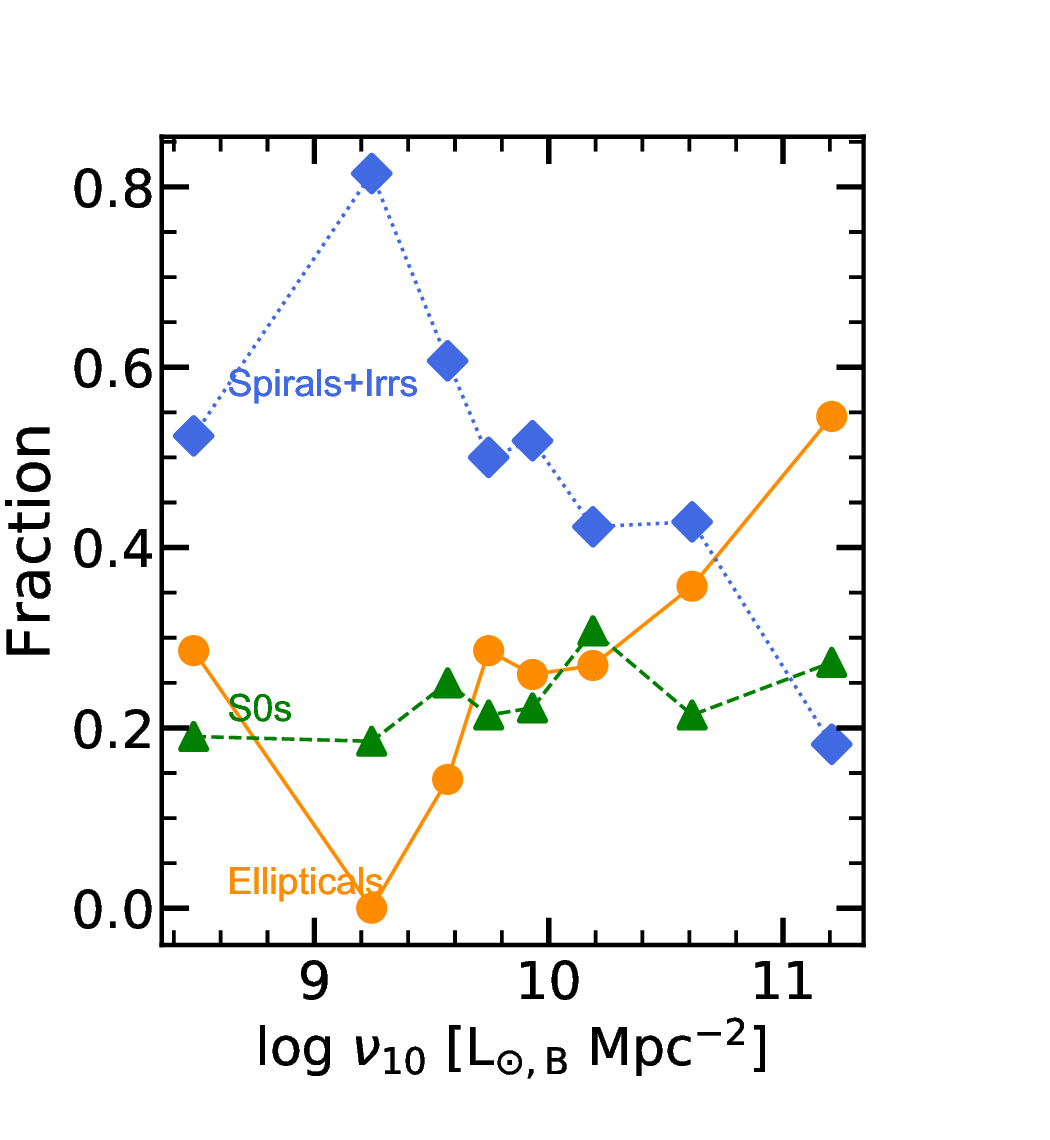}
\vspace{-4.28407cm}
\caption{Morphology-density relation. The fraction of elliptical
  (circles), S0 (triangles) and spiral+Irr galaxies (diamonds) plotted
  against our $B$-band luminosity surface density $\nu_{10}$ for 205
  (=173+32) nearby galaxies \citep[][plus this
  work]{2014MNRAS.444.2700D,2019ApJ...886...80D}. }
   \label{Morph_density_reln}
\label{Fig2}
\end{figure}

\newpage
\newpage

\setcounter{section}{1}
\renewcommand{\thesection}{A\arabic{section}}
\section{Data Tables}\label{DataTables}
Table~\ref{TableD} provides global and central properties of the
sample galaxies including distance, morphological classification,
velocity dispersion, bulge and galaxy stellar masses, optical and
radio luminosities, ellipticity, isophote shape parameter and
logarithmic slope of the inner light profiles of our sample galaxies.

\setcounter{table}{0}
\renewcommand{\thetable}{A\arabic{table}}

\begin{center}
\begin{table*} 
\begin{sideways}
\setlength{\tabcolsep}{0.0479in}
\begin {minipage}{242mm}
\caption{LeMMINGs data. }
\label{TableD}
\setlength{\tabcolsep}{2.775pt}
\begin{tabular}{@{}llcccccccccccccccccccccc@{}}
  \hline
  \hline
 Galaxy&Type&$D$&Class&  $\sigma$ &$M_{\rm V,bulge}$&$M_{\rm V,glxy}$&log$M_{\rm *,bulge}$&log$M_{\rm *,glxy}$ &log~$M_{\rm BH}$  &log$\rm L_{\rm R,core}$ &$\langle \epsilon \rangle$& Med ($B_{4}$)&$\gamma$&Det&AGN&Sequence&log$\nu_{10}$&log$\nu_{3}$& log$M_{\rm halo}$\\
        &&(Mpc)&&(km s$^{-1}$)&(mag)&(mag)&($\rm M_{\sun}$)&($\rm M_{\sun}$)&($\rm M_{\sun}$)&(erg s$^{-1}$)&&&&&&&($\rm L_{\sun}$ Mpc$^{-2}$)&($\rm L_{\sun}$ Mpc$^{-2}$)&($\rm M_{\sun}$)\\
  (1)&(2)&(3)&(4)&(5)&(6)&(7)&(8)&(9)&(10)&(11)&(12)&(13)&(14)&(15)&(16)&(17)&(18)&(19)&(20)\\   
  \hline     
  I0239    & SAB(rs)c &   9.9 & H {\sc{ii}}     &      92.3          &$-$16.82$\pm0.29$   &$-$18.80 &  9.04$\pm 0.14$    &9.84 &--- &<35.0    &0.18&0.003 &0.03&U   &--- &---&9.5 &11.2 &13.1 &\\
 I0356   & SA(s)abp &  11.7 & H {\sc{ii}}    &     156.6          &$-$16.63$\pm 0.41$ &$-$19.03 & 10.57$\pm 0.29$ &11.53 &--- &35.5      &0.18&0.006 &0.40&I    &  A &---&9.6 &11.0 &14.6\\
  I0520    & SAB(rs)a &  50.9 & L   	       &      138.1 	&$-$20.74$\pm 0.52$ &$-$22.62&  10.71$\pm 0.21$ &11.47  &--- &<36.1    &0.07&0.005 &0.62&U  &--- &---&10.3&10.7&13.6 \\
  I2574\textsuperscript{$\dagger$}      & SAB(s)m  &   2.1 & H {\sc{ii}}    &      33.9 &     --- &          $-$19.87 &    --- &                9.46 &      --- &        <33.7 &      ---&---& ---&U&   ---  &BS&<9.8&<10.7&12.3\\
  N0147&E5 pec& 0.8$^{\rm k}$&ALG &22.0                        &$-$15.76$\pm0.17$  &$-$15.76  &9.05$\pm0.12$	&9.05   &---   &<32.4   &0.30&0.002&0.00&U&---&---&<7.7&<7.3&13.6\\
  N0205&E5 pec& 0.8$^{\rm k}$ &ALG &23.3			&$-$18.22$\pm0.27$  &$-$18.22 & 9.53$\pm0.13$	& 9.53  &3.83$^{+1.18}_{-1.83}$&<32.4 &0.16&0.001&0.07&U&---&---&<7.7&<7.1&11.2\\
  N0221&compact E2& 0.8$^{\rm k}$&ALG &72.1 		&$-$15.52$\pm0.18$  &$-$16.48 &8.61$\pm0.11$	&8.99   &---   &<32.3    &0.24&0.006&0.24&U&---&---&<7.7&<7.2&11.2\\
  N0266&SB(rs)ab&  62.9&L &229.6					& $-$22.93$\pm0.20$ &$-$23.36 &11.80$\pm0.12$	&11.97 &---   &36.9      &0.18&0.003&0.39&I&A&---&11.8&12.3&13.6\\
  N0278&SAB(rs)b&5.3 &H {\sc{ii}} &47.6					&$-$16.04$\pm0.21$  &$-$17.65 &8.82$\pm0.12$	& 9.46  &---   &34.8      &0.10&0.001&0.15&I&A&---&<9.4&<9.1	&12.5\\
  N0315& cD&67.0 &L/RL &303.7					&$-$23.59$\pm0.27$  &$-$23.79 &11.69$\pm0.14$	&11.77 &8.99$^{+0.32}_{-0.32}$&39.6 &0.29&$-$0.021& 0.03&I&A&---&10.5&10.5&	13.5\\
  N0404&SA0-(s)&3.1$^{\rm k}$ &L &40.0 			&$-$17.39$\pm0.23$  &$-$17.40 &9.33$\pm0.13$	&9.33   &5.65$^{+0.25}_{-0.25}$&<33.3 & 0.14&0.005&0.31&U&---&---&<8.1&<8.0&14.4\\
  N0410& cD&72.4&L&299.7 					         &$-$20.72$\pm0.13$ &$-$21.51 &11.76$\pm0.10$	&12.09 &---   &37.3      &0.17&$-$0.020&$-$0.02&I&A&---&12.7&12.5&14.4\\
  N0507&SA00(r)&63.7 &ALG&292.0					&$-$22.56$\pm0.27$  &$-$23.48 &11.70$\pm0.13$	&12.07 &---   &36.9      &0.25&0.002&0.07&I&A&---&10.9&	11.3&	14.4 \\
  N0598&SA(s)cd&1.0$^{\rm b}$&H {\sc{ii}}  &21.0 				&$-$11.84$\pm0.21$  &$-$18.59 &6.89$\pm0.12$	&9.58   &---   &<32.4    &0.10&0.001&1.21&U&---&---&<8.1&<7.9&13.0\\
  N0672\textsuperscript{$\dagger$}  &SB(s)cd&7.2$^{\rm k}$ &H {\sc{ii}}  &64.3 &---  & $-$19.51  &--- &9.76 &---&<34.6 &---&---&---&U&---&---&9.6&9.1&	13.7\\
  N0777&E1&68.8 &L/RL &324.1 					&$-$20.68$\pm 0.22$ & $-$20.68 & 11.75$\pm 0.12$ & 11.75&---&36.8 &0.14&$-$0.073&0.07&I&A&---&10.2&10.8&13.5\\
  N0784\textsuperscript{$\dagger$}  &SBdm&5.0$^{\rm k}$ &H {\sc{ii}}  &35.5 &--- &$-$16.73 &--- &8.65 &---&<34.2 &---&---&---&U&---&---&<8.2&<9.0&12.9\\
  N0841&(R')SAB(s)ab&62.2 &L&159.2				&$-$21.34$\pm0.22$  &$-$22.23 &10.27$\pm0.12$ 	&10.62&---&<36.2 &0.42&0.005&0.01&U&---&---&9.7&11.3&	13.2\\
  N0890&SAB0-(r)?&54.0&ALG &210.9 				&$-$20.12$\pm0.19$  &$-$20.46 &11.41$\pm0.12$ 	&11.55&---&<36.0 &0.31&$-$0.013&0.50&U&---&---&9.2&10.4&	14.4\\ 
 N0959    & Sdm?     &   5.4 & H {\sc{ii}}     &      43.6 		&$-$14.68$\pm0.14$  &$-$16.12 & 	8.08$\pm0.09$ 	&8.65 &---  &<34.7 &0.07&0.016&0.12&U&   --- &---&<10.0&<10.5&	13.7   \\
  N1003    & SA(s)cd  &   6.1 & H {\sc{ii}}   &     --- 		&$-$11.25$\pm0.02$  &$-$18.12 & 6.97$\pm0.08$ 	&9.72 &---  &<34.6 &0.77&$-$0.185&0.16&U&   ---&---&9.5&11.2&14.0    \\
  N1023    & SB0-(rs)&   6.2 & ALG  &     204.5 			&$-$19.35$\pm0.26$  &$-$19.94 & 10.86$\pm0.13$   &11.10& 7.38$^{+0.04}_{-0.04}$  &<34.6 &    0.26&0.007&0.52&U&   ---&---&9.4&	11.1&	13.8   \\
N1058    & SA(rs)c  &   4.5 & L     &      31.0 			&$-$13.36$\pm0.34$  &$-$16.34 & 7.36$\pm0.15$ 	&8.55 & --- &<34.4 &0.16&0.006&0.01&U&   --- &---&	<9.7&<10.9&	14.0 \\
  N1156\textsuperscript{$\dagger$}      & IB(s)m   &   2.5 & H {\sc{ii}}      &      35.9 &         ---    &$-$16.81 &       ---&          8.51 &    --- &       <34.2 &      ---& --- &---&U&   ---&---&<9.4&<9.3&12.9   \\
  N1161    & S0       &  25.6 & L/RL     &     258.4 		       &$-$22.08$\pm0.07$  &$-$22.25 & 11.43$\pm0.10$ &    11.49 &     ---      &36.6 &0.35&0.006 &0.13&I& A&---&8.0&9.3&13.6 \\
  N1167    & SA0-    &  68.0 & L/RL     &     216.9	               &$-$19.34$\pm0.27$  &$-$20.23 & 11.57$\pm0.12$ &     11.84&    ---      &39.7 &0.08&$-$0.001&0.23 &I& A &---&10.0&10.8&	13.1\\
  \hline
  error   & ---&   --- &  ---   &    ---   &   ---  &   0.33& --- &0.17 dex   &  ---&     1.0 dex& 20\%&  35\%&10\%& ---& ---& ---&0.6 dex &0.6 dex &0.5 dex \\
  \hline
\end{tabular} 
{\it Notes.} (1) galaxy name. (2) morphological classification from RC3
\citep{1991rc3..book.....D}.  (3) distance ($D$) are primarily from
the NASA/IPAC Extragalactic Database (NED;
\url{http://nedwww.ipac.caltech.edu}), other sources are
\citet[][k]{2004AJ....127.2031K} and \citet[][b]{2006ApJ...652..313B}.
(4) optical spectral class from \citet{2021MNRAS.500.4749B}: \mbox{H
  {\sc{ii}}}, L = LINER, S = Seyfert and ALG = Absorption Line Galaxy. Radio-loud galaxies (RL) are shown (Section~\ref{Sec735}). 
(5) the central velocity dispersion ($\sigma$) from
\citet{2009ApJS..183....1H}.  (6)--(7) $V$-band bulge and galaxy
magnitudes calculated by integrating the best-fit S\'ersic, Gaussian
or core-S\'ersic functions.  (8)--(9) logarithm of the stellar masses
of the bulge and galaxy.  (10) logarithm of the SMBH mass
($M_{\rm BH}$) for galaxies with measured $M_{\rm BH}$ from
\citet{2016ApJ...831..134V}, except for \mbox{NGC 205} which is from
\citet{2019ApJ...872..104N}. The BH masses are adjusted here to our
distance.  (11) logarithm of the radio core luminosity
($L_{\rm R,core}$).  (12) average ellipticity of the galaxy inside
$R_{\rm e}$ after omitting the PSF-affected region
($\langle \epsilon \rangle$).  (13) median of the isophote shape
parameter inside $R_{\rm e}$ (Med ($B_{4}$)).  (14) negative, inner
logarithmic slope of the bulge light profile.  (15) radio detection of
the galaxies based on \citet{2021MNRAS.500.4749B} and following their
nomenclature `I' = detected and core identified; `unI?' = detected but
core unidentified; `U' = undetected; `I+unI?' = detected and core
identified having additional unknown source(s) in the FOV. Col.\ (16)
radio morphologies following \citet{2021MNRAS.500.4749B}: A =
core/core-jet; B = one-sided jet; C = triple; D = doubled-lobed ; E =
jet+complex.  (17) sample galaxies with {\it GALEX} NUV, FUV band and
{\it Spitzer} 3.6~$\upmu$m data in \citet[][]{2018ApJS..234...18B}
separated into `red sequence' (RS), `blue sequence' (BS) and `green
valley' (GV) based on the \mbox{(FUV -- NUV)$-$(NUV -- [3.6])} colour
$-$ colour diagram (\citet[][their eqs.\ 1--3]{2018ApJS..234...18B}.
(18)--(19) luminosity surface densities of a galaxy inside a cylinder
containing 10 and 3 nearest neighbours, $\nu_{10}$ and $\nu_{3}$. Note
that for (23/173) sample galaxies at distances $D \la 5$ Mpc our
values of $\nu_{10}$ and $\nu_{3}$ are upper limits.  (20) halo mass
($M_{\rm halo}$) based on the HDC catalogue from
\citet{2007ApJ...655..790C}, see the text for details.
\end {minipage}
\end{sideways}
\end{table*}
\end{center}

\begin{center}
\begin{table*} 
\begin{sideways}
\setlength{\tabcolsep}{0.0479in}
\begin {minipage}{242mm}
\setcounter{table}{0} 
\caption{(\it continued)}
\label{Tab40}
\setlength{\tabcolsep}{2.775pt}
\begin{tabular}{@{}llcccccccccccccccccccccc@{}}
\hline
\hline
Galaxy&Type& $D$&Class&  $\sigma$ &$M_{\rm V,bulge}$&$M_{\rm V,glxy}$&log$M_{\rm *,bulge}$&log$M_{\rm *,glxy}$ &log~$M_{\rm BH}$  &log$L_{\rm R,core}$ &$\langle \epsilon \rangle$& Med ($B_{4}$)&$\gamma$&Det&AGN&Sequence&log$\nu_{10}$&log$\nu_{3}$& log$M_{\rm halo}$\\
&&(Mpc)&&(km s$^{-1}$)&(mag)&(mag)&($\rm M_{\sun}$)&($\rm M_{\sun}$)&($\rm M_{\sun}$)&(erg s$^{-1}$)&&&&&&&($\rm L_{\sun}$ Mpc$^{-2}$)&($\rm L_{\sun}$ Mpc$^{-2}$)&($\rm M_{\sun}$)\\
(1)&(2)&(3)&(4)&(5)&(6)&(7)&(8)&(9)&(10)&(11)&(12)&(13)&(14)&(15)&(16)&(17)&(18)&(19)\\          
\hline   
N1275    & cDpec       &  73.9   & L/RL  		&258.9&   $-$23.50$\pm 0.43$ &       $-$23.77 &      11.47$\pm0.18$  &             11.57&       9.00$^{+0.20}_{-0.20}$  &              41.0&     0.23&$-$0.006&0.24&I&A &---&9.5&11.2&13.1 &\\
N1961    & SAB(rs)c   &  56.4   & L/RL  		&241.3&   $-$22.96$\pm 0.47$ &       $-$23.49 &      11.28$\pm 0.20$ &             11.49&       8.35$^{+0.35}_{-0.35}$  &              37.2&     0.21&0.014&0.30&I&A &---&9.7&11.4&13.4 \\
N2273    & SB(r)a?     &  26.9   & S       		&148.9&   $-$20.74$\pm 0.35$&        $-$21.27 &      10.73$\pm0.16$ &              10.94&       6.89$^{+0.04}_{-0.04}$  &             36.6&     0.37&0.005 &0.32&I&E&---&9.0&10.3&13.9  \\ 
N2276    & SAB(rs)c   &  34.3   & H {\sc{ii}}	&83.5&    	$-$19.13$\pm 0.08$&        $-$21.43 &       9.30$\pm0.09$  &              10.22&        ---&       					   <35.9&    0.39&$-$0.017&0.17&U&   --- &---&9.2&9.2&	13.9 \\  
N2300    &SA0\^0       &  25.7   &ALG		&266.0&	$-$21.33$\pm 0.20$&	 $-$21.94&	11.20$\pm0.18$ & 	       11.40&        ---&	     					   36.2&     0.18&$-$0.007&0.08&I&A&---&9.8&	10.5&	13.9 \\
N2342    & Spec         &  78.6   & H {\sc{ii}} 	&147.3&   $-$19.78$\pm 0.75$&         $-$22.52&        10.02$\pm0.31$ &             10.71&       ---&        				   	   36.5&    0.22&0.012&0.58&I&E&---&9.3&	9.2&	11.4  \\   
N2366    & IB(s)m       &  1.6     & H {\sc{ii}} 	&---&   	 $-$9.21$\pm 0.44$&          $-$16.87&        5.36$\pm0.19$   &             8.43&         ---&      				    	   <33.6&    0.11&0.000 &1.71&U&   ---&---&<9.0&<9.2&	13.3  \\   
N2403    & SAB(s)cd   &  2.6    & H {\sc{ii}} 	&68.4&     $-$19.58$\pm 0.15$&          $-$20.46&       9.45$\pm0.11$   &             9.80&        ---&      					   <34.0&   0.32&$-$0.010&0.00&U&   ---&---&<9.2&<9.8&	13.7  \\  
N2500    & SB(rs)d      &  9.0    & H {\sc{ii}} 	&47.1&     $-$14.39$\pm 0.49$&  	   $-$18.09&        7.82$\pm0.20$   &             9.30&         ---&       					   <34.7&    0.22&0.008 &0.03&U&   ---&BS&9.0&9.3&14.2  \\   
N2537\textsuperscript{$\dagger$}     & SB(s)mpe  &  8.3    & H {\sc{ii}} 	&63.0&   	 	   ---&   	  $-$19.33		 &       --- 			&              10.10&       ---&       		  <34.6&    ---&--- &---&U&   ---&BS	&9.2&	9.8&12.5 \\
N2541    & SA(s)cd     &  9.8    & H {\sc{ii}} 	&53.0&    $-$12.92$\pm0.80$&     	    $-$19.15 &       6.92$\pm0.32$&             9.41 &         ---&    				 	   	   <34.7&    0.33&0.001 &0.88&U&   ---&BS&9.0&9.0&12.5  \\
N2549    & SA0\^0(r)   &  16.6  & ALG  		&142.6&   $-$21.07$\pm0.33$&        $-$21.73 &      10.94$\pm0.14$&            11.21&       7.28$^{+0.37}_{-0.37}$  &          	   <35.3&     0.30&0.006 &0.35&U&   ---&---&9.5&10.2&13.8  \\
N2634    & E1?            &  33.0  & ALG   		&181.1&   $-$19.68$\pm0.38$&       $-$20.53 &      10.19$\pm0.16$&             10.54&       ---&       				           35.8&     0.07&0.004 &0.37&I&A &---&9.8&11.1&13.1\\
N2639    & (R)SA(r)     &  50.3  & L/RL    		&179.3&   $-$20.74$\pm0.37$&       $-$21.97 &      11.22$\pm0.15$&             11.71&        ---&      					  	  37.6&     0.28&0.009  &0.37&I&C&---&9.2&	10.1&	14.2\\
N2655    & SAB0a(s)   &  20.3  & L/RL     		&159.8&   $-$21.08$\pm0.38$&       $-$21.72 &      10.95$\pm0.16$&             11.20&        ---&      					  	  37.6 &     0.33&$-$0.011&0.32&I& E&RS&9.6&10.5&	12.0 \\  
N2681    &SAB(rs)0a   &12.1    &L			&121.0&   $-$19.00$\pm0.63$&	        $-$20.22&      9.90$\pm0.26$&		     10.30&        ---&				           	  35.5&      0.20&$-$0.013 &0.68&I&C&GV&9.5&	10.2&13.9 \\
N2683    & SA(rs)b      &  9.1    & L     		&130.2&   $-$19.64$\pm0.73$&       $-$21.51 &      10.53$\pm0.30$&             11.28 &        ---&  				           	 34.5 &   0.37&0.003 &0.16&I&A&---&9.7&9.6&	12.3 \\
N2685    & (R)SB0\^+ &  14.4   & L     		&93.8&     $-$18.48$\pm0.13$&       $-$19.88 &      10.07$\pm0.07$&             10.63&       6.65$^{+0.41}_{-0.41}$ &        	<34.8 &   0.51&0.024&0.20&U&   ---&BS&9.6&	9.3&	13.0 \\
N2748  & SAbc         &  21.6  & H {\sc{ii}}         &83.0&    $-$18.47$\pm0.78$&        $-$21.18 &    9.72$\pm0.32$&             10.80 &       7.62$^{+0.24}_{-0.24}$  &        	<35.4 &   0.31&0.010&0.24&U&   ---&BS&11.3&11.7&	12.9   \\
N2768    & E6?           &  21.6   & L/RL              &181.8&    $-$21.09$\pm0.59$&        $-$21.27 &      11.42$\pm0.24$&              11.50&       ---&      					  	 37.1 &    0.36&$-$0.006&0.41&I&  A&RS&9.8&	10.2&	12.8\\
N2770    & SA(s)c?     &  31.4   & H {\sc{ii}}       &81.0&     $-$18.62$\pm0.93$&         $-$22.46 &       9.15$\pm0.38$&               10.69&      ---&       					   	<35.5 &    0.48&$-$0.117 &0.02&U&   --- &BS&9.3&9.6&	12.3 \\
N2782    & SAB(rs)a   &  39.7   & H {\sc{ii}}       &183.1&    $-$20.38$\pm0.49$ &        $-$21.51&      10.44$\pm0.20$&              10.89&      ---&       					  36.8 &    0.31&0.004 &0.37&I&A&BS&9.5&10.1&	13.4 \\
N2787    & SB0\^+(r)  &  11.0    & L   	  		&202.0&   $-$18.10$\pm0.31$ &        $-$19.34 &      10.29$\pm0.13$&             10.80&      7.78$^{+0.09}_{-0.09}$  &             36.3 &     0.21&0.006&0.06&I&A &RS&9.7&	9.8&	13.7\\
N2832    & cD2?         & 105.0  & L    		&334.0&   $-$24.58$\pm0.26$ &       $-$24.69 &      12.38$\pm0.11$&              12.43&      ---&       				 	 36.8 &     0.18&$-$0.001 &0.05&I+unI&A&---&10.5&10.7&	13.7 \\
N2841    & SA(r)b?     &  11.6   & L		   	&222.0&   $-$21.08$\pm0.19$ &        $-$21.98 &      11.22$\pm0.09$&              11.58&     ---&      					 34.8 &     0.27&$-$0.001 &0.48&I& C&BS&9.1&8.7&13.5\\
N2859    & (R)SB0\^+ &  27.8  & L         		&188.2&    $-$20.15$\pm0.08$&      $-$20.87 &      11.12$\pm0.05$  &              11.41&     --- &     					 <35.3 &     0.14&0.000&0.51&U&   ---&RS&9.6&9.5&	13.7  \\
N2903    & SAB(rs)b   &  12.1  & H {\sc{ii}}          &89.0&    $-$15.04$\pm1.01$&        $-$22.03 &       8.19$\pm0.41$  &               10.99&    7.13$^{+0.28}_{-0.28}$  &       	 <34.3 &     0.42&0.001 &0.81&U&   ---&BS &9.2&8.9&	12.5 \\
N2950    & (R)SB0\^0 &  20.9  & ALG   		&163.0 &   $-$19.57$\pm0.41$&      $-$20.45 &      10.78$\pm0.17$  &              11.13&     ---&      					 <35.5 &     0.25&$-$0.003 &0.27&U&   ---&---&10.1&9.8&13.0  \\
N2964    & SAB(r)bc   &  22.9  & H {\sc{ii}}     	&109.4 &   $-$17.31$\pm0.22$&       $-$20.85 &       9.27$\pm0.10$  &                10.68&     6.80$^{+0.61}_{-0.61}$  &            36.0 &     0.26&$-$0.002&0.00&I& E&BS&9.6&10.2&13.6 \\
N2976    & SAcpec     &   1.3   & H {\sc{ii}}     	&36.0 &     $-$8.94$\pm0.77$ &        $-$16.64 &       5.67$\pm0.31$  &                8.75&       ---&      					 <33.2 &     0.27&0.006  &1.24&U&   ---&BS&<9.2&<11.1&	12.3   \\
N2985    & (R')SA(r     &  19.8  & L   			&140.8 &   $-$20.70$\pm0.78$&       $-$21.13 &      10.79$\pm0.32$ &              10.96&     ---&    						 35.8 &     0.14&$-$0.003&0.40&I& A&---&9.8&10.1&	12.8   \\
N3031    & SA(s)ab     &   0.7   & L     		&161.6 &   $-$16.62$\pm0.42$&      $-$17.63 &       9.52$\pm0.17$  &                9.92 &       7.09$^{+0.13}_{-0.13}$ &            35.5 &     0.19&0.002 &0.36&I&A&BS&	<9.3&		<9.3&	12.3   \\
N3034    & I0edge-on  &  4.0   & H {\sc{ii}}     	&129.5 &   $-$15.67$\pm0.45$&      $-$21.10 &       8.75$\pm0.19$  &               10.92&      ---&      					 34.3 &     0.19&0.033 &0.37&unI&   ---   &GV&<8.9&<10.5&	12.3 \\
N3073    & SAB0\^-     &  19.1 & H {\sc{ii}}   	&35.6 &    $-$18.14$\pm0.16$&       $-$18.70 &       9.65$\pm0.07$ &                9.87&       ---&  						 <35.3 &     0.34&0.002&0.33&U&   ---&GV&9.9&10.2&	11.5  \\
N3077\textsuperscript{$\dagger$} & I0pec &1.4 & H {\sc{ii}} &32.4 &  --- &    				$-$16.24&      --- &     						 8.98&     ---&       					 33.3 & 	---&---&---& I+unI & A&---&<9.2&<11.6&	12.3 \\
N3079    & SB(s)c edge-on &  18.3 & L           	&182.3 &    $-$20.34$\pm0.32$&       $-$22.33&       10.20$\pm0.12$&               11.00&       6.46$^{+0.05}_{-0.05}$  &           37.3 &     0.24&$-$0.030&0.17&I& C&BS&9.7&	9.7&	11.5    \\
N3184    & SAB(rs)c &  11.4 & H {\sc{ii}}     	&43.3 &     $-$16.62$\pm0.35$&         $-$20.24&       8.72$\pm0.15$&               10.17&      --- &        					<34.6 &      0.30&$-$0.006&0.24&U&   --- & --- & 9.8&9.7&	13.0  \\
\hline
error   & ---&   --- &  ---   &    ---   &   ---  &   0.33& --- &0.17 dex   &  ---&     1.0 dex& 20\%&  35\%&10\%& ---& ---& ---&0.6 dex &0.6 dex &0.5 dex \\
\hline
\end{tabular} 
\end {minipage}
\end{sideways}
\end{table*}
\end{center}

\begin{center}
\begin{table*} 
\begin{sideways}
\setlength{\tabcolsep}{0.079in}
\begin {minipage}{242mm}
\setcounter{table}{0} 
\caption{(\it continued)}
\setlength{\tabcolsep}{2.775pt}
\begin{tabular}{@{}llcccccccccccccccccccccc@{}}
\hline
\hline
Galaxy&Type&$D$&Class&  $\sigma$ &$M_{\rm V,bulge}$&$M_{\rm V,glxy}$&log$M_{\rm *,bulge}$&log$M_{\rm *,glxy}$ &log~$M_{\rm BH}$  &log$L_{\rm R,core}$ &$\langle \epsilon \rangle$& Med ($B_{4}$)&$\gamma$&Det&AGN&Sequence&log$\nu_{10}$&log$\nu_{3}$& log$M_{\rm halo}$\\
&&(Mpc)&&(km s$^{-1}$)&(mag)&(mag)&($M_{\sun}$)&($\rm M_{\sun}$)&($\rm M_{\sun}$)&(erg s$^{-1}$)&&&&&&&($L_{\sun}$ Mpc$^{-2}$)&($L_{\sun}$ Mpc$^{-2}$)&($\rm M_{\sun}$)\\
(1)&(2)&(3)&(4)&(5)&(6)&(7)&(8)&(9)&(10)&(11)&(12)&(13)&(14)&(15)&(16)&(17)&(18)&(19)&(20)\\    
\hline  
N3185    & (R)SB(r) &  22.0 & S     &      79.3 &     $-$19.16$\pm0.39$&         $-$19.94&      10.31$\pm0.17$ &            10.63 &      ---  &       		         <35.2 &      0.20&0.001 &0.37&U&   ---&BS&10.0&	10.6&	12.4    \\
N3190    & SA(s)ape &  21.8 & L     &     188.1&    $-$19.42$\pm0.02$&         $-$21.79 &     10.79$\pm0.04$  &           11.74 &      ---  &      		         <35.3 &  	 0.20&0.010  &0.16&U&   ---  &RS&10.1&	11.1&	12.4  \\
N3193    & E2       &  24.3 & L     &         194.3 &    $-$21.79$\pm0.20$&         $-$21.93 &      11.37$\pm0.09$  &           11.42 &      --- &  	      		 <35.2 &      0.19&0.004 &-0.02&U&   ---&RS&9.9&	10.6&	12.4   \\
N3198    & SB(rs)c  &  12.6 & H {\sc{ii}}  &46.1 &   $-$20.74$\pm0.47$&         $-$21.51 &      10.49$\pm0.19$ &       	 10.79&       ---  &     		         35.0 & 	0.32&0.005&0.33&I&  A&---&9.7&	9.8&	13.0\\
N3245    & SA0\^0(r) &  22.3 & H {\sc{ii}}&209.9 & $-$20.74$\pm0.18$&        $-$21.39 &      10.57$\pm0.08$ &           10.83& 8.40$^{+0.11}_{-0.11}$ & 35.7 & 	0.27&0.003 &0.44&I& E  &---&9.7&9.6&	11.8\\
N3319    & SB(rs)cd &  14.1 & L     &        87.4&     $-$19.43$\pm0.74$&        $-$20.56 &       9.49$\pm0.30$&             9.94 &     ---  &      			 <34.8 &      0.47&0.007 &0.05&U&   ---  &BS&9.7&9.9&	14.0 \\
N3344    & (R)SAB(r &  12.3 & H {\sc{ii}} &73.6 &   $-$17.83$\pm0.79$&       $-$20.82 &       9.25$\pm0.32$&             10.45 &       ---  &      			  <34.1 & 	0.07&0.004 &0.56&U&   --- &BS&9.7&	9.3&	13.1  \\
N3348    & E0       &  41.8 & ALG   &     236.4 &     $-$21.31$\pm0.24$&        $-$21.31 &      11.39$\pm0.11$ &            11.39 &      --- &        			  36.6 &    0.07&0.002 &0.04&I& C&---&9.7&10.1&	12.5\\
N3414    & S0pec    &  24.4 & L   &     236.8 &       $-$20.59$\pm0.15$&         $-$20.77 &      11.33$\pm0.07$ &           11.39 & 8.39$^{+0.07}_{-0.07}$& 36.2 & 	0.20&0.002 &0.64&I& C&RS&10.0&9.9&	13.0 \\
N3448    & I0       &  21.0 & H {\sc{ii}}  &   50.7 &    $-$20.54$\pm0.26$&         $-$21.39 &      10.29$\pm0.11$ &           10.63 &       --- &        		  35.5 & 	0.37&$-$0.016&0.04&I&   A&BS&9.8&9.7&14.0\\
N3486    & SAB(r)c  &  14.1 & S     &      65.0 &     $-$18.58$\pm0.47$&         $-$20.27 &       9.33$\pm0.19$ &            10.01 &        ---  &       		  <34.3 &     0.14&$-$0.002 &0.37&U&   ---&BS&9.3&9.4&	13.2  \\
N3504    & (R)SAB(s) & 26.2 & H {\sc{ii}} &119.3&$-$20.43$\pm0.38$&         $-$20.90 &      10.62$\pm0.16$&           10.81 &     ---  &        		  37.3 &	 0.17&0.005  &0.44&I&A&BS&9.8&	9.8&	13.2  \\
N3516    & (R)SB0\^0 &  37.5 & S     &  181.0 &    $-$21.01$\pm0.39$&         $-$22.36 &      11.05$\pm0.16$ &           11.59 &7.37$^{+0.16}_{-0.16}$&  36.8 &      0.17&$-$0.003 &0.47&I&C&---&9.8&10.1&	13.5 \\
N3600    & Sa?      &  13.2 & H {\sc{ii}} &  49.8 &   $-$16.23$\pm0.56$&        $-$17.57 &       9.22$\pm0.23$  &            9.76 &       ---  &        			 <34.7 &       0.56&$-$0.010  &0.05&U&   --- &BS&10.0&9.8&	12.3  \\
N3610    & E5?      &  25.6 & ALG   &     161.2 &    $-$20.79$\pm0.64$&        $-$20.88&      10.55$\pm0.26$  &           10.59&      --- &       			 <35.6 &      0.33&0.003 &0.38&U&   --- &---&10.0&	10.8&	14.0  \\
N3613    & E6       &  31.7 & ALG   &     220.1 &     $-$21.39$\pm0.34$&        $-$21.99&      10.93$\pm0.14$  &           11.17&       ---   &    			 <35.7 &      0.27&$-$0.002  &0.09&U&   --- & ---&9.8&	10.2&	14.0\\
N3631    & SA(s)c   &  19.2 & H {\sc{ii}} &43.9 &    $-$20.13$\pm0.22$ &        $-$20.32 &      10.15$\pm0.10$  &         10.23&        ---  &    			  <35.2 &   0.05&0.002 &0.27&U&   ---&BS&10.1&	10.4&13.1  \\
N3642    & SA(r)bc? &  24.9 & L    &      85.0   &     $-$18.85$\pm0.03$ &        $-$20.63&       9.41$\pm0.05$  &           10.12 & 7.48$^{+0.04}_{-0.04}$ &<35.5 &    0.14&0.000   &0.61&U&   --- &BS&10.0&	10.7&	14.0  \\
N3665& SA0\^0(s) &  32.1 &H {\sc{ii}}/RL &236.8&$-$24.81$\pm0.69$ &         $-$25.12  &      12.50$\pm0.28$  &        12.60 & 8.73$^{+0.09}_{-0.09}$& 36.8 &   0.23&0.001 &-0.02&I& B&---&9.0&10.4&	12.3 \\
N3675    & SA(s)b   &  13.8 & L    &     108.0 &       $-$15.54$\pm0.71$ &         $-$20.93 &       9.42$\pm0.29$  &          11.57 &  7.31$^{+0.29}_{-0.29}$& 35.0 &    0.30&$-$0.022  &0.61&I&A&---&10.2&9.9&13.5  \\
N3718    & SB(s)a &  16.9 & L/RL    &     158.1 &     $-$19.62$\pm0.31$ &        $-$20.15 &      10.47$\pm0.13$  &         10.68 &       ---  &     			   36.8 &   0.28&$-$0.020   &0.18&I&A &---&10.7&	10.8&	13.1 \\
N3729    & SB(r)ape &  17.8 & H {\sc{ii}}  &76.2 &    $-$16.22$\pm0.37$ &       $-$18.86 &       8.88$\pm0.15$  &           9.94&   ---  & 				   35.8 &   0.24&0.006  &0.21&I& A &BS&10.6&10.7&	13.1\\
N3756    & SAB(rs)b &  21.4 & H {\sc{ii}} &47.6 &    $-$17.81$\pm0.78$ &       $-$20.57 &       9.25$\pm0.32$  &           10.35&    --- &      			   <35.3 &    0.19&0.015   &0.33&U&   ---   &BS&10.7&10.6&	13.9\\
N3838    & SA0a?    &  21.0 & ALG  & 141.4 &  	     $-$18.69$\pm0.39$&         $-$22.11&       10.00$\pm0.16$ &            11.36 &       --- & 		        	  35.4 &   0.36&0.019 &0.74& unI & --- &---&10.4&10.5&	12.7   \\
N3884    &SA(r)0/a&	107.0&L&208.3&  		     $-$21.92$\pm0.29$ &        $-$23.11 &	    11.55$\pm0.13$     &         12.03&&					   37.8&0.15&0.005 & 0.43&I&A&---&10.9&10.9&	14.9\\
N3898    & SA(s)ab  &  19.2 & L  & 206.5 &             $-$20.25$\pm0.59$ &       $-$20.59 &      11.00$\pm0.24$  &            11.13 &        ---  &     	           35.8 &     0.28&0.004   &0.44&I&  A&BS&10.5&10.3&	11.7 \\
N3900    & SA0\^+(r) &  30.2 & ALG   &139.2 &      $-$20.83$\pm0.04$ &       $-$21.61 &      10.74$\pm0.07$  &            11.05&      ---  &     			  <35.6 &0.33&0.006    &0.72&U&   --- &BS&9.0&	8.9&	13.0  \\
N3945    &(R)SB0\^+(rs)   &20.3&L &191.5 &	     $-$17.68$\pm0.38$ &      $-$20.20  &     10.03$\pm0.16$&  		   11.04&6.96$^{+0.47}_{-0.47}$ &35.8& 0.20&0.018& 0.53&I&C&---&10.1&	10.5&	13.4\\
N3949    & SA(s)bc? &  14.5 & H {\sc{ii}}&82.0 &    $-$19.92$\pm0.29$ &      $-$20.43 &       9.62$\pm0.13$  &               9.82 &        --- &       		   <35.0 &  0.35&$-$0.004&0.06&U&   --- &BS&11.0 &10.8&	15.2    \\
N3982    & SAB(r)b? &  18.3 & S     &      73.0 &    $-$19.40$\pm0.34$ &        $-$20.77 &       8.66$\pm0.15$  &              9.21  & 7.01$^{+0.26}_{-0.26}$& 36.2 &     0.15&$-$0.005&0.27&I& A  &---&10.8&10.8&	11.7 \\
N3992    & SB(rs)bc &  17.6 & L   &     148.4 &      $-$18.50$\pm0.39$ &        $-$21.11 &       9.82$\pm0.16$  &             10.86 &  7.57$^{+0.28}_{-0.28}$&  <35.1 &      0.19&0.004   &0.54&U&   ---&---& 10.7&	10.8&	11.7  \\
N3998    & SA0\^0(r) &  17.4 & L/RL &304.6 &      $-$19.27$\pm0.47$ &        $-$20.04 &      10.85$\pm0.19$ &     	   11.16&  9.02$^{+0.05}_{-0.05}$& 38.0 &   0.12&0.003   &0.36&I&A  &---&10.8&10.9&	11.7\\
N4026    & S0edge-on &  16.9 & L  &     177.2 &    $-$22.37$\pm0.41$ &       $-$23.11 &      11.32$\pm0.17$  &             11.61 &   8.36$^{+0.12}_{-0.12}$& <35.1 &0.28&0.017   &0.43&U&   ---&---&11.1&11.1&	13.6   \\
N4036    & S0\^-     &  21.7 & L     &     215.1 &     $-$19.23$\pm0.08$ &        $-$21.45 &      10.50$\pm0.09$  &              11.39 &  7.95$^{+0.36}_{-0.36}$& 36.0 & 0.28&0.007  &0.15&I& C  &---&10.1&10.5&13.4\\
N4041    & SA(rs)bc &  19.5 & H {\sc{ii}} &95.0 &    $-$19.32$\pm0.11$ &      $-$20.14 &       9.85$\pm0.09$  &              10.18&   6.00$^{+0.20}_{-0.20}$& 35.5 &     0.34&$-$0.006   &0.01&I& E&BS&10.4&10.5&12.7\\
N4062    & SA(s)c   &  14.9 & H {\sc{ii}} &  93.2 &   $-$15.35$\pm0.22$ &      $-$20.31&       8.38$\pm0.10$  &              10.36 &    ---  &       			     <34.6 &    0.32&0.004 &0.00&U&   ---   &BS&10.2&9.9&	13.3 \\
N4096    & SAB(rs)c &  11.1 & H {\sc{ii}}& 79.5 &   $-$17.49$\pm0.49$ &      $-$20.94 &       8.99$\pm0.20$  &              10.37 &    ---  &    		             <34.5 &   0.36&$-$0.006  &0.04&U&   ---  &---&10.9&	11.0	&15.2   \\
\hline 
error   & ---&   --- &  ---   &    ---   &   ---  &   0.33& --- &0.17 dex   &  ---&     1.0 dex& 20\%&  35\%&10\%& ---& ---&---&0.6 dex &0.6 dex &0.5 dex \\
\hline
\end{tabular} 
\end {minipage}
\end{sideways}
\end{table*}
\end{center}

\begin{center}
\begin{table*} 
\begin{sideways}
\setlength{\tabcolsep}{0.079in}
\begin {minipage}{242mm}
\setcounter{table}{0} 
\caption{(\it continued)}
\setlength{\tabcolsep}{2.775pt}
\begin{tabular}{@{}llcccccccccccccccccccccc@{}}
\hline
\hline
Galaxy&Type&$D$&Class&  $\sigma$ &$M_{\rm V,bulge}$&$M_{\rm V,glxy}$&log$M_{\rm *,bulge}$&log$M_{\rm *,glxy}$ &log~$M_{\rm BH}$  &log$L_{\rm R,core}$ &$\langle \epsilon \rangle$& Med ($B_{4}$)&$\gamma$&Det&AGN&Sequence&log$\nu_{10}$&log$\nu_{3}$& log$M_{\rm halo}$\\
&&(Mpc)&&(km s$^{-1}$)&(mag)&(mag)&($\rm M_{\sun}$)&($\rm M_{\sun}$)&($\rm M_{\sun}$)&(erg s$^{-1})$&&&&&&&($L_{\sun}$ Mpc$^{-2}$)&($L_{\sun}$ Mpc$^{-2}$)&($\rm M_{\sun}$)\\
(1)&(2)&(3)&(4)&(5)&(6)&(7)&(8)&(9)&(10)&(11)&(12)&(13)&(14)&(15)&(16)&(17)&(18)&(19)&(20)\\    
\hline  
N4102    & SAB(s)b? &  14.7 & H {\sc{ii}} &  174.3 &  $-$18.30$\pm0.44$&         $-$18.77 &       10.60$\pm0.21$  &           10.79 &       ---  &        				    35.9 &    0.13&0.008 &0.36&I&  E&BS&11.1&11.1&	13.6 \\
N4125    & E6pec    &  21.0 & L    &     226.7 &   	 $-$21.14$\pm0.23$&        $-$21.14 &    11.26$\pm0.10$  &         11.26&        --- &        				   <35.4 &    0.24&0.006     &0.28&U&   ---&---&10.0&10.2&	12.7 \\
N4138    & SA0\^+(r) &  16.0 & L     &     120.9 &  	 $-$20.01$\pm0.39$&        $-$20.58 &       10.56$\pm0.16$  &           10.79 &    --- &        				     <35.3 &     0.37&0.003 &0.41&U&   --- &BS&10.5&10.8&	14.5   \\
N4143    & SAB0\^0(s &  16.9 & L     &     204.9 &       $-$20.45$\pm0.34$&        $-$21.22 &      10.92$\pm0.14$  &           11.23&       7.98$^{+0.37}_{-0.37}$  &        36.1 &  0.16&0.005  &0.52&I& B&---&10.2&10.5&14.5\\
N4144    & SAB(s)cd &   6.8 & H {\sc{ii}}  & 64.3 &     $-$10.08$\pm1.10$&         $-$19.38 &       5.54$\pm0.44$  &             9.26&       ---  &        				     <34.0 &    0.46&0.017 &0.57&U&   --- &BS& 9.9&10.5&15.2   \\
N4150    & SA0\^0(r) &   7.1 & L     &      87.0 &    	$-$15.80$\pm0.14$&         $-$17.77 &         9.03$\pm0.08$  &              9.81 &       5.68$^{+0.44}_{-0.44}$&            <34.6 &0.33&$-$0.008  &0.07&U&   --- &---&9.4&9.6&	12.7   \\
N4151    & (R')SAB( &  17.8 & S     &      97.0 &    	$-$19.22$\pm0.84$&         $-$20.85 &       10.14$\pm0.34$  &            10.79 &       7.76$^{+0.08}_{-0.08}$ &         37.8 &     0.11&0.000   &0.81&I& C&---&10.0&10.0&14.1 \\
N4183\textsuperscript{$\dagger$}  & SA(s)cd? &  16.6 & H {\sc{ii}}  &34.4 &   --- &$-$20.54 &      ---  &          				 9.84 &      ---  &       			             <35.2 &    ---&---&---&U&   --- &BS&9.3&10.2&14.5   \\
N4203   &SAB0\^-     &   19.5  &L&    167.0& 		  $-$19.59$\pm0.27$ &        $-$20.44&	     10.90$\pm0.11$&    	11.24&---&						     36.1&0.07&0.005& 0.36&I&A&BS&9.9&9.3&13.9 \\
N4217    & Sb-edge on&  17.7 & H {\sc{ii}}  &  91.3 &    $-$19.88$\pm0.18$ &      $-$21.16 &      10.60$\pm0.08$  &            11.24&        ---  &  		                      35.2 &  0.58&0.009&0.01&I& C &BS&10.3&9.9&13.8 \\
N4220    & SA0\^+(r) &  16.0 & L     &     105.5 &    	   $-$18.25$\pm0.14$ &       $-$19.96 &      10.06$\pm0.08$  &            10.75&        ---  &      				     35.2 & 0.44&$-$0.008&0.38&I&A  &BS&10.6&10.5&	13.8 \\
N4242    & SAB(s)dm &  10.3 & H {\sc{ii}}    &     --- &    $-$11.74$\pm0.19$ &       $-$18.80 &       6.68$\pm0.09$  &            9.50&       ---  &        				      34.4 &  0.71&$-$0.012 &1.68& unI&---&BS& 10.7&10.7&	13.8  \\
N4244    & SA(s)cd? &   7.1 & H {\sc{ii}}      &  36.8 &    $-$14.79$\pm0.09$ &       $-$21.99 &       7.65$\pm0.06$  &            10.53&       ---  &       				      33.8 & 0.40&0.018 &0.78&I& A&BS&10.0&	9.9&	14.1  \\
N4245    & SB0a?(r) &  16.7 & H {\sc{ii}}      &82.7 &    $-$19.30$\pm0.33$ &        $-$20.18 &      10.20$\pm0.14$  &        	10.57&       7.25$^{+0.48}_{-0.48}$ &       <34.6 &  0.32&$-$0.014&0.16&U&   --- &BS&10.5&11.6&	13.2     \\
N4258    & SAB(s)bc &   9.4 & S     &     148.0 & 	   $-$19.94$\pm0.31$ &       $-$21.69 &      10.85$\pm0.13$  &        	11.53&    7.69$^{+0.03}_{-0.03}$ &           34.9 & 0.38&0.008&0.39&I& A &BS&10.0&	10.8&13.8 \\
N4274    & (R)SB(r) &  17.4 & L   &      96.6 &  		  $-$17.93$\pm0.16$ &         $-$20.10 &       9.58$\pm0.10$  &            10.44&       ---  &      				       <34.6 & 0.45&0.004 &0.05&U&   --- &BS&10.3&11.2&	13.2     \\
N4278 &E1-2&15.6&L/RL&237.0					   &$-$20.91$\pm0.27$&	   $-$20.91& 	    11.00$\pm0.12$ &  		 11.00&7.98$^{+0.27}_{-0.27}$&		       37.6&0.16&0.000&0.22&I&A&BS& 9.8&10.7&	13.2\\
N4291 &E&25.5&ALG&293.0 					    &$-$20.71$\pm0.19$& 	   $-$20.71& 	   10.90$\pm0.09$ & 		 10.90&8.97$^{+0.16}_{-0.16}$&		       <35.5&0.24&$-$0.006&0.10&U&   ---&---& 10.0&10.6&13.3   \\
N4314    & SB(rs)a  &  17.8 & L   &     117.0 & 		    $-$20.11$\pm0.22$ &       $-$20.91 &      10.74$\pm0.10$  &             11.12&       6.97$^{+0.30}_{-0.30}$&         <34.7 & 0.19&0.009&0.16&U&   --- &BS&9.5&10.9&13.2   \\
N4414    & SA(rs)c? &  14.2 & L   &     117.0 &		    $-$19.00$\pm0.79$&        $-$20.83&       10.39$\pm0.32$  &             10.77&       ---  &   				       <34.7 &0.30&0.004&0.42&U&   ---&BS&10.5&10.7&	13.9   \\
N4448    & SB(r)ab  &  13.5 & H {\sc{ii}}      &119.8 &    $-$18.89$\pm0.02$ &        $-$19.38 &      10.56$\pm0.05$  &            10.76&      ---  &      				       <34.5 & 0.18&0.012 &0.19&U&   --- &BS&10.0&10.7&13.2  \\
N4449    & IBm      &   6.1 &H {\sc{ii}}    &      17.8 &     $-$16.61$\pm0.54$ &         $-$20.10 &       8.37$\pm0.22$  &         	  9.76&       --- &        			  	        <33.6 & 0.26&$-$0.010 &0.46&U&   ---  &---&9.9&10.6&	14.1  \\
N4485    & IB(s)mpe &  10.3 & H {\sc{ii}}     &  52.2 &   $-$17.58$\pm0.32$ &         $-$18.09 &       8.72$\pm0.14$  &              8.92&       ---  &  			                <34.3 & 0.49&$-$0.018 &0.00&U&   ---&BS&10.7&10.7&14.5     \\
N4490    & SB(s)dpe &  11.4 & H {\sc{ii}}     & 45.1 &    $-$20.05$\pm0.14$ &         $-$21.94 &       9.45 $\pm0.07$ &        	 10.20&     ---  &      				      <34.3 & 0.38&0.003  &0.00&U&   ---  &---&10.3&9.9&	14.5  \\
N4559    & SAB(rs)c &  15.6 & H {\sc{ii}}     & 49.2 &    $-$16.16$\pm0.40$ &         $-$19. 64&       8.06$\pm0.15$  &             9.45&     --- &      				       <34.5 & 0.13&0.035 &0.61&U&   --- &---&10.5&10.0&	14.0  \\
N4565    & SA(s)b?e &  21.8 & S     &     136.0 		&$-$21.71$\pm0.25$ &    	     $-$23.62&      11.16$\pm0.11$&        	  11.93&   --- &        				       35.2 & 0.36&$-$0.002 &0.69&I&A  &BS&9.4&10.0&14.0\\
N4589    &E2	&	 29.2 &L/RL&     224.3		&$-$21.04$\pm0.25$&           $-$21.04&      11.02$\pm0.11$  &       	  11.02&    ---&					        37.5&0.34&0.003&0.30&I&C&---&9.7&	10.0&14.0\\
N4605    & SB(s)cpe &   3.7 & H {\sc{ii}}   &   26.1 &    $-$17.72$\pm0.29$&           $-$18.48&       8.99$\pm0.12$ &         	  9.29 &    ---  &    			 	        <33.8 & 0.49&0.050&0.00&U&   --- &BS&<9.7&<9.5&	13.2  \\
N4648    & E3       &  21.0 & ALG   &     224.5 &  	  $-$18.66$\pm0.39$ &          $-$19.58 &      10.05$\pm0.16$  &            10.42 &  ---  &         				         <35.5 & 0.16&$-$0.001 &0.53&U&   --- &---&9.5&10.3&	14.0 \\
N4656\textsuperscript{$\dagger$}   & SB(s)mpe &  13.0 & H {\sc{ii}} & 70.4 &    --- &$-$21.47 &        ---  &    					     9.77&  ---  &         									<34.3 &    ---&---&---&U&   ---&BS&  10.5&10.5&	14.0  \\
N4736    & (R)SA(r) &   7.6 & L  &     112.0 &    		   $-$19.78$\pm0.77$ &           $-$20.34 &      10.50$\pm0.12$  &            11.13 &       7.00$^{+0.12}_{-0.12}$&       34.8 &  0.11&$-$0.007 &0.01&I&A &BS&10.4&10.7&13.1 \\
N4750    & (R)SA(rs &  24.1 & L     &     136.0 &    	  $-$19.57$\pm0.56$ &            $-$20.89 &     10.23$\pm0.23$  &            10.76 &      --- &       			 	 35.8 &     0.23&0.010&0.37&I& A &BS&9.7&10.1&13.5 \\
N4800    & SA(rs)b  &  15.6 & H {\sc{ii}}    & 111.0 &    $-$19.24$\pm0.26$ &    	          $-$19.41 &      10.36$\pm0.11$  &           10.41 &       7.12$^{+0.53}_{-0.53}$ &      <34.9 &    0.18&$-$0.003  &0.04&U&   --- &BS&10.0&9.3&	13.6  \\
N4826    & (R)SA(rs &  10.0 & L   &  96.0 &    		  $-$19.63$\pm0.33$ &       	 $-$21.45 &      10.50$\pm0.14$  &          11.23 &       6.33$^{+0.13}_{-0.13}$ &      33.9  & 0.28&$-$0.007 &0.25&I& A &BS&9.9&9.5&	13.5 \\
N4914    & cD       &  70.8 & ALG   &224.7 & 		  $-$20.94$\pm0.25$ &       	 $-$20.94  &     11.60$\pm0.11$  &           11.60&        --- &        				<36.2 &   0.31&0.006 &$-$0.01&U&   ---   &---&9.9&10.1&	13.1\\
N5005    & SAB(rs)b &  16.8 & L     &172.0 &  		  $-$21.30$\pm0.64$ &      	$-$21.47 &       11.03$\pm0.26$  &           11.10 &       8.33$^{+0.23}_{-0.23}$ &       36.3 &       0.45&$-$0.005&0.21&I&D&BS&9.5&10.3&	12.7   \\
N5033    & SA(s)c   &  15.8 & L     & 151.0 &  		  $-$20.00$\pm0.44$ &    		$-$22.31 &        9.90$\pm0.18$  &           10.83&        ---  &        				36.0 &   0.31&$-$0.002&0.40&I& A &BS&9.6&10.8&	12.7 \\
N5055    & SA(rs)bc &   9.9 & L    & 117.0 &    		 $-$17.70$\pm0.33$ &       	$-$21.56&         9.51$\pm0.15$  &           11.05 &       8.97$^{+0.11}_{-0.11}$ &       <34.4 &      0.24&0.001 &0.06&U&   ---  &BS&10.2&	10.2&	14.1 \\
\hline  
error   & ---&   --- &  ---   &    ---   &   ---  &   0.33& --- &0.17 dex   &  ---&     1.0 dex& 20\%&  35\%&10\%& ---& ---&---&0.6 dex &0.6 dex &0.5 dex \\
\hline
\end{tabular} 
\end {minipage}
\end{sideways}
\end{table*}
\end{center}

\begin{center}
\begin{table*} 
\begin{sideways}
\setlength{\tabcolsep}{0.079in}
\begin {minipage}{242mm}
\setcounter{table}{0} 
\caption{(\it continued)}
\setlength{\tabcolsep}{2.775pt}
\begin{tabular}{@{}llcccccccccccccccccccccc@{}}
\hline
\hline
Galaxy&Type& $D$&Class&  $\sigma$ &$M_{\rm V,bulge}$&$M_{\rm V,glxy}$&log$M_{\rm *,bulge}$&log$M_{\rm *,glxy}$ &log $M_{\rm BH}$  &log$L_{\rm R,core}$ &$\langle \epsilon \rangle$ & Med ($B_{4}$)&$\gamma$&Det&AGN&Sequence&log$\nu_{10}$&log$\nu_{3}$& log$M_{\rm halo}$\\
&&(Mpc)&&(km s$^{-1}$)&(mag)&(mag)&($\rm M_{\sun}$)&($\rm M_{\sun}$)&($\rm M_{\sun}$)&(erg s$^{-1}$)&&&&&&&($L_{\sun}$ Mpc$^{-2}$)&($L_{\sun}$ Mpc$^{-2}$)&($\rm M_{\sun}$)\\
(1)&(2)&(3)&(4)&(5)&(6)&(7)&(8)&(9)&(10)&(11)&(12)&(13)&(14)&(15)&(16)&(17)&(18)&(19)&(20)\\    
\hline   	
N5112\textsuperscript{$\dagger$}     & SB(rs)cd &  17.0 & H {\sc{ii}}      &      60.8 &    --- 		 &    			        $-$19.51 &       ---  &        				      9.51 &        ---  &      				  <35.4 &---&--- &---&U&   ---&---&9.6&10.8&13.8     \\
N5204    & SA(s)m   &   4.6 & H {\sc{ii}}      &      39.9 &    $-$17.75$\pm0.28$  &       $-$18.31 &       8.67$\pm0.12$  &              8.89 &       --- &        				  <34.1 &0.45&$-$0.011&0.01&U&   --- &BS&<9.7&<10.0&14.2    \\
N5273    & SA0\^0(s) &  18.6 & S     &      		71.0 &      $-$18.86$\pm0.09$  &       $-$20.24 &      10.05$\pm0.06$  &           10.60&       6.69$^{+0.27}_{-0.27}$ &        35.3 &0.11&0.000 &0.55&U&   ---  &RS&9.0&9.2&14.1 \\
N5308    & S0\^-edge on &  30.1 & ALG   &     249.0 &       $-$19.99$\pm0.49$  &  	 $-$21.08&      10.84$\pm0.20$  &            11.27 &       ---  &         				  <35.8 &0.46&0.009  &0.12&U&   --- &---& 10.4&10.7&13.7 \\
N5322 &E3-4 & 27.0&L/RL & 230.0&					$-$21.77$\pm0.24$ &	$-$22.10&	   11.30$\pm0.11$ &  	      11.50&---&						  37.0&0.32&$-$0.004 &0.17&I& B&---&9.8&10.7&13.7\\
N5353    & S0edge-on &  36.1 & L/RL     &     286.4 &    	$-$18.04$\pm0.23$     &   $-$19.72 &      10.41$\pm0.10$  &            11.10 &       --- &       				  37.6 &0.37&$-$0.011 &0.11&I&A &RS&11.4&12.3&	13.6  \\
N5354    & S0edge-on &  39.8 & L/RL     &     217.4 &     	$-$20.51$\pm0.18$  &      $-$21.34 &      10.88$\pm0.08$  &            11.21 &    ---  &      				  36.9 &0.17&$-$0.005 &0.26&I&  A&RS&11.3&10.8&	13.6 \\
N5377    & (R)SB(s) &  28.0 & L     &     169.7 &  		 $-$20.99$\pm0.31$  &      $-$21.77&      11.18$\pm0.15$  &	      11.49 &     ---  & 				          35.7 & 0.23&$-$0.008  &0.22&I&  C&BS&9.9&	10.0&14.4\\
N5448    & (R)SAB(r &  31.1 & L     &     124.5 &    		$-$19.77$\pm0.22$  &      $-$21.84&       9.48$\pm0.10$  &   	      10.93 &   --- &   					  35.7 &0.32&$-$0.008  &0.18&I&  C&BS&	10.0&10.0&14.4 \\
N5457    & SAB(rs)c &   5.2 & H {\sc{ii}}      &      23.6 &     $-$15.72$\pm0.81$  &      $-$17.42&       8.28$\pm0.33$  &              8.96 &       6.41$^{+0.08}_{-0.08}$ &        <34.3 & 0.31&0.002 &0.57&U&   ---&BS&<9.8&<9.6&13.7    \\
N5474    &SA(s)cd pec	 &5.7&H {\sc{ii}} &    29.0& 			$-$16.04$\pm0.39$&        $-$17.58&       8.39$\pm0.16$&     	      9.01&&							 <34.4& 0.29&$-$0.054&  0.00&U&---&BS&9.8&9.9&	11.7\\
N5548    & (R')SA0a &  77.6 & S     &     291.0 &	        $-$21.45$\pm0.42$  &       $-$22.31&      10.76$\pm0.17$  &             11.10 &       7.72$^{+0.13}_{-0.13}$ &     37.0 &0.07&$-$0.005&0.33&I&  B&---&9.7&10.7&	13.1  \\
N5557   &E1&46.4&ALG&259.0 &					$-$22.39$\pm0.32$ &	 $-$22.39& 	   11.40$\pm0.14$ & 	      11.40&---&						 <35.9&0.13&0.000&0.19&U& --- &---&9.6&10.8&13.4\\
N5585    & SAB(s)d  &   5.6 & H {\sc{ii}}      &      42.0 &    $-$14.11$\pm0.33$ &     	 $-$17.19 &      7.35$\pm0.13$  &              8.58&       --- &        				 <34.5 &0.24&0.007&0.01&U&   ---&BS&9.6&9.8&	12.8    \\
N5631    & SA0\^0(s) &  29.3 & L    &     168.1 &   		 $-$20.84$\pm0.19$ &      $-$20.90 &      11.37$\pm0.09$  &            11.40 &       ---    & 			         <35.9 &0.12&0.000 &$-$0.01&U&   ---  &RS& 10.1&10.0&12.8   \\
N5866    & SA0\^+edge &  11.8 & L/RL     &     169.1 & 	$-$19.58$\pm0.32$  &      $-$20.57&      10.39$\pm0.14$  &            10.79 &       ---  &    			         36.5 &0.33&$-$0.001 &0.08&I&D&---&9.4&10.3&	12.9 \\
N5879    & SA(rs)bc &  12.0 & L     &      73.9 &    		$-$18.20$\pm0.95$ &       $-$20.29 &       9.38$\pm0.38$  &            10.21& 6.67$^{+0.28}_{-0.28}$ &           35.3 &0.39&0.008 &0.40&I&  A &BS&9.2&10.7&12.4\\
N5907\textsuperscript{$\dagger$}      & SA(s)c?e &  10.4 & H {\sc{ii}}    &120.2 &---  &    	 $-$22.12&     --- &        					      11.16&        ---  &      35.0 &        ---&---&---& unI &  ---&BS&9.4&10.5&	12.4  \\
N5982    &E3     &44.0&ALG&239.4					&$-$22.08$\pm0.27$&	$-$22.08&	    11.37$\pm0.11$ & 	    11.37&---& 						<35.8&0.28&0.003&0.09&U&---&---&8.9&10.3&13.7\\
N5985    & SAB(r)b  &  36.7 & L     &     157.6 &   		 $-$21.22$\pm0.34$ &      $-$22.70 &      10.78$\pm0.14$&            11.37 &      --- &      				35.8 &0.27&0.013   &0.32&I&D &BS&8.9&9.7&13.7 \\
N6140    & SB(s)cd &  12.9 & H {\sc{ii}}     &      49.4 &   	 $-$17.18$\pm0.19$ &       $-$18.36 &       8.83$\pm0.09$&             9.30&  ---  &  				        <35.2 &0.58&0.000&0.00&U&   ---   &BS&9.6&	9.5&	14.4    \\
N6207    & SA(s)c   &  12.3 & H {\sc{ii}}      &      92.1 &   	 $-$17.36$\pm0.51$  &      $-$20.20 &       8.79$\pm0.21$&              9.92 &       ---  & 				        <35.2 &0.28&0.009&0.10&U&   ---&---&8.4&9.2&	14.2      \\
N6217    & (R)SB(rs &  19.1 & H {\sc{ii}}      &      70.3 &     $-$18.40$\pm0.71$  &       $-$19.85 &       9.48$\pm0.29$&             10.06 &       --- &   				       35.7 &0.18&$-$0.005 &0.02&I&A &---&9.7&9.4&12.4  \\
N6340    & SA0a(s)  &  16.6 & L     &     143.9 &   		 $-$19.78$\pm0.39$  &      $-$20.20 &      10.73$\pm0.16$&          10.90 &        --- &   				       35.5 &0.05&0.001 &0.60&I& A&RS&9.8&10.6&12.8   \\
N6412    & SA(s)c   &  18.1 & H {\sc{ii}}   &      49.9 &    	 $-$17.81$\pm0.19$ &       $-$19.35 &       9.03$\pm0.09$&              9.64&      --- &     				       <35.4 &0.34&0.003 &0.10&U&   --- &BS&10.1&10.0&13.2 \\
N6482    & E?       &  55.7 & L   &     310.4 &    			 $-$22.30$\pm0.19$ &       $-$22.53 &      11.25$\pm0.09$&             11.34 &        --- &     			       36.4 & 0.17&0.006 &0.07&I&  A&---&8.7&8.8&13.1  \\
N6503    & SA(s)cd  &   4.9 & L   &      46.0 &   			 $-$13.75$\pm0.37$ &        $-$19.20 &       7.70$\pm0.15$&             9.88 &       6.27$^{+0.11}_{-0.11}$ &     <34.3 &0.31&$-$0.001 &0.45&U&   ---&---&<9.1&<9.4&	13.9     \\
N6654    & (R')SB0a &  25.0 & ALG   &     172.2 &    		$-$18.61$\pm0.04$  &        $-$20.19 &      10.45$\pm0.05$&            11.08 &     --- &      			      <35.6 &0.12&0.001   &0.38&U&   ---&---&8.5&9.1&	13.2  \\
N6946    & SAB(rs)c &   5.0 & H {\sc{ii}}    &      55.8 &    	$-$16.38$\pm0.31$  &       $-$18.79&         8.27$\pm0.13$&              9.25 & --- &       				      34.4 &0.32&$-$0.019  &0.12&I&E &---&<8.2&<8.7&12.9 \\
N6951    & SAB(rs)b &  18.2 & L    &     127.8 &   		 $-$20.25$\pm0.55$  &      $-$21.51 &       10.33$\pm0.22$&            10.85 & 6.99$^{+0.20}_{-0.20}$ &        35.4 &0.26&0.001 &0.27&I&  C&---& 9.3&8.8&13.1 \\
N7217    & (R)SA(r)ab &   9.0 & L     &     141.4 &    		$-$19.55$\pm0.18$  &       $-$19.86 &       10.60$\pm0.08$&            10.73 &  --- &  				      35.1 &0.08&0.001  &0.22&I&C &---&9.7&10.2&13.6  \\
N7331    & SA(s)b   &   7.0 & L    &     137.2 &   			 $-$17.62$\pm0.47$  &       $-$19.29 &      10.25$\pm0.19$&             10.92 & 7.78$^{+0.18}_{-0.18}$ &       <35.0 &0.34&0.020  &0.44&U&   ---  &---&9.3&9.5&14.1 \\
N7457    & SA0\^-(rs) &   7.2 & ALG   &      69.4 &   		 $-$16.84$\pm0.72$  &       $-$18.11 &       9.58$\pm0.30$&              10.09& 6.71$^{+0.30}_{-0.30}$ &        <34.9 &0.24&0.003 &0.48&U&   ---    &---&9.5&10.2&13.6  \\
N7640    & SB(s)c   &   0.9 &H {\sc{ii}}     &      48.1 &  	 $-$14.45$\pm0.33$  &      $-$16.09 &       7.21$\pm0.14$&              7.86&       --- &       			       <34.5 &0.52&0.011 &0.01&U&   ---   &---&<8.4&<8.0&13.8  \\
N7741    & SB(s)cd  &   5.8 & H {\sc{ii}}     &      29.4 &   	 $-$15.33$\pm0.37$  &       $-$16.00 &       7.87$\pm0.15$&              8.13 &        ---  &     			      <34.7 &0.77&$-$0.005 &0.00&U&   ---   &BS&9.2&9.2&	14.6   \\
\hline
error   & ---&   --- &  ---   &    ---   &   ---  &   0.33& --- &0.17 dex   &  ---&     1.0 dex& 20\%&  35\%&10\%& ---& ---&---&0.6 dex &0.6 dex &0.5 dex \\
\hline
\end{tabular} 
\end {minipage}
\end{sideways}
\end{table*}
\end{center}

\label{lastpage}
\end{document}